\documentclass[prd,onecolumn,notitlepage,
nofootinbib,aps,tightenlines,
preprintnumbers,amsmath,amssymb,amsfonts,showpacs,superscriptaddress]{revtex4-1}

\usepackage[hyperindex,breaklinks]{hyperref}
\usepackage{graphicx,epstopdf}

\usepackage{tikz}
\usetikzlibrary{shapes,arrows}

\usepackage{natbib,bm}
\usepackage{slashed}    
\usepackage{hyperref}
\usepackage{breakurl}
\usepackage{multirow}
\usepackage{bbold}
\usepackage{listings}

\def\be{\begin{equation}}
\def\ee{\end{equation}}
\def\ba{\begin{eqnarray}}
\def\ea{\end{eqnarray}}
\def\ge{\mathrel{\raise.3ex\hbox{$>$\kern-.75em\lower1ex\hbox{$\sim$}}}}
\def\la{\mathrel{\raise.3ex\hbox{$<$\kern-.75em\lower1ex\hbox{$\sim$}}}}

\def\simgt{\mathrel{\raise.3ex\hbox{$>$\kern-.75em\lower1ex\hbox{$\sim$}}}}
\def\simlt{\mathrel{\raise.3ex\hbox{$<$\kern-.75em\lower1ex\hbox{$\sim$}}}}

\newcommand{\nc}{\newcommand}

\nc{\gone}{\bar g_{\pi NN}^{(1)}}
\nc{\gzero}{\bar g_{\pi NN}^{(0)}}
\nc{\al}{\alpha}
\nc{\ga}{\gamma}
\nc{\de}{\delta}
\nc{\ep}{\epsilon}
\nc{\ze}{\zeta}
\nc{\et}{\eta}
\nc{\ka}{\kappa}
\nc{\rh}{\rho}
\nc{\si}{\sigma}
\nc{\ta}{\tau}
\nc{\up}{\upsilon}
\nc{\ph}{\phi}
\nc{\ch}{\chi}
\nc{\ps}{\psi}
\nc{\om}{\omega}
\nc{\Ga}{\Gamma}
\nc{\De}{\Delta}
\nc{\La}{\Lambda}
\nc{\Si}{\Sigma}
\nc{\Up}{\Upsilon}
\nc{\Ph}{\Phi}
\nc{\Ps}{\Psi}
\nc{\Om}{\Omega}
\nc{\ptl}{\partial}
\nc{\del}{\nabla}
\nc{\ov}{\overline}
\nc{\newcaption}[1]{\centerline{\parbox{15cm}{\caption{#1}}}}
\nc{\us}{U(1)$_S$}
\newcommand{\code}[1]{\texttt{#1}}
\mathchardef\mhyphen="2D

\def\beq{\begin{equation}}
\def\eeq{\end{equation}}
\def\bmat{\begin{displaymath}}
\def\emat{\end{displaymath}}
\def\bear{\begin{eqnarray}}
\def\eear{\end{eqnarray}}
\def\ba{\begin{eqnarray}}
\def\ea{\end{eqnarray}}
\def\bery{\begin{array}}
\def\ery{\end{array}}
\def\bit{\begin{itemize}}
\def\eit{\end{itemize}}
\def\ben{\begin{enumerate}}
\def\een{\end{enumerate}}
\def\btab{\begin{tabular}}
\def\etab{\end{tabular}}
\def\btbl{\begin{table}}
\def\etbl{\end{table}}
\def\bfig{\begin{figure}[htb]}
\def\efig{\end{figure}}
\def\bpic{\begin{picture}}
\def\epic{\end{picture}}

\def\ga{\mathrel{\raise.3ex\hbox{$>$\kern-.75em\lower1ex\hbox{$\sim$}}}}
\def\la{\mathrel{\raise.3ex\hbox{$<$\kern-.75em\lower1ex\hbox{$\sim$}}}}
\def\gappeq{\mathrel{\rlap {\raise.5ex\hbox{$>$}}
{\lower.5ex\hbox{$\sim$}}}}
\def\lappeq{\mathrel{\rlap{\raise.5ex\hbox{$<$}}
{\lower.5ex\hbox{$\sim$}}}}

\def\gyr{{\rm \, G\kern-0.125em yr}}
\def\mev{{\rm \, Me\kern-0.125em V}}
\def\gev{{\rm \, Ge\kern-0.125em V}}
\def\tev{{\rm \, Te\kern-0.125em V}}

\begin{document}
\title{Light dark matter in neutrino beams: production modelling and 
scattering signatures at MiniBooNE, T2K and SHiP}

\author{Patrick deNiverville}
\affiliation{Department of Physics and Astronomy, University of Victoria, 
Victoria, BC V8P 5C2, Canada}

\author{Chien-Yi Chen}
\affiliation{Department of Physics and Astronomy, University of Victoria, 
Victoria, BC V8P 5C2, Canada}
\affiliation{Perimeter Institute for Theoretical Physics, Waterloo, ON N2J 2W9, 
Canada}

\author{Maxim Pospelov}
\affiliation{Department of Physics and Astronomy, University of Victoria, 
Victoria, BC V8P 5C2, Canada}
\affiliation{Perimeter Institute for Theoretical Physics, Waterloo, ON N2J 2W9, 
Canada}

\author{Adam Ritz}
\affiliation{Department of Physics and Astronomy, University of Victoria, 
Victoria, BC V8P 5C2, Canada}

\date{September 2016}

\begin{abstract}
\noindent 
We analyze the prospects for detection of light sub-GeV dark matter produced in  experiments designed to study the properties of neutrinos, such as MiniBooNE, T2K, SHiP, DUNE etc. We present an improved production model, when dark matter couples to hadronic states via a dark photon or baryonic vector mediator, incorporating bremsstrahlung of the dark vector. In addition to elastic scattering, we also study signatures of light dark matter undergoing deep inelastic or quasi-elastic NC$\pi^0$-like scattering in the detector producing neutral pions, which for certain experiments may provide the best sensitivity.

An extensive appendix provides documentation for a publicly available simulation tool {\tt BdNMC} that can be applied to determine the hidden sector dark matter production and scattering rate at a range of proton fixed target experiments.
\end{abstract}
\maketitle

\section{Introduction}
\label{sec:intro}

A variety of gravitational signatures, over a range of distance scales, suggest the presence of dark matter (DM). Arguably the simplest realization is in terms of one or more species of weakly-interacting massive particles (WIMPs), and there is currently a broad experimental program aiming to detect dark matter through its non-gravitational interactions. High luminosity fixed target experiments provide a potentially interesting probe of light sub-GeV WIMP dark matter, which is less accessible to underground direct detection.  The use of proton-beam fixed target neutrino experiments \cite{Batell:2009di,deNiverville:2011it,deNiverville:2012ij,Kahn:2014sra,Adams:2013qkq,Soper:2014ska,Dobrescu:2014ita,Coloma:2015pih} and electron-beam fixed target experiments \cite{Izaguirre:2013uxa,Diamond:2013oda,Izaguirre:2014dua,Batell:2014mga} has been highlighted as a way to efficiently probe the parameter space of light dark matter \cite{Alexander:2016aln}. A dedicated beam-dump run was carried out by the MiniBooNE experiment in 2014, based on the proposal \cite{Dharmapalan:2012xp}, and the final results are anticipated soon. (See also \cite{Hewett:2012ns,Kronfeld:2013uoa,Essig:2013lka,pospelov2008,Batell:2009yf,Essig:2009nc,Reece:2009un,Bjorken:2009mm,Freytsis:2009bh,Batell:2009jf,Freytsis:2009ct,Essig:2010xa,Essig:2010gu,McDonald:2010fe,Williams:2011qb,Abrahamyan:2011gv,Archilli:2011zc,Lees:2012ra,Davoudiasl:2012ag,Kahn:2012br,Andreas:2012mt,Essig:2013vha,Davoudiasl:2013jma,Morrissey:2014yma,Babusci:2014sta,Izaguirre:2015yja} for studies of related hidden sectors.) Models of sub-GeV WIMP dark matter generally require light force mediators to ensure efficient annihilation channels that avoid overproduction in the early universe \cite{Boehm:2003hm}, and these new `dark' forces provide the primary production mode in fixed target experiments. 
 
In this paper, we will extend these earlier analyses in two ways. First, we present an improved model for the production of light dark matter in fixed target neutrino experiments such as MiniBooNE and T2K, and potential future experiments such as SBND, SHiP, DUNE and others. We incorporate bremsstrahlung of the light mediator from nucleons, which provides an important production channel when dark matter is too heavy to be produced through the decay of light pseudoscalar mesons. Second, we expand on the potential scattering signatures of light dark matter in the detector, by considering quasi-elastic NC$\,1\pi^0$-like scattering, which is of interest in neutrino oscillation experiments as one of the main backgrounds for $\nu_e$ appearance, and also deep inelastic scattering at higher energy facilities. We will focus our attention on what is regularly taken as the benchmark model of sub-GeV dark matter, which incorporates a dark photon mediator, coupled to the Standard Model (SM) via kinetic mixing with hypercharge (see e.g.~\cite{Pospelov:2007mp,deNiverville:2011it,deNiverville:2012ij,Izaguirre:2015yja}). For comparison, we also consider a leptophobic vector mediator, obtained by gauging the baryon current \cite{Batell:2014yra}.
 
The rest of this paper is organized as follows. In Section \ref{sec:ldm}, we briefly summarize the benchmark sub-GeV dark matter models of interest. In Section \ref{sec:prod}, we present an updated analysis of various production modes for dark matter in proton fixed target experiments, including the associated momentum and angular distributions. In Section \ref{sec:scatter}, we analyze various scattering signatures in the near/far detectors of long-baseline neutrino experiments, including for the first time quasi-elastic single pion production, which may provide a highly efficient channel with lower neutrino backgrounds. Section \ref{sec:sens} provides a comprehensive summary of yields and estimated sensitivities at a range of experiments. We focus on MiniBooNE, T2K, and SHiP, as they appear to provide the best prospects for covering large regions of light dark matter parameter space. We also contrast this reach with other approaches, including underground direct detection that is now reaching the sub-GeV range with recent nucleon scattering limits from CRESST-II \cite{Angloher:2015ewa} and electron scattering limits from XENON10 \cite{Essig:2011nj,Essig:2012yx}. The results are obtained using a comprehensive and flexible production, propagation, and scattering Monte Carlo simulation code {\tt BdNMC}, which has now been made publicly available;\footnote{{\tt BdNMC} software package available at: https://github.com/pgdeniverville/BdNMC/releases} full documentation is provided in the Appendix. We finish with some concluding remarks in Section \ref{sec:outlook}.

\section{Light dark matter}
\label{sec:ldm}

Light sub-GeV dark matter, if assumed to be a thermal relic, requires annihilation channels through light mediators to avoid over-production in the early universe. The three renormalizable `portal' couplings for a SM-gauge neutral field are the natural focal points for model building on effective field theory grounds. To be concrete, we focus on a simple benchmark model in this paper, utilizing the vector portal with a light dark photon mediator, but there are other possibilities, and we will also analyze a leptophobic mediator for comparison. 

\subsection{Vector portal light dark matter benchmark}

A spontaneously broken $U(1)'$ gauge symmetry in a hidden sector, leading to a massive dark photon $V_\mu$, provides a minimal light dark matter model with SM interactions obtained via kinetic mixing with the hypercharge gauge boson~(see e.g.~\cite{Holdom:1985ag,Pospelov:2007mp,Batell:2009di,deNiverville:2011it}). The dark matter candidate $\ch$ can then be taken as a scalar or fermionic field charged under $U(1)'$,
\begin{align}
\label{eq:L1}
{\cal L} &=  {\cal L}_\chi - \frac{1}{4}V_{\mu\nu}V^{\mu\nu} + \frac{1}{2}m_V^2 V_\mu V^\mu - \frac{\ep}{2} V^{\mu\nu} F_{\mu\nu} + q_Bg' V_\mu J_B^\mu +\cdots \;\;\;\;\; 
\end{align}
with
\begin{align}
{\cal L}_\chi & = 
\begin{cases}
i \bar \chi \not \!\! D \chi - m_\chi \bar \chi \chi,  ~~~~~~~ ({\rm Dirac ~ fermion ~ DM})\\
|D_\mu \chi|^2 - m^2_\chi |\chi|^2,~~~~({\rm Complex ~ scalar ~ DM})
\end{cases} \nonumber 
\end{align}
where the $U(1)'$ covariant derivative takes the form $D_\mu = \partial_\mu - i q' g' V_\mu$, with $g'$ the $U(1)'$ gauge coupling (and throughout we set the charge $q'=1$ for the DM field $\chi$). We have allowed for the possibility that the $U(1)'$ charge is identified with baryon number, giving two mediation channels:
\begin{enumerate}
\item {\it Vector Portal} - the pure vector portal is obtained by setting $q_B=0$ and $q'=1$ in the Lagrangian above. The couplings to the mediator then take the form, ${\cal L} \supset V^\mu (g' J_\mu^{\chi}-\ep e J_\mu^{\rm EM})$ with $J_\mu^{\rm EM} = \sum_{f} Q_f {\bar \psi}_f \gamma_\mu {\psi}_f$ summed over the fermions with electric charge $Q_f$. This mediation channel is entirely UV complete.
\item {\it Baryonic Vector Mediator} - the limit of a gauged baryon number current is obtained by setting $\ep=0$, $g'\rightarrow g_B$, with $q_B=1$ (which assigns charge 1 to nucleons and 1/3 to quarks) and $q'=1$ as before. The couplings to the mediator then take the form ${\cal L} \supset V^\mu ( g_B J_\mu^\chi - g_B J_\mu^B)$ with $J_\mu^B = \frac{1}{3} \sum_i {\bar q}_i \gamma_\mu {q}_i$ given by a sum over all quark species.  This mediation channel requires an additional UV completion to ensure anomaly cancellation.
\end{enumerate}
For simplicity, we will focus on scalar DM here, as this is less constrained by astrophysical considerations, such as annihilation-induced distortion of the CMB. However, the fermionic DM model can be modified to avoid these constraints by adding dark Higgs couplings that provide a sufficient mass splitting between the pseudo-Dirac components. (See Refs. \cite{Beacom:2004pe,Madhavacheril:2013cna,Wilkinson:2016gsy} for a more extensive discussion of the astrophysical constraints on light dark matter.) A number of terrestrial particle physics constraints on this model will be summarized in more detail in later sections.
We will focus just on the two extreme mediation cases independently below, with either $q_B=0$ (pure vector portal), or $\ep=0$, $g'=g_B$ (leptophobic mediator). Note that, generically, even if one sets $\ep=0$ at a high scale, it will be generated through loops when $q_B\neq 0$, but generally at a subleading level.

The minimal vector portal model provides a natural thermal relic dark matter candidate, and thus the $s$-channel annihilation rate in the early universe, $\langle \sigma_{\rm ann} v\rangle (\chi + \chi^\dagger \rightarrow f + \bar{f})$, determines the freeze-out abundance. This imposes a bound on the parameter space to ensure that the relic abundance is not too large, and a one-dimensional constraint on the parameters $\{\ep,\al'\equiv g'^2/(4\pi),m_V,m_\chi\}$ when the relic abundance matches the observed DM abundance, given by $\langle \si_{\rm ann} v \rangle \sim 1\,$pb. Details of the calculation of the annihilation rate in this model are given in \cite{deNiverville:2012ij}. For the baryonic mediator, further model building is required for a consistent model of dark matter \cite{Batell:2014yra}, and so we will instead use a weak-scale benchmark by equating $4\pi \al_B/m_V^2$ to $G_F$ (where $\al_B\equiv g_B^2/(4\pi)$) in order to quantify the performance of different 
experiments. In practice, this model is useful for distinguishing the leptonic vs hadronic sensitivity of different detection channels and experiments.

\section{Production in proton fixed target experiments}
\label{sec:prod}

\noindent We will account for a number of light dark matter production modes:

\begin{itemize}
 \item $\pi^0/\eta$ decay in flight - relevant for low $m_V$.
 \item Bremsstrahlung, with resonant vector meson mixing - relevant for intermediate $m_V$
 \item Direct production from quark and gluon constituents - relevant for higher $m_V>1\,$GeV.
\end{itemize}

\begin{figure}[t]
 \centerline{\includegraphics[width=0.8\textwidth]{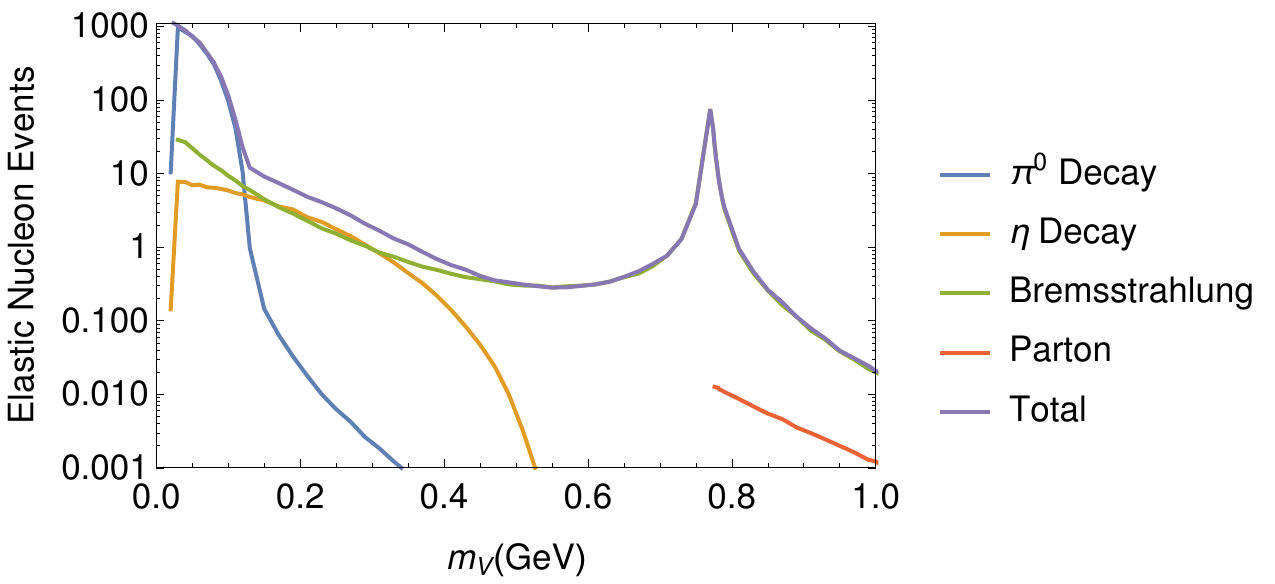}}
 \caption{A plot illustrating the distinct contributions to DM production (coupled through the vector portal), as discussed in the text, using the 9~GeV proton beam at MiniBooNE as an example. The rate of elastic scattering events on nucleons is plotted versus the vector mediator mass. From smaller to larger values of $m_V$, the dominant channels are $\pi^0$ decays, $\eta$ decay, bremsstrahlung, which becomes resonant near the $\rho/\omega$ mass region, and finally direct parton-level production. The plot uses $m_\chi = 0.01$ GeV, $\ep = 10^{-3}$ and $\alpha^\prime = 0.1$.}
 \label{fig:prodn}
\end{figure}

\noindent We summarize these channels below, while Fig.~\ref{fig:prodn} shows the relative contributions at MiniBooNE.

\subsection{$\pi^0/\eta$ decay in flight}

\noindent Radiative $\varphi$ = $\pi^0$, $\eta$ decay provides the dominant production channel for sufficiently light DM and mediators,
\be
 \pi^0/\eta \longrightarrow \gamma + V^{(*)} \longrightarrow \gamma + \chi^\dagger + \chi.
\ee
If kinematically allowed, on-shell $V$-production is dominant, but we also include off-shell $V^*\rightarrow \chi^\dagger \chi$ decays, 
\be \label{eq:offshell}
  \Gamma_{\varphi\rightarrow \gamma \ch^\dagger \ch} = \frac{1}{4\pi m_\varphi} \int d\Pi_{\varphi\rightarrow \gamma V} d\Pi_{V\rightarrow \ch^\dagger \ch} dq^2  |{\cal M}|^2.
  \ee
Here $d\Pi$ is the 2-body phase space, and \cite{Kahn:2014sra}
  \be
  |{\cal M}|^2 =  \frac{c^\varphi_{V,B} k^{(1)}_{V,B}\al f(q^2,p\cdot k_1,p\cdot k_2)}{\pi f_\varphi^2 [(q^2-m_V^2)^2+m_V^2\Gamma_V^2]},
\ee
where $f{=}(q^2{-}4m_\ch^2)(m_{\varphi}^2{-}q^2)^2{-}4q^2(p\cdot k_1{-}p\cdot k_1)^2$, $c^\varphi_V= c^\pi_B = 1$, $c^\et_B=0.61$ (with $q_B=1$), and for later convenience we have defined the coupling combination,
\be
 k^{(n)}_{V,B} = \left\{ \begin{array}{ll} \ep^2\al(\al')^n & {\rm for}\,U(1)' \\ \al_B^{n+1} & {\rm for}\,U(1)_B \end{array}\right.. \label{kVB}
\ee 
In these expressions $p$ is the photon momentum, $q$ the momentum of $V$, and $k_{1,2}$ the momenta of the dark sector particles in the final state, so that $q=k_1+k_2$. It should be noted that, in this approximation, the mesons are treated as elementary but a form factor could be incorporated to account for the virtuality dependence.

\begin{figure}[t]
 \centerline{\includegraphics[width=0.43\textwidth]{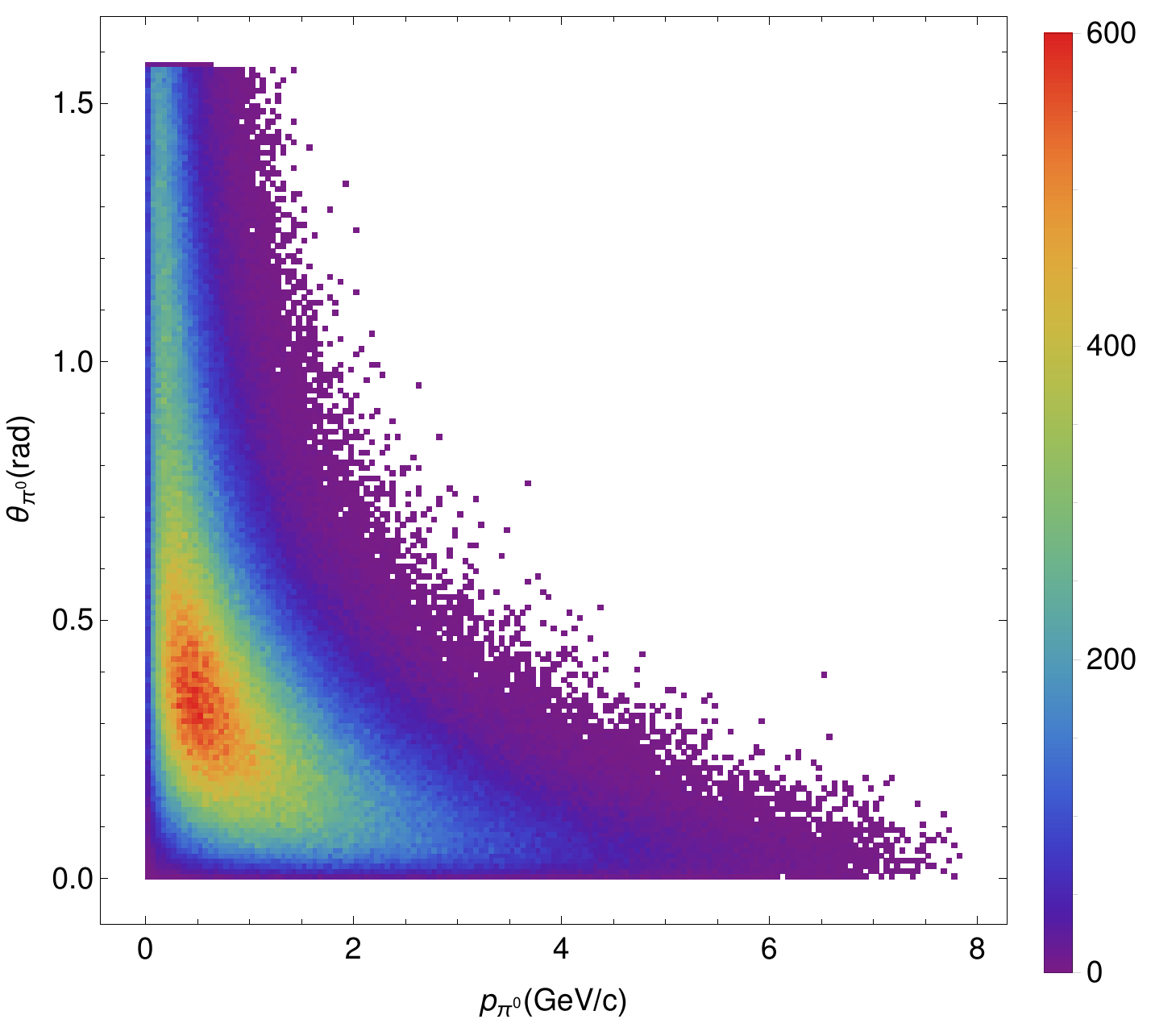}\hspace*{1cm}\includegraphics[width=0.45\textwidth]{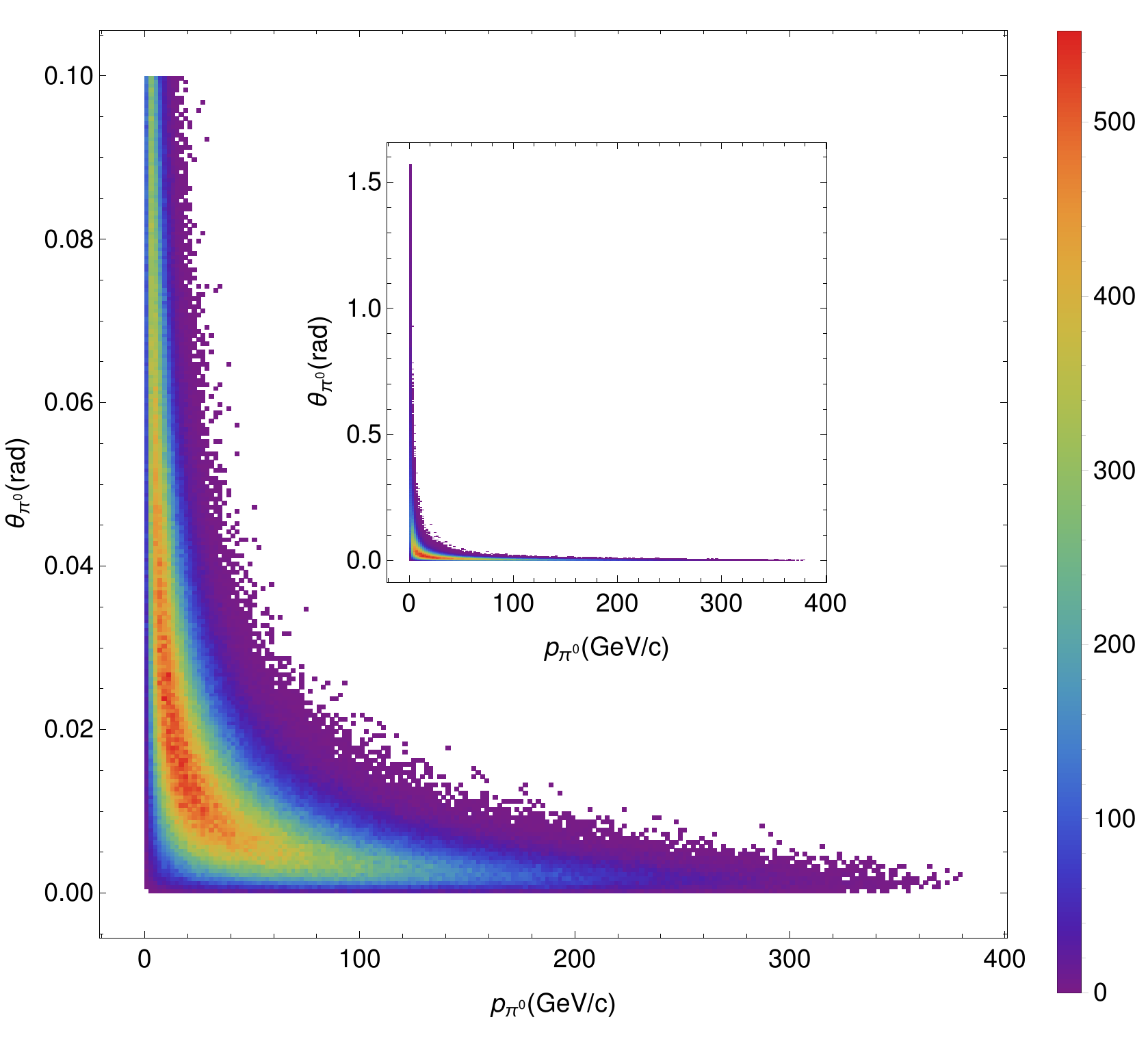}}
 \caption{Plots showing two-dimensional momentum-angle distributions for $\pi^0/\eta$ production at the MiniBooNE beam energy of 9~GeV  (left, Sanford-Wang), and at the SPS energy of 400~GeV (right, BMPT). The colour bars indicate the relative scale.}
 \label{fig:dist}
\end{figure}

The production distributions vary depending on beam energy, and we make use of the Sanford-Wang distribution for the 9~GeV beam at MiniBooNE \cite{sanfordwang_miniboone} (more precisely the average of the $\pi^+$ and $\pi^-$ distributions), and the distribution determined in \cite{Bonesini:2001iz} (denoted BMPT) for experiments with higher beam energies (see Fig. \ref{fig:dist}).
We use similar angular distributions for $\pi^0$ and $\eta$, but account for the lower production rate, $N_{\pi^0} \approx 30 N_{\eta}$ \cite{Teis:1996kx} for the beam energies considered here. 

\subsection{Bremsstrahlung and resonant vector meson mixing}

\noindent For higher mass mediators and dark matter, additional production channels become important. Particularly for on-axis detectors, the 
forward-peaked emission of $V$ particles through proton bremsstrahlung,
\begin{equation}
p +N \to p + N + V  \nonumber,
\end{equation}
with $N=p$ or $n$, could provide a significant dark matter source with very little angular spread. The treatment here follows and elaborates upon previous work on vector boson production via proton bremsstrahlung \cite{Blumlein:2013cua,Gorbunov:2014wqa}. In particular, we account for the timelike form-factor which naturally incorporates off-shell mixing with vector mesons such as $\rho$ and $\omega$. This contribution was added by hand  in the on-shell limit in our earlier analysis \cite{Batell:2014yra}. The approach \cite{Blumlein:2013cua} rests upon the assumption of $p-N$ scattering being dominated by exchange in the $1^{--}$ channel, which is consistent with pomeron exchange, but with validity limited only to high energies.  

We define the four-vectors for the incident proton and emitted $V$ as $p = (E_p,0,0,P)$ with $E_p =P+m_p^2/(2P)$, and $p_V = (E_V, p_\perp \cos(\phi), p_\perp \sin(\phi), zP)$ with $E_V=zP + (p_\perp^2+m_V^2)/(2Pz)$, where $p$ is the momentum of the incident proton, $p_V$ is the momentum of the outgoing $V$ boson, $z$ is the fraction of the beam momentum carried by the outgoing $V$ and 
$p_\perp$ is the $V$ momentum perpendicular to the beam momentum.
The differential $V$ production rate can be written as
\begin{equation}
\label{eq:bremrate}
	\frac{d^2 N_V}{dz dp^2_\perp} = \frac{\sigma_{pA}(s^\prime)}{\sigma_{pA}(s)} F^2_{1,N}(q^2) w_{ba}(z,p_\perp^2),
\end{equation}
where $s^\prime=2m_p(E_p - E_{V})$, $s=2m_p E_p$ and the photon splitting function is \cite{Blumlein:2013cua}
\begin{equation}
\begin{aligned}
\label{eq:wba}
	w_{ba}(z,p^2_\perp) =& \frac{k^{(0)}_{V,B}}{2\pi H}\Bigg[ \frac{1+(1-z)^2}{z} -2z(1-z)\left(\frac{2m_p^2+m_V^2}{H}-z^2\frac{2m_p^4}{H^2}\right) \nonumber\\
	 &\qquad\qquad+2z(1-z)(z+(1-z)^2)\frac{m_p^2 m_V^2}{H^2}+2z(1-z)^2\frac{m_V^4}{H^2}\Bigg],
\end{aligned}
\end{equation}
with $H = p_\perp^2 + (1-z)m_V^2+z^2 m_p^2$, and $k^{(n)}_{V,B}$ was defined above in (\ref{kVB}).

The radiated $V$ has timelike momenta, and the timelike form factor $F_{1,N}(q^2)$ will necessarily incorporate mixing with vector mesons in the appropriate kinematic region. Utilizing just the leading charge form factor, only $F_{1,p}$ contributes for the vector portal, while both $F_{1,p}$ and $F_{1,n}$ contribute for the baryonic channel. Existing fits in the low invariant mass regime are limited, and we make use of the form-factor given by \cite{Faessler:2009tn} which does not fully resolve the $\rho$ and $\omega$ contributions. For the vector portal, the form factor $F_{1,p}(q^2)$ incorporates both isovector ($\rho$-like) and isoscalar ($\om$-like) Breit-Wigner components, directly following \cite{Faessler:2009tn}. For the baryonic portal, we retain only the isoscalar components, as described in \cite{Batell:2014mga} in the spacelike regime.
 The fit to data is good on the side-bands of the resonance peak, but appears to be dominated primarily by $\omega$ very close to the peak of the resonance region, which is somewhat unphysical for the vector portal (although it works well for the baryonic mediator where only $\omega$ contributes). Comparing to the production rate from on-shell vector meson decay considered previously, we observe a slight under-estimate of the rate away from the resonance peak, and likely an overestimate in a very small band around the peak of the form-factor (see Fig.~3). The form factor suppresses the rate for virtualities much above 1~GeV, where the internal nucleon structure is probed, and direct parton-level production takes over for higher mass vectors.
  
\begin{figure}[tb]
\label{fig:bremform}
\centerline{\includegraphics[width=0.5\textwidth]{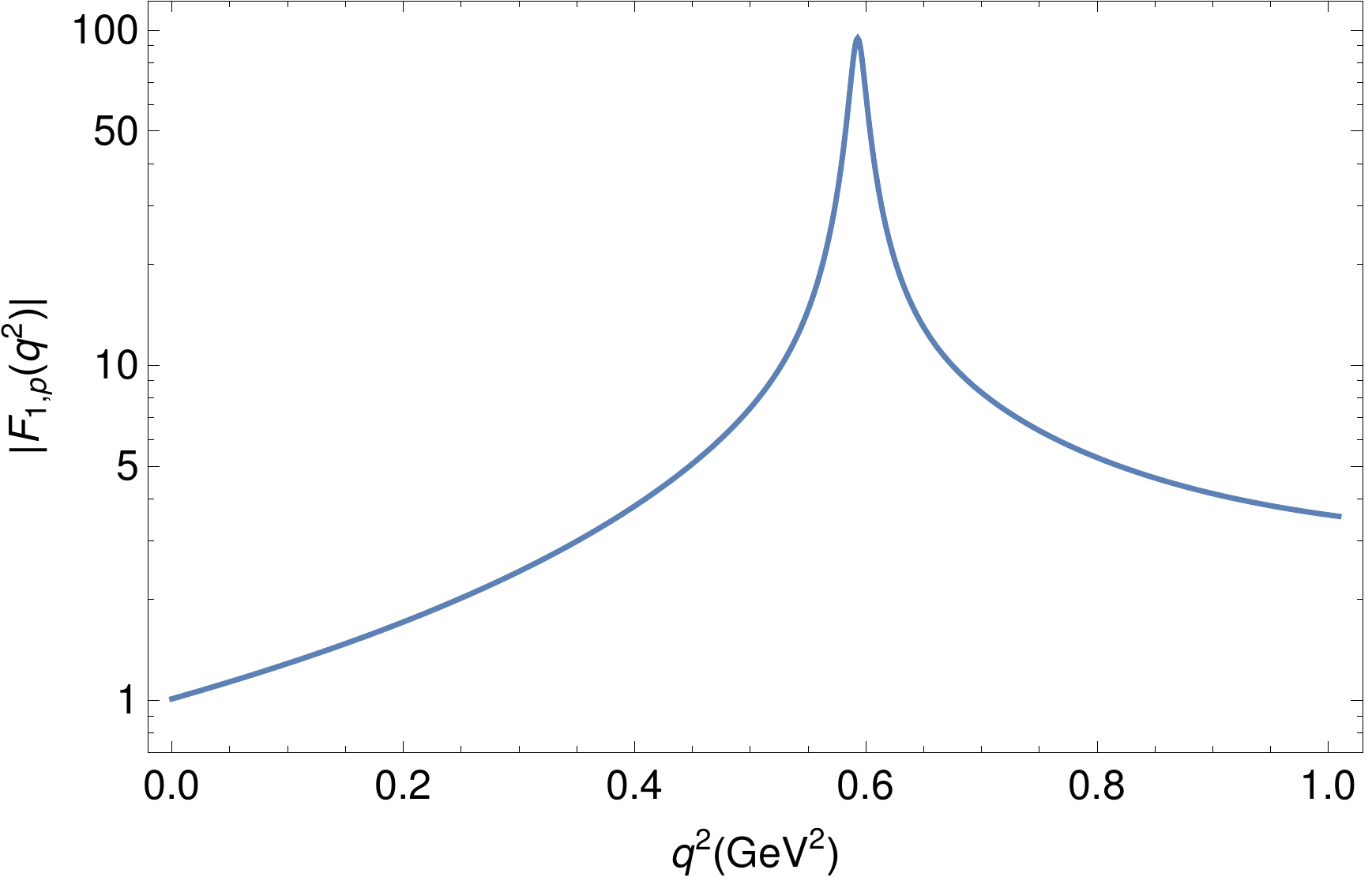}}
\caption{The timelike form factor $F_{1,p}(q^2)$ from \cite{Faessler:2009tn}. The resonant enhancement around the $\rho/\omega$ region is not fully resolved in the fit. }
\end{figure}

Note that several kinematic conditions must hold to make use of this calculation, summarized in \cite{Blumlein:2013cua} as
\begin{equation}
\label{eq:bremcond}
E_p,E_V,E_p-E_V \gg m_p, m_V, |p_\perp|.
\end{equation}
In order to calculate the total $V$ production rate, and therefore the dark matter signal $\chi$, (\ref{eq:bremrate}) must be integrated over $z$ and $p_\perp$ with ranges chosen such that (\ref{eq:bremcond}) are satisfied. These conditions are easily satisfied by $z\in[0.1,0.9]$ and $|p_\perp|<1$ GeV for high energy experiments such as SHiP, but 
less so at lower beam energies. For MiniBooNE, the energy of the incident and outgoing proton, 
as well as the energy of the emitted $V$, $E_V$, must be significantly larger than $1\,\mathrm{GeV}$. We have chosen a range $z\in[0.3,0.7]$ for MiniBooNE and $|p_\perp| < 0.2$ GeV, which should allow some margin for error. A more complete calculation for MiniBooNE could be based on the meson-nucleon 
exchange model \cite{Serot:1984ey}.

\subsection{Direct production}
\label{ssec:parton}

\noindent Direct production corresponds to parton-level processes such as
\begin{equation}
	p + N \to V^* \to \chi^\dagger \chi
\end{equation}
and following \cite{deNiverville:2012ij} we use the narrow width approximation, so that
\begin{equation}
	\sigma(pN \to V^* \to \chi^\dagger \chi) = \sigma(pN \to V)\mathrm{Br}(V\to \chi^\dagger \chi).
\end{equation}
In this expression
\begin{align}
\sigma\left(pN\to V\right)&=\int_{\tau}^1dx\ \frac{d\sigma\left(pN\to V\right)}{dx}
\nonumber
\\
&=\frac{4\pi^2}{m_V^2}\sum_q Q^2_q k^{(0)}_{(V,B)}\int_{\tau}^1 \frac{dx}{x}\ \tau\left[f_{q/p}\left(x\right)f_{\bar q/p(n)}\left(\frac{\tau}{x}\right)+f_{\bar q/p}\left(x\right)f_{q/p(n)}\left(\frac{\tau}{x}\right)\right],
\label{eq:sigmaV}
\end{align}
where $k^{(n)}_{V,B}$ was defined above in (\ref{kVB}), $Q_q$ is the quark electric charge (or 1/3 in the baryonic case), $\tau=m_V^2/s$, and $\sqrt s$ is the hadron-level center-of-mass energy. 
To obtain estimates, we use CTEQ6.6 PDFs~\cite{cteq} and set $Q=m_V$; varying $Q$ in between $m_V/2$ and $2m_V$ resulted in an uncertainty in the production cross section of less than $\sim 30\%$ for $m_V>1~{\rm GeV}$ at T2K and SHiP beam energies.  However, higher-order QCD corrections are larger for the lowest mass considered for direct production, $M_V \sim 1$~GeV, introducing an error that can potentially be ${\cal O}(1)$.

The production cross section as a function of the DM lab frame energy, $E_\chi$, and the angle between its lab frame momentum and the beam direction, $\theta$, can be related to the differential cross section in Eq.~(\ref{eq:sigmaV}) through
\begin{align}
\frac{d\sigma\left(pN\to V\to\bar\chi\chi\right)}{dE_\chi d\cos\theta}=\left[\frac{\partial(x,\cos\hat\theta)}{\partial(E_\chi,\cos\theta)}\right]\frac{d\sigma\left(pN\to V\right)}{dx}~{\rm Br}\left(V\to\bar\chi\chi\right)g\left(\cos\hat\theta\right),
\label{eq:com_to_lab}
\end{align}
where $\hat\theta$ is the angle between the momentum of $\chi$ and the beam in the $V$ rest frame and the quantity in square brackets is the Jacobian associated with this variable change. The function $g$ describes the angular distribution of the DM in the $V$ rest frame. For scalar DM produced through a vector mediator, this is $g\left(\cos\hat\theta\right)=\frac{3}{4}\left(1-\cos^2\hat\theta\right)$.

\section{Elastic and $NC\pi^0$-like scattering}
\label{sec:scatter}

\noindent We will analyze a number of scattering signatures, listed below, including for the first time the inelastic $NC\pi^0$-like scattering of light dark matter off nucleons:

\begin{itemize}
\item Electron scattering 
\item Elastic $NC$-like scattering off nucleons
\item Inelastic $NC1\pi^0$-like scattering 
\item Deep inelastic scattering off nucleons
\end{itemize}

\noindent We summarize the scattering cross-sections for all these signatures below.

\begin{figure}[t]
\centerline{\includegraphics[viewport=160 520 450 720, clip=true, scale=0.4,width=0.3\textwidth]{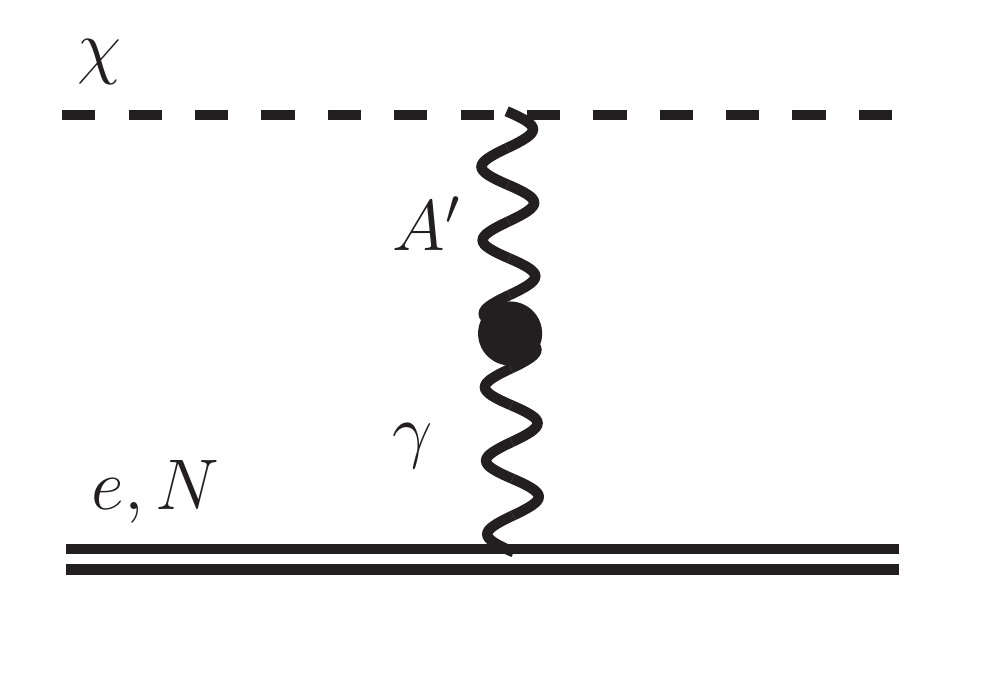}\includegraphics[viewport=160 520 450 720, clip=true, scale=0.4,width=0.3\textwidth]{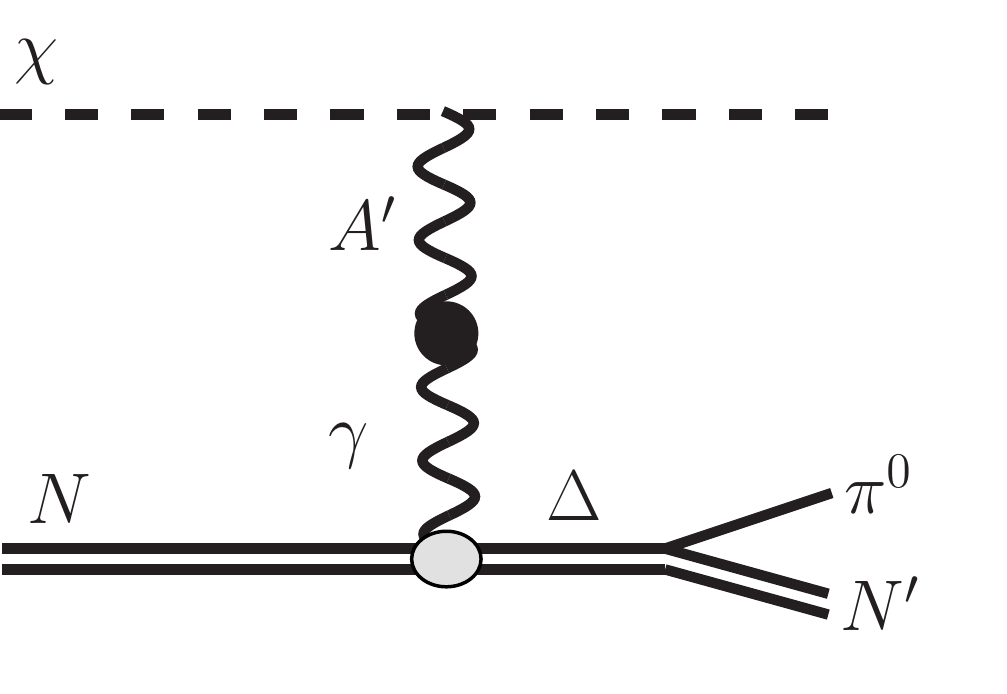}\includegraphics[viewport=160 520 450 720, clip=true, scale=0.4,width=0.3\textwidth]{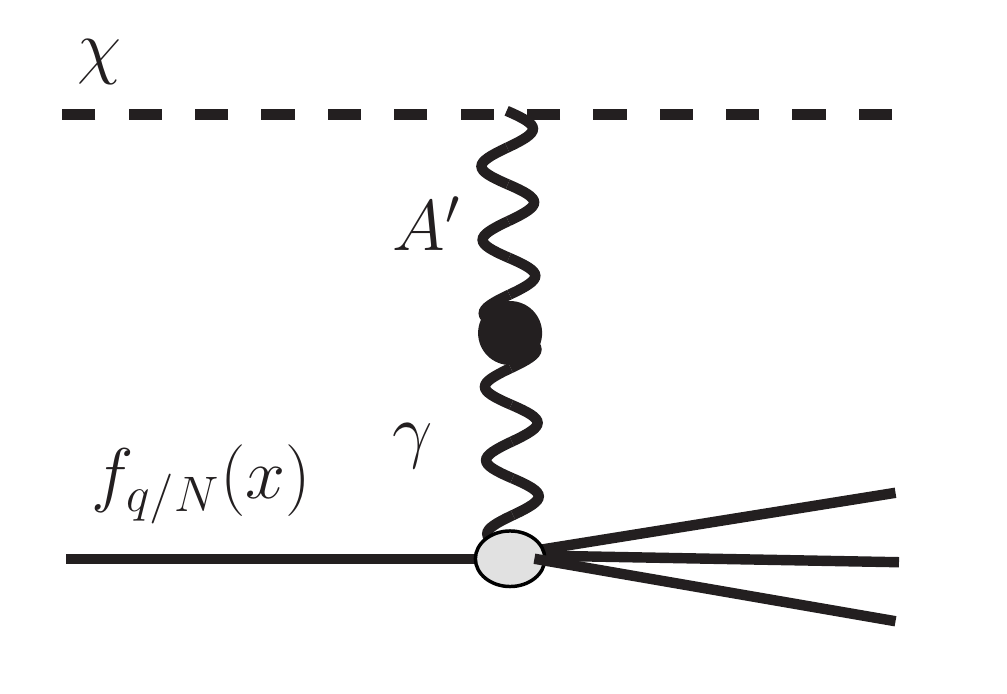}}
  \caption{Scattering channels analyzed: from left to right, elastic scattering on electrons or nucleons, quasielastic (incoherent) single pion production, and deep inelastic scattering. 
  }
  \label{fig:scatter}
\end{figure}

\subsection{Elastic electron scattering}

For elastic scattering off electrons, which is relevant only for the vector portal with $\ep\neq 0$, the cross section has the form \cite{deNiverville:2011it}
\be \label{eq:e-scatter}
 \frac{d\si_{\ch e }}{dE_{e}} = 4\pi k_V^{(1)} \frac{2m_e E_\ch^2{-}(2m_e E_\ch + m_\ch^2) (E{-}m_e)}{(E^2-m_\ch^2)(m_V^2+2m_e E_e-2m_e^2)^2},
 \ee
 where $E_e$ is the energy of the recoiling electron, and $E_\chi$ is the energy of the incoming DM particle.

\subsection{Elastic $NC$-like nucleon scattering}

For incoherent scattering off nucleons, the leading contribution to the cross section has the form,
\be \label{eq:scatter}
 \frac{d\si_{\ch N}}{dE_{\ch}} = 4\pi k_{V,B}^{(1)} Q^2_N G_D(Q^2) \frac{2m_N E E_\ch{-}m_\ch^2 (E{-}E_\ch)}{(E^2-m_\ch^2)(m_V^2+Q^2)^2}{+}\cdots
 \ee
where again $Q_N$ is the nucleon electric charge (or unity in the baryonic case), $E_\ch$ the energy of the recoiling DM particle, $Q^2=2m_N(E-E_\ch)$ is the momentum transfer, 
and $G_D(Q^2)$ is the Sachs form-factor, $G_D(Q^2) = 1/(1+Q^2/M^2)^2$ with $M=0.843\,$GeV. 
Further dipole form factor terms, which are generally subleading (for protons), are suppressed to simplify the presentation, although they are included in the final results  (see \cite{deNiverville:2011it,Batell:2014yra} for full details). The resulting nuclear scattering cross section will be discussed later in Section~5.

\subsection{Inelastic $NC\pi^0$-like nucleon scattering}

Of primary interest here is the possibility of inelastic scattering where there is sufficient momentum transfer to produce a neutral pion which subsequently decays producing a two-photon signature. This is one of the main backgrounds for $\nu_\mu \rightarrow \nu_e$ appearance, and thus of interest for all long baseline experiments. We will focus on what is referred to as `incoherent' $NC\pi^0$, in which the pion emerges via the production of a $\De(1232)$ resonance in the scattering process,
\be
 \ch + N \rightarrow \ch + \Delta (\rightarrow N + \pi^0),
\ee
with the subsequent decay $\Delta \to N\pi^0$ having an ${\cal O}(1)$ branching ratio \cite{pdg}. 
 As for neutrino scattering, there is also the possibility for `coherent' single pion production where a sufficiently low momentum (off-shell) $V$ interacts coherently with the nucleus, but this is expected to be subleading at MiniBooNE energies \cite{MBpi0}. It would be interesting to analyze this channel further in the future.

\begin{figure}[t]
\centerline{\includegraphics[width=0.5\textwidth]{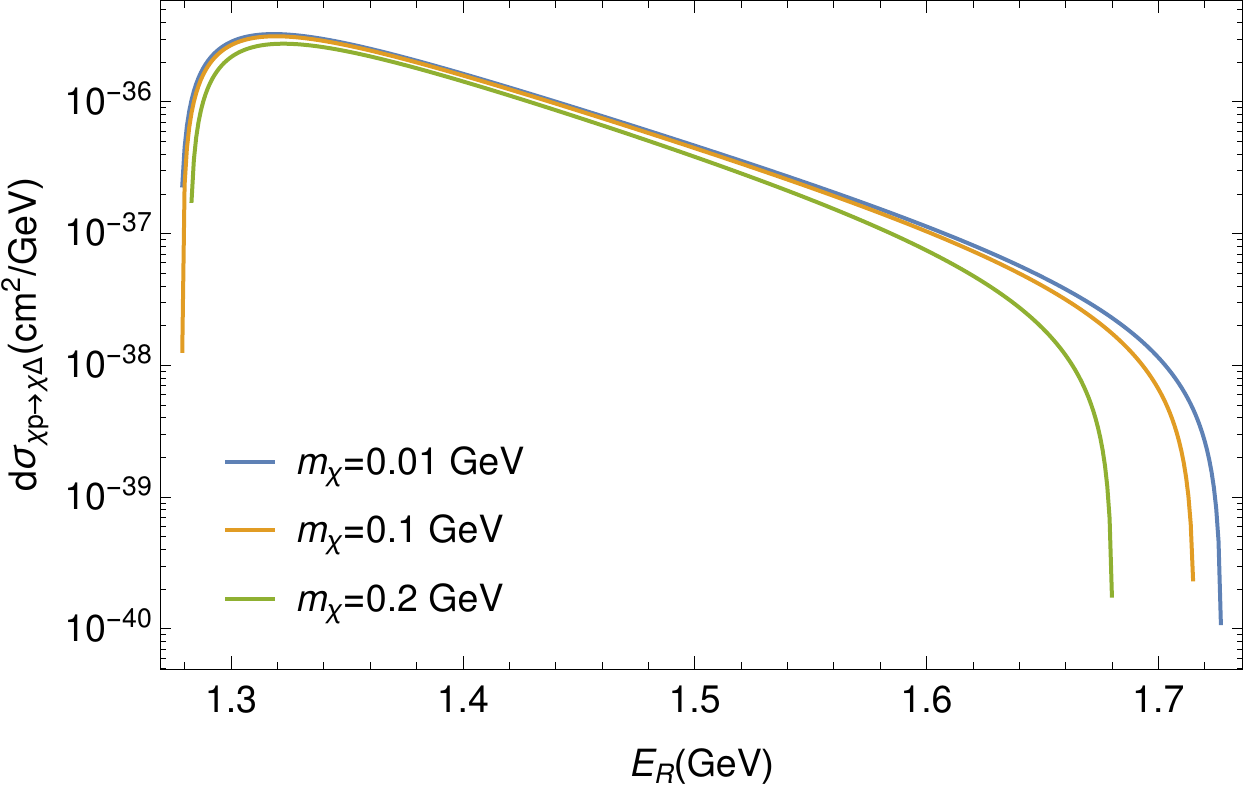}}
  \caption{The differential $\chi N \rightarrow \chi \Delta$ per nucleon scattering cross section of (\ref{eq:delta_scatter}) with $m_V=500$ MeV, $\ep=10^{-3}$, $\alpha^\prime=0.1$ and $E_\chi= 1$ GeV for three dark matter masses. Note that the $\Delta$ cannot be produced at rest due to momentum conservation.}
  \label{fig:delta_scatter}
\end{figure}

Since $\Delta$ is a spin-3/2 particle, the corresponding Rarita-Schwinger field takes the product form
$\psi_{3/2}^\mu(p,s) = \psi_{1/2}(p,s) \epsilon^\mu(p)$. The general $\Delta-\gamma-N$ vertex can be written as a combination of magnetic dipole, electric dipole and quadrupole terms \cite{JonesScadron}.
It can be shown that for transitions between an S-wave nucleon and $\Delta$, the latter two are suppressed \cite{Ramalho:2008ra}. The vertex $\Gamma_{\beta\mu}$ then takes the form,
\be
 \Ga_{\beta\mu}(P,q) = - \frac{3 (m_\Delta+m_N)}{2m_N\left((M+m)^2-q^2\right)}\epsilon_{\beta \mu\sigma \tau} P_\sigma q_\tau \times e \sqrt{\frac{2}{3}} G_M^*(q^2) +\cdots
\ee
where $q=p_N-p_\Delta$ is the momentum transfer, $P = \frac{p_N+p_\Delta}{2}$, and the factor of $e \sqrt{\frac{2}{3}}$ reflects the normalization conventions \cite{Faessler2002}. $G_M^*$ is the magnetic dipole form factor of the nucleon $N$, for which we adopt a spectator quark model \cite{Ramalho:2008ra,Gross:2006fg,Ramalho:2015qna}. Including higher orbitals in the calculation allows for further multipole contributions, but these are expected to add corrections of only a few percent.

For completeness, we note that the squared amplitude takes the form
\begin{align}
\label{eq:delta_scatter_amplitude2}
	\frac{1}{2}\sum |\mathcal{M}|^2 &= - \frac{(4\pi)^2 k^{(1)}_{V,B} (p+p^\prime)_\mu(p+p^\prime)_\nu}{2\left(m_V^2-q^2\right)^2} \mathrm{Tr}\left[P^{\rho \sigma}_\Delta(p) \Gamma^{\rho \mu}(P,q) (\slashed{p}_N+m_N) \Gamma^{\sigma \nu}(P,q) \right],
	\end{align}
where $s$, $t$, and $u$ are the standard Mandelstam variables, and $P^{\mu \nu}_\Delta(p)=\sum_s u^\mu(p,s) \bar u^\nu(p,s)$ is the spin-3/2 sum for positive parity states \cite{Aliev:2004ju}. The final differential cross section in the lab frame is
\begin{equation}
\label{eq:delta_scatter}
	\frac{d\sigma}{dE_R}(\ch +N \rightarrow \ch+ \De) = -\frac{1}{8\pi} \frac{m_N \sum |{\cal M}|^2}{\lambda(s,m_\chi^2,m_N^2)},
\end{equation}
where $E_R$ is the energy of the recoiling $\Delta$, and $\lambda(m_1, m_2, m_3) = m_1^2+m_2^2+m_3^2 - 2 m_1 m_2 - 2 m_2 m_3 - 2 m_1 m_3$.	

\begin{figure}[tp]
 \centerline{\includegraphics[width=0.6\textwidth]{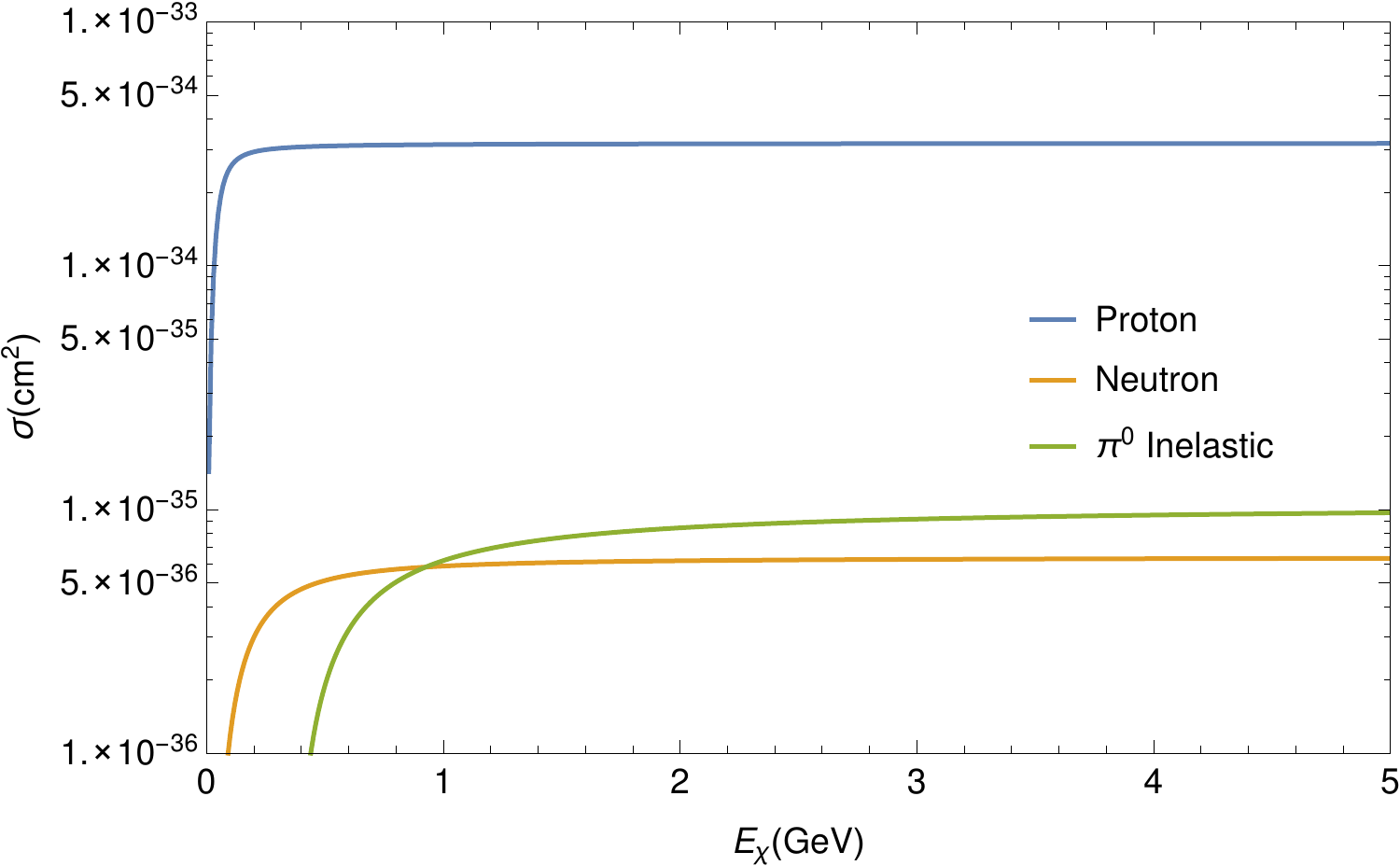}}
 \caption{A comparison of the integrated per nucleon cross section for neutral-current elastic nucleon scattering and inelastic $\De$ production with $m_\chi = 0.01$ GeV, $m_V = 0.1$ GeV, $\ep = 10^{-3}$ and $\alpha^\prime = 0.1$.}
 \label{fig:delta_scatter2}
\end{figure}

This process produces a $\Delta^+$ ($\Delta^0$) for proton (neutron) scattering. The produced $\Delta$'s decay to $\pi N$ nearly 100\% of the time, but we must refer to the isospin of the $\Delta^+$, $\Delta^0$ and that of their possible decay products to calculate the $\pi^0$ production rate \cite{compnuc}. Conveniently, both the $\Delta^+$ and $\Delta^0$ decay to a $\pi^0$ with a branching ratio of $\frac{2}{3}$. When plotting the scattering rates for $\chi N \to \chi \Delta$ resulting in inelastic pion production (see Figs.~5,6), this branching fraction is included.

\subsection{Deep inelastic scattering}

At sufficiently high momentum, the majority of DM scattering in the detector becomes deeply inelastic, with a cross section on nucleons of the form
\be
 \frac{d\si}{dxdQ^2} (\ch+N \rightarrow \chi +X)= \sum_{q,\bar{q}} \left( \frac{d\hat\si}{dxdQ^2} (\chi + q/\bar{q} \rightarrow \chi + X)\right) q_{q/\bar{q}} f_{q/\bar{q}}(x)
\ee
where as usual $x=Q^2/(2m_N (E-E'))$, in terms of the incoming ($E$) and outgoing ($E'$) dark matter energy and the mediator momentum squared $q^2=-Q^2$.

We study deep inelastic DM scattering on nucleons in the two scenarios mentioned in Section \ref{sec:ldm} with scalar DM. (See also \cite{Soper:2014ska,Coloma:2015pih}.)
The parton-level production cross sections are obtained using the matrix element generator {\sc CalcHEP}~\cite{Belyaev:2012qa} with
CTEQ6L parton distribution functions $f(x)$~\cite{Pumplin:2002vw}. We also impose the restriction that $Q^2 > 2$ GeV$^2$. For much of the later analysis, we will fix the mass of DM to $M_{A'}/3$. 
However, in general as long as $M_{A'} > 2 m_\chi$ the qualitative scattering rates do not change much.

\section{Signatures at MiniBooNE, T2K and SHiP}
\label{sec:sens}

Having analyzed the production channels for light dark matter in a proton fixed target experiment, and the scattering channels of interest at detectors downstream from the target, in this section we determine the signal rate at specific experimental neutrino facilities. We begin by outlining the simulation used to determine the signal rate (further details are contained in the Appendix), summarize existing constraints on the models, and then turn to the results for MiniBooNE, T2K and SHiP.

\subsection{Simulation}

The total number of signal events can be written generically in the form
\begin{align}
\nonumber   N_{A\chi \to A^{(*)}\chi} &= n_A \epsilon_{\rm eff}  \sum_{\substack{{\rm prod.}\\{\rm chans.}}}\left(\frac{N_\chi}{2N_{\rm{trials}}}\sum_i L_i \sigma_{A\chi,i} \right).
\end{align}
In this expression, $n_A$ denotes the number density of target atoms in the detector, $\ep_{\rm eff}$ is the corresponding detection efficiency, and the outer sum refers to the relevant production channels.  The total number of dark matter particles $N_\chi$ is determined according to the relevant production channel, while $N_{\rm trials}$ refers to the number of trajectories generated by the production Monte Carlo. The inner sum is over all dark matter 4-momenta $p_i$ generated by the production Monte Carlo, and $L_i$ is the length of the intersection between the dark matter trajectory (with momentum $p_i$) and the detector. Finally, the scattering cross section was computed in the form
\be
 \si_{A\ch} (E) = \int_{E_\chi^{\rm min}}^{E_\chi^{\rm max}} dE_\chi \sum_{\al=p,n,e} f_\al \frac{d\si_{\chi, \al}}{d E_\chi},
\ee
where $E$ is the incoming kinetic energy, and $E_\chi^{\rm max/min}$ are determined by the experimental cuts on the nucleon/electron recoil momentum $q=\sqrt{2M(E-E_\chi)}$, while the sum is over the relevant scattering channels. For elastic or quasi-elastic nucleon scattering, we take $f_{p,n} = Z,A-Z$ for the vector portal and $f_{p,n}=A$ for the baryonic portal, since scattering is incoherent for the momentum transfers $q^2 > (50\,{\rm MeV})^2$ of interest, and nuclear binding effects (e.g. Pauli blocking) are subleading for the cuts on recoil energy (and thus momentum transfer) that are relevant for the experiments studied here.  For electron scattering, we necessarily take $f_e=Z$.

The results presented below were computed using the Monte Carlo simulation tool {\tt BdNMC}, developed by one of the authors (P.dN.). 
It is now publicly available, and full documentation is provided in the Appendix.

\subsection{Constraints}

The benchmark models and parameter regimes are constructed so that they ameliorate a number of potential astrophysical constraints, e.g. light scalar dark matter annihilation has a $p$-wave suppression at low velocities that limits sensitivity from CMB distortion or galactic limits. However, there are a number of terrestrial constraints, which are continually being improved.  The sensitivity plots below exhibit a number of existing constraints, that are summarized briefly below (see e.g.~\cite{deNiverville:2012ij,Batell:2014mga} and references below for further details):

\begin{itemize}
	\item {\it E137} - a 20 GeV electron beam dump experiment carried out at SLAC \cite{e137,e137_exp}, which was sensitive to light DM scattering.
	\item {\it LSND} - an 800 MeV neutrino experiment, with a $\nu-e$ scattering analysis that was recast as a constraint on light dark matter \cite{deNiverville:2011it,Kahn:2014sra,lsnd2001}.
	\item {\it BaBar} - a mono-photon search provides stringent constants on higher mass invisibly-decaying dark vectors \cite{babar1}. 
	\item $K^+ \to \pi^+ \nu \bar \nu$ - the result of E949 at Brookhaven can be recast as a constraint on $K^+ \to \pi^+ V$ \cite{Pospelov:2008zw,Artamonov:2009sz,Batell:2014mga}.
	\item $J/\Psi\to${\it invisible} - a constraint is provided by the limit Br($J/\Psi\to$invisible)$<7 \times 10^{-4}$ placed by the BES collaboration \cite{Ablikim:2007ek}.
	\item $\pi^0 \to \gamma +${\it invisible} - a limit Br$(\pi^0 \to \gamma V) < 5 \times 10^{-4}$ from the Brookhaven alternating gradient synchrotron \cite{pi0invis}.
	\item $\Delta m_Z$ {\it and EW fit} - a limit due to the induced shift in the $Z$ mass and electroweak precision fits \cite{deltamz}.
	\item {\it CDF constraints on Monojets} - limits from searches for $pp \to \mathrm{jet} +$missing energy \cite{monojet,monojet2}. 	
	\item {\it Lepton} $g-2$ -  the blue band is where agreement with muon $g-2$ is improved to within $3\sigma$. All parameter space is excluded which increases the disagreement of either muon or electron $g-2$ to more than $5\sigma$ \cite{g2_1,g2_2,g2_3}.
	\item {\it Direct Detection} - the strongest current low mass limits are from CRESST-II \cite{Angloher:2015ewa} and CDMS-Lite \cite{cdmslite2015}.
	\item {\it Angular Dependence in Neutron Scattering} - a constraint on baryonic vectors $\alpha_B < 3.4 \times 10^{-11} \left(\frac{m_V}{\rm MeV}\right)^4$ \cite{neutronlimit1,neutronlimit2}.
\end{itemize}

\begin{table}[t]
\centering
\begin{tabular}{|l|r|r|r|r|r|r|r|}
\hline
Name & Energy & POT & Detector Mass & Material & Distance & Angle & Efficiency \\
\hline
MiniBooNE-Beam Dump & 8 GeV & $2\times10^{20}$ & 400 tons & CH$_2$ & 490 m & 0 & 0.35 \\
\hline
T2K-ND280 (P0D) & 30 GeV & $5\times10^{21}$ & 6 tons & H$_2$O,Plastic & 280 m & $2.5^o$ & 0.35 \\
\hline
T2K-Super-K & 30 GeV & $5\times10^{21}$ & 50 kilotons & H$_2$O & 295 km & $2.5^o$ & 0.66 \\
\hline
SHiP & 400 GeV & $2\times10^{20}$ & 10 tons & LAr & 100 m & 0 & 0.5\\
\hline
\end{tabular}
\label{tab:pars}
\caption{A summary of the relevant characteristics of the experiments considered. The listed detector mass is the fiducial mass, when available. Note that SHiP is still in the proposal and planning stage, and the design has not been finalized, so the detector material and mass have been chosen for illustration (the final fiducial mass may be larger).}
\end{table}

\subsection{Sensitivity}

There are a number of different short- and long-baseline neutrino facilities either in, or recently in, operation around the world (MiniBooNE, MicroBooNE, T2K, MINOS, NOvA, OPERA,\ldots). There are also several future facilities that may have interesting sensitivity to this class of light dark matter models (DUNE, SBN, SHiP,\ldots). The sensitivity depends on a combination of the effective yield, e.g. through having a sufficiently energetic beam, and a sizable angular acceptance for the detector, along with a viable means of detecting scattering and mitigating the neutrino backgrounds. Based on these criteria, we have focussed our analysis on MiniBooNE, T2K and SHiP, which currently appear to provide the best (potential) sensitivity in different ranges for the mediator mass. The relevant parameters for these experiments, beam energy, number of protons on target, detector geometry and material, etc, are listed in Table~I. It is worth noting that future facilities, such as DUNE, or T2HK with $\nu$PRISM, could provide excellent sensitivity if additional near detectors were to be installed. 

\begin{figure}[t]
 \centerline{\includegraphics[width=0.43\textwidth]{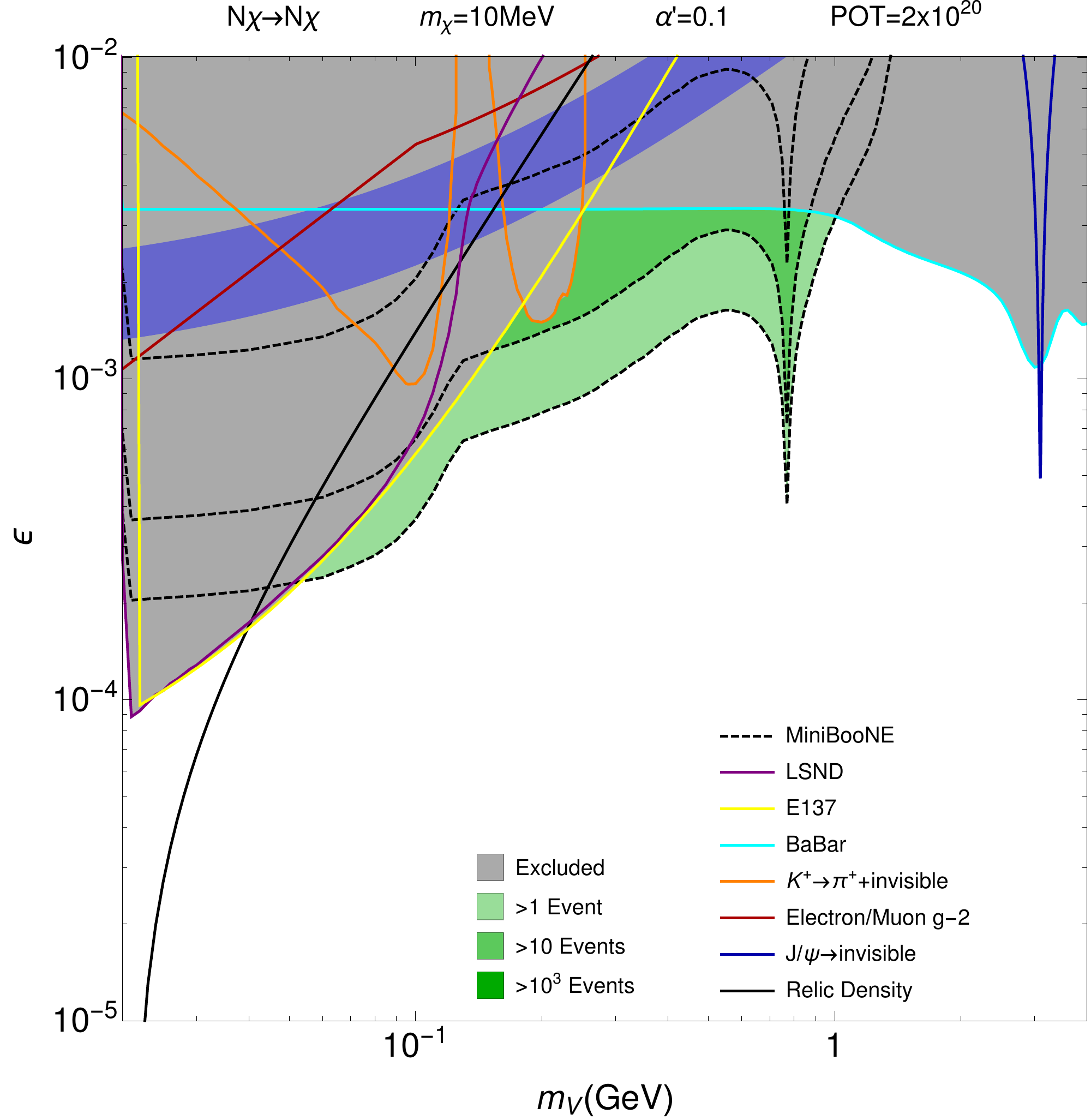}\hspace*{1cm}\includegraphics[width=0.43\textwidth]{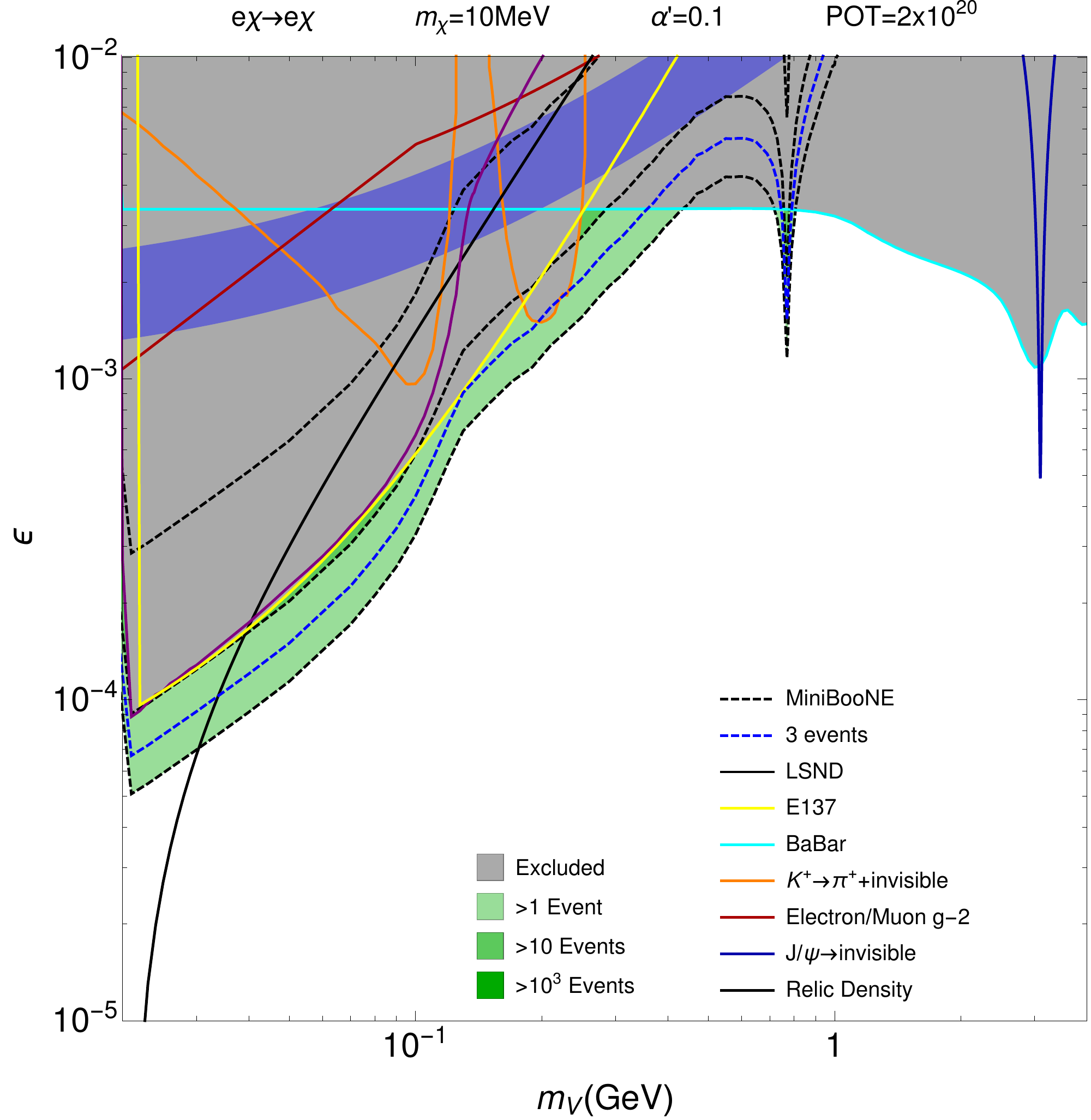}}
 \caption{Plots showing the MiniBooNE yield of light dark matter scattering events in nucleon and electron elastic scattering channels as labelled. In this plot and the others to follow, the gray regions are excluded by existing constraints, while the green contours indicate 1, 10 and 1000 events. The plot on the right also shows a 3 event contour, corresponding to the estimated sensitivity for electron scattering.}
 \label{fig:MB1}
\end{figure}

\begin{figure}[t]
 \centerline{\includegraphics[width=0.43\textwidth]{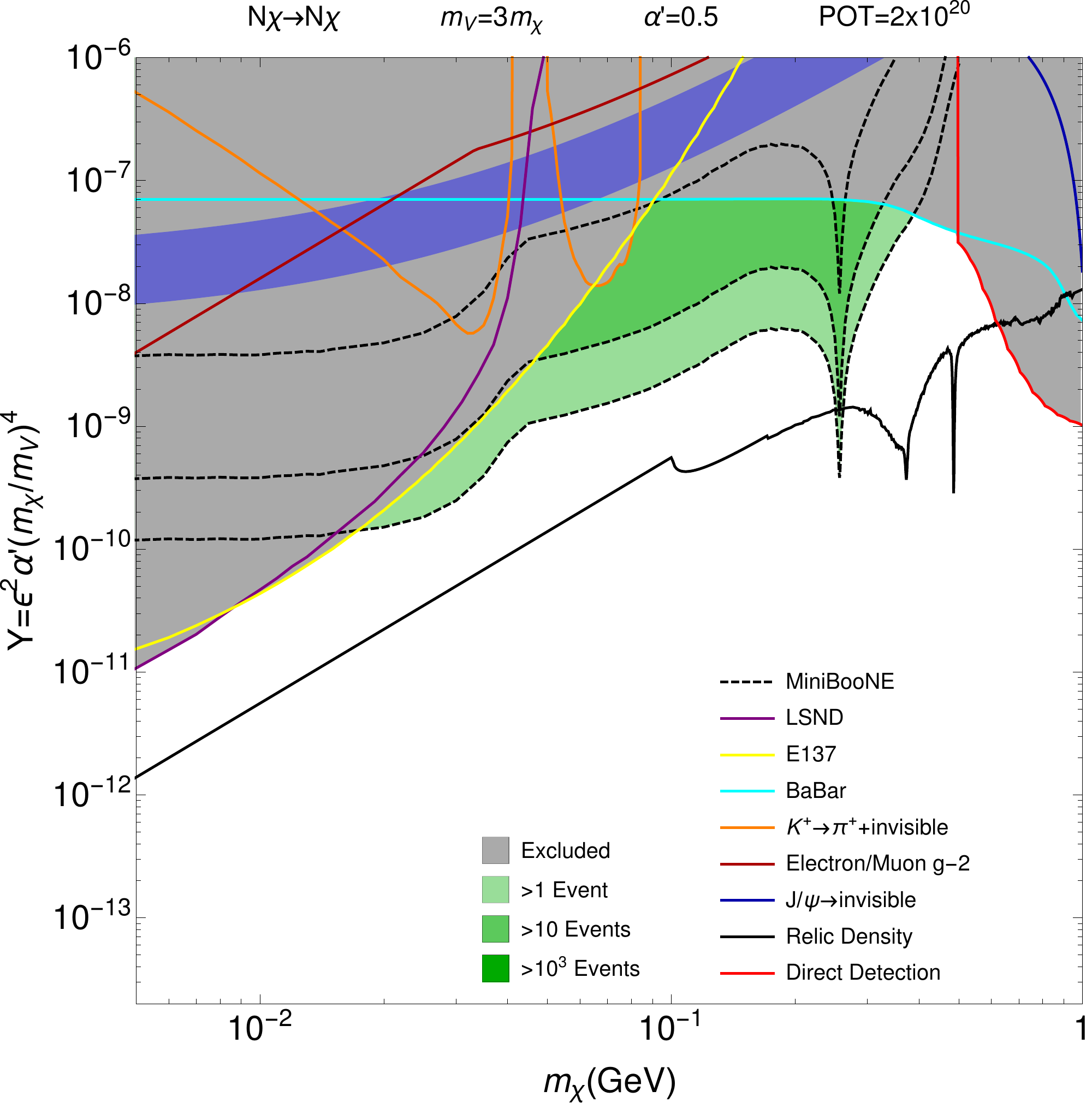}\hspace*{1cm}\includegraphics[width=0.43\textwidth]{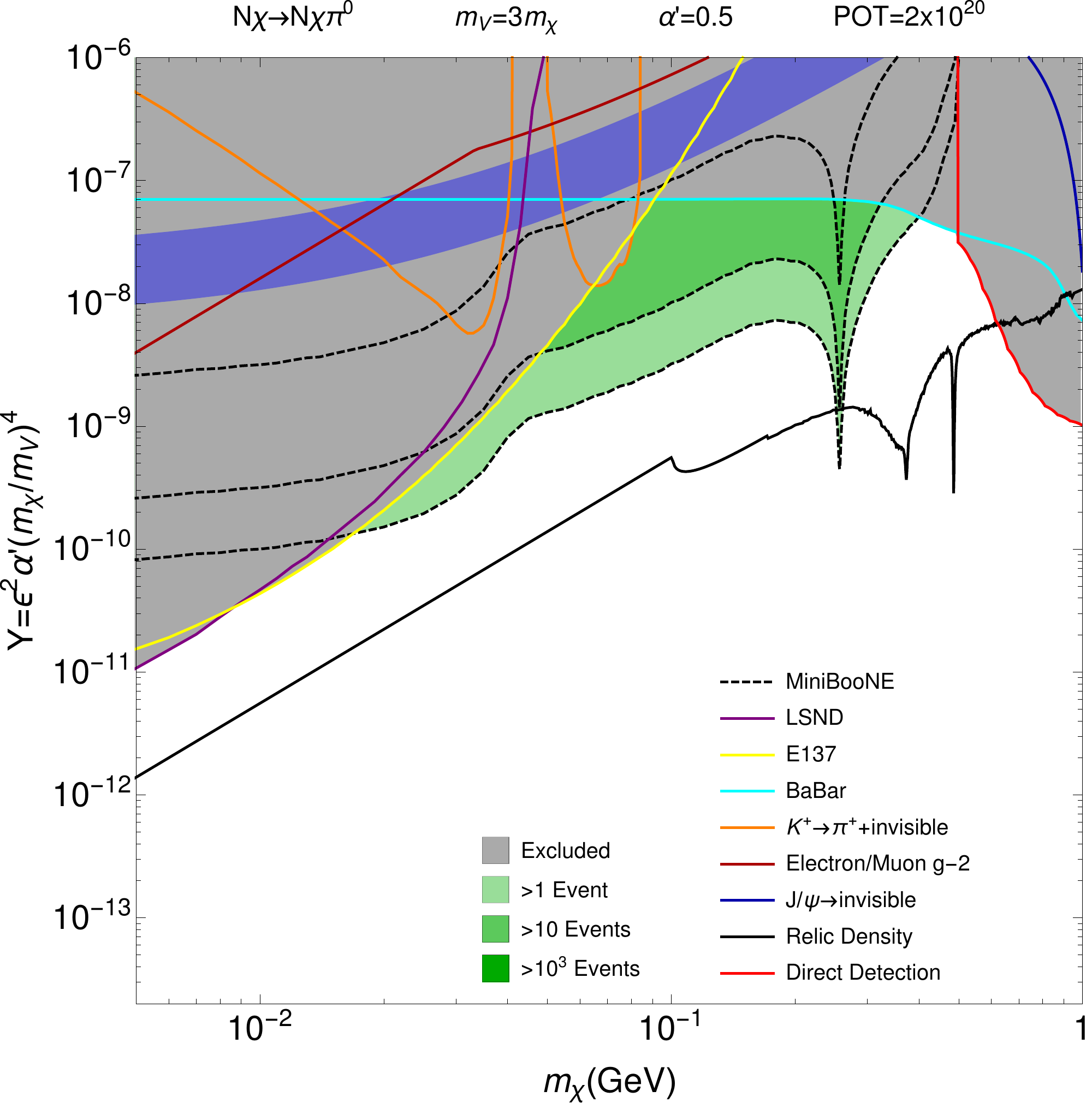}}
 \vspace*{0.4cm}
 \centerline{\includegraphics[width=0.43\textwidth]{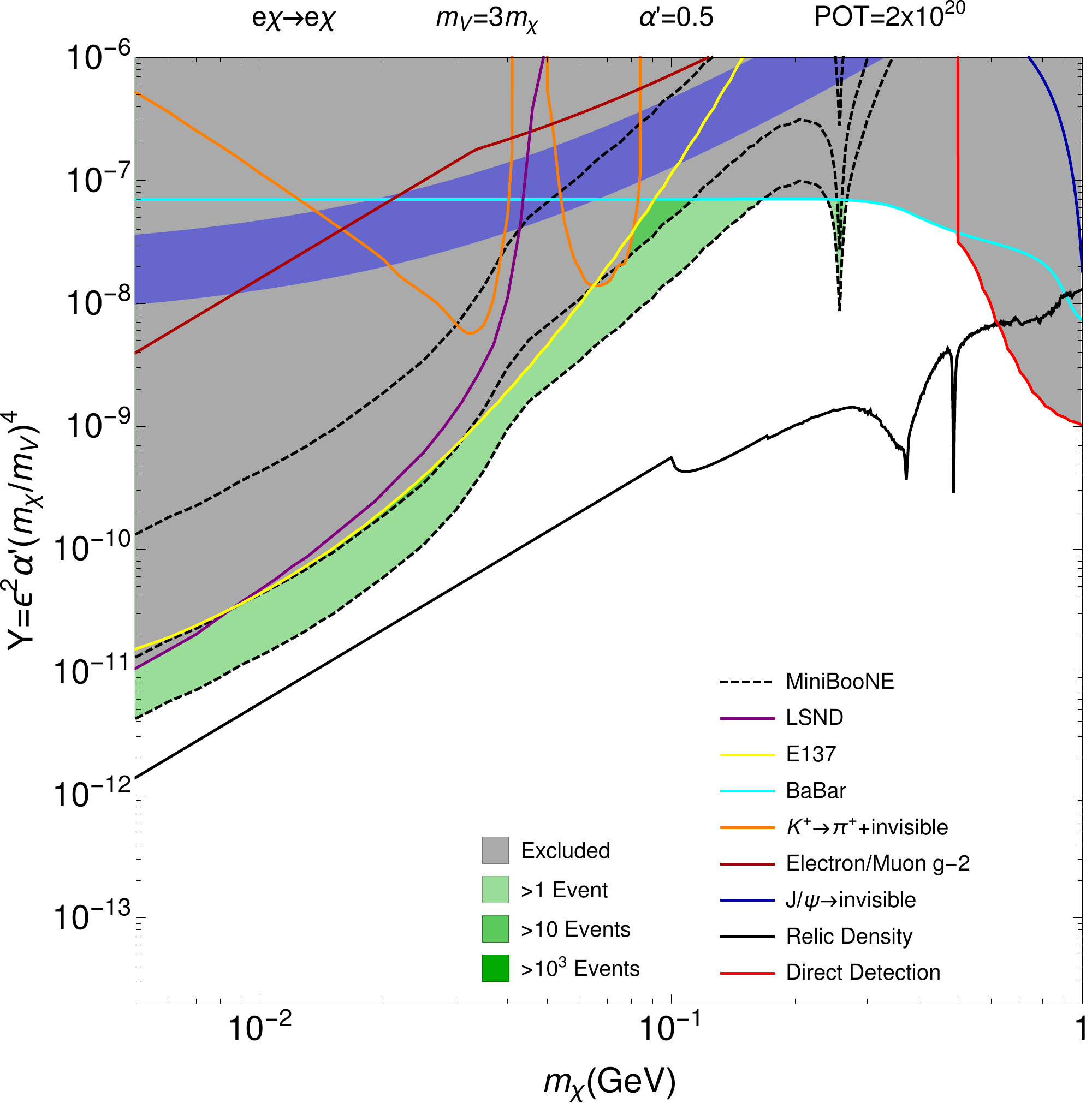}\hspace*{1cm}\includegraphics[width=0.43\textwidth]{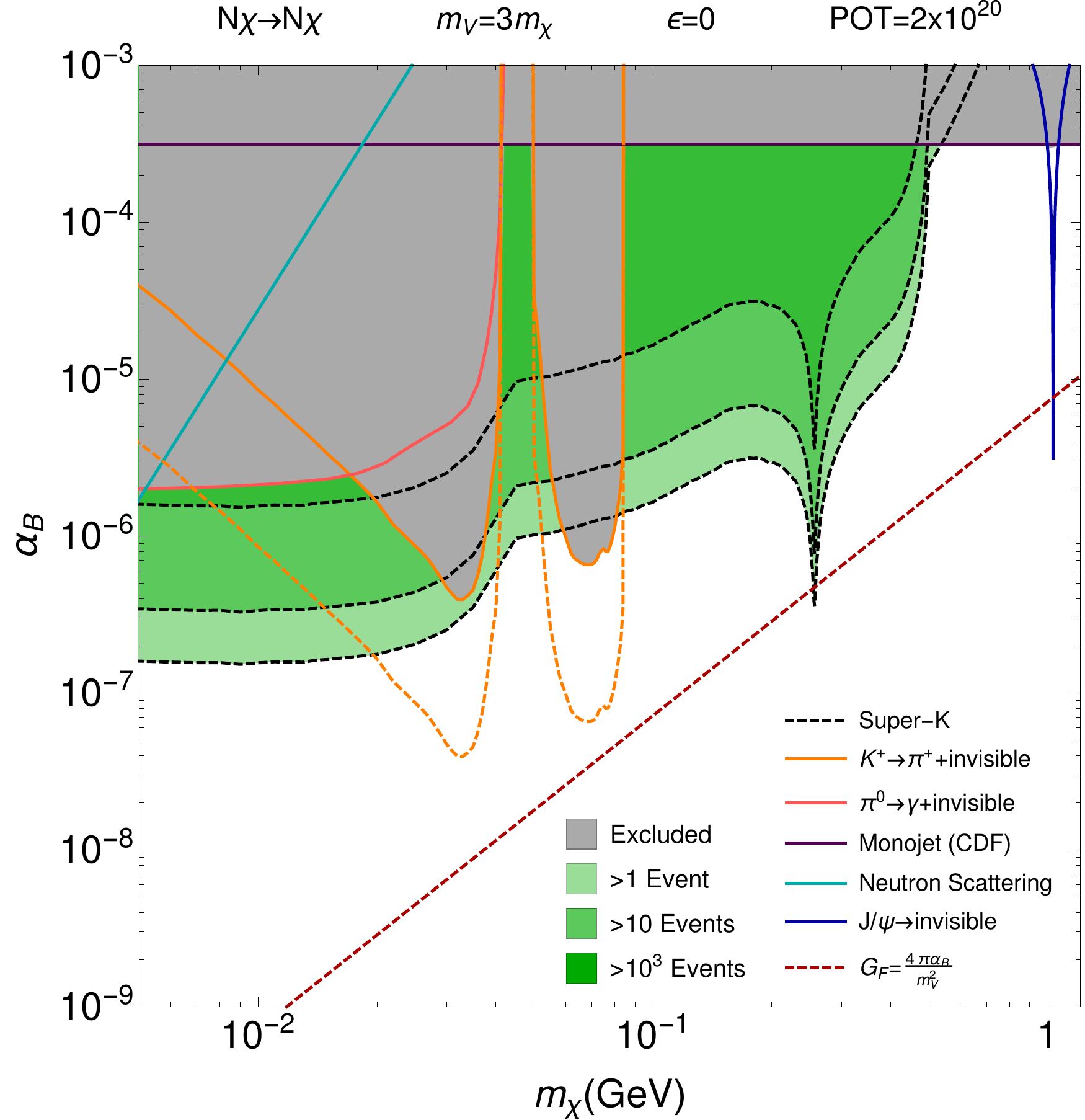}}
 \caption{Further plots showing the MiniBooNE yield of light dark matter scattering events in various channels, now fixing $m_V=3m_\chi$ with $\alpha'=0.5$, and using the variable $Y$ for the vertical scale (see Eq.~(\ref{Y})). The strongest low-mass direct detection constraint is from CRESST-II \cite{Angloher:2015ewa}, shown as the red contour. In these plots and below the black dotted line shows the parameters required to achieve the dark matter relic density, so smaller values of $Y$ are excluded due to over-production of dark matter. Note that the lower right plot shows the sensitivity in the case of a baryonic vector mediator.}
 \label{fig:MB2}
\end{figure}

\begin{figure}[t]
 \centerline{\includegraphics[width=0.43\textwidth]{Y_minibooneN.pdf}\hspace*{1cm}\includegraphics[width=0.43\textwidth]{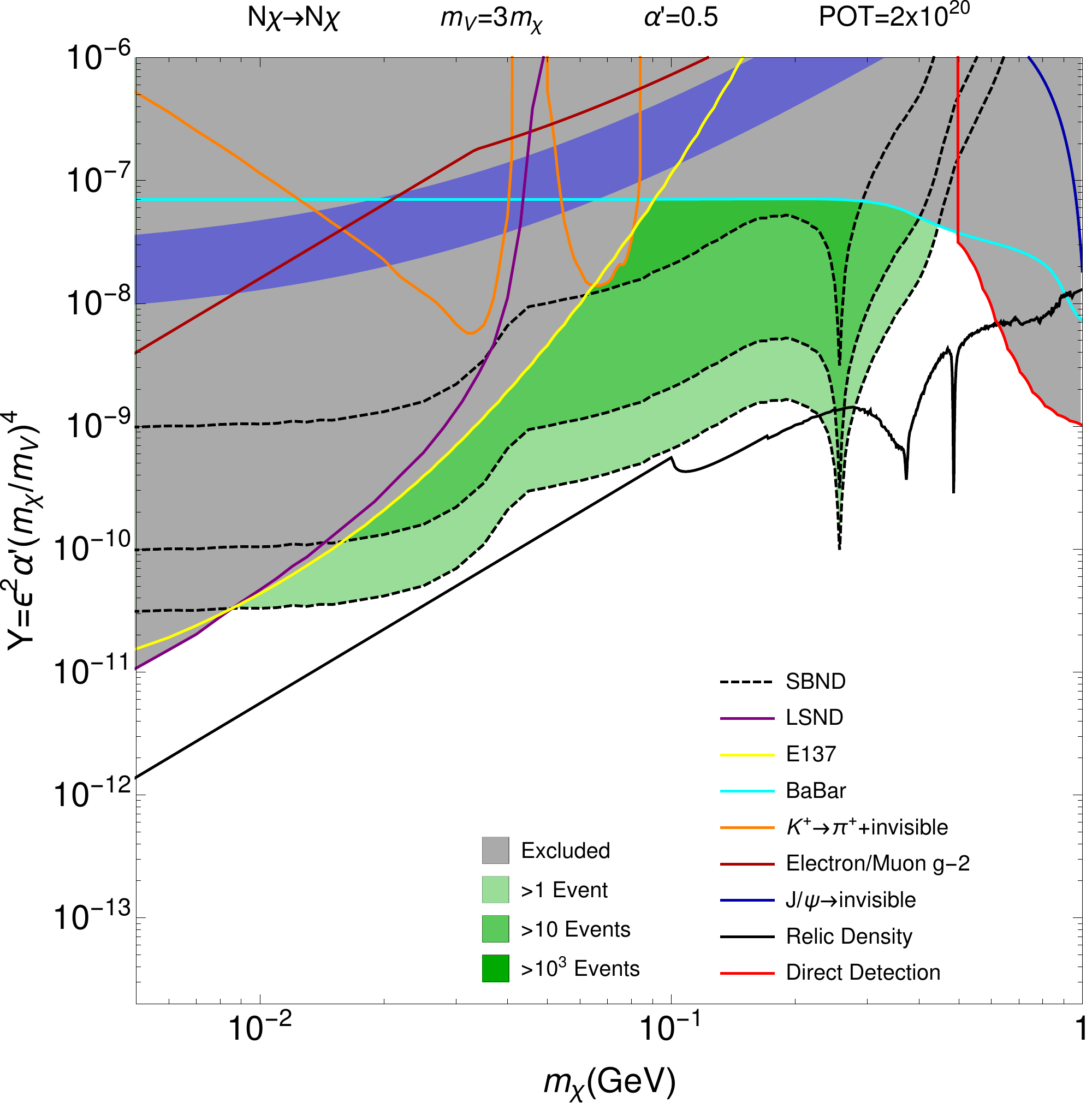}}
 \caption{Plots showing the comparative yield of MiniBooNE and the proposed SBND experiment in the nucleon elastic scattering channel.}
 \label{fig:SBND}
\end{figure}

\begin{figure}[t]
 \centerline{\includegraphics[width=0.43\textwidth]{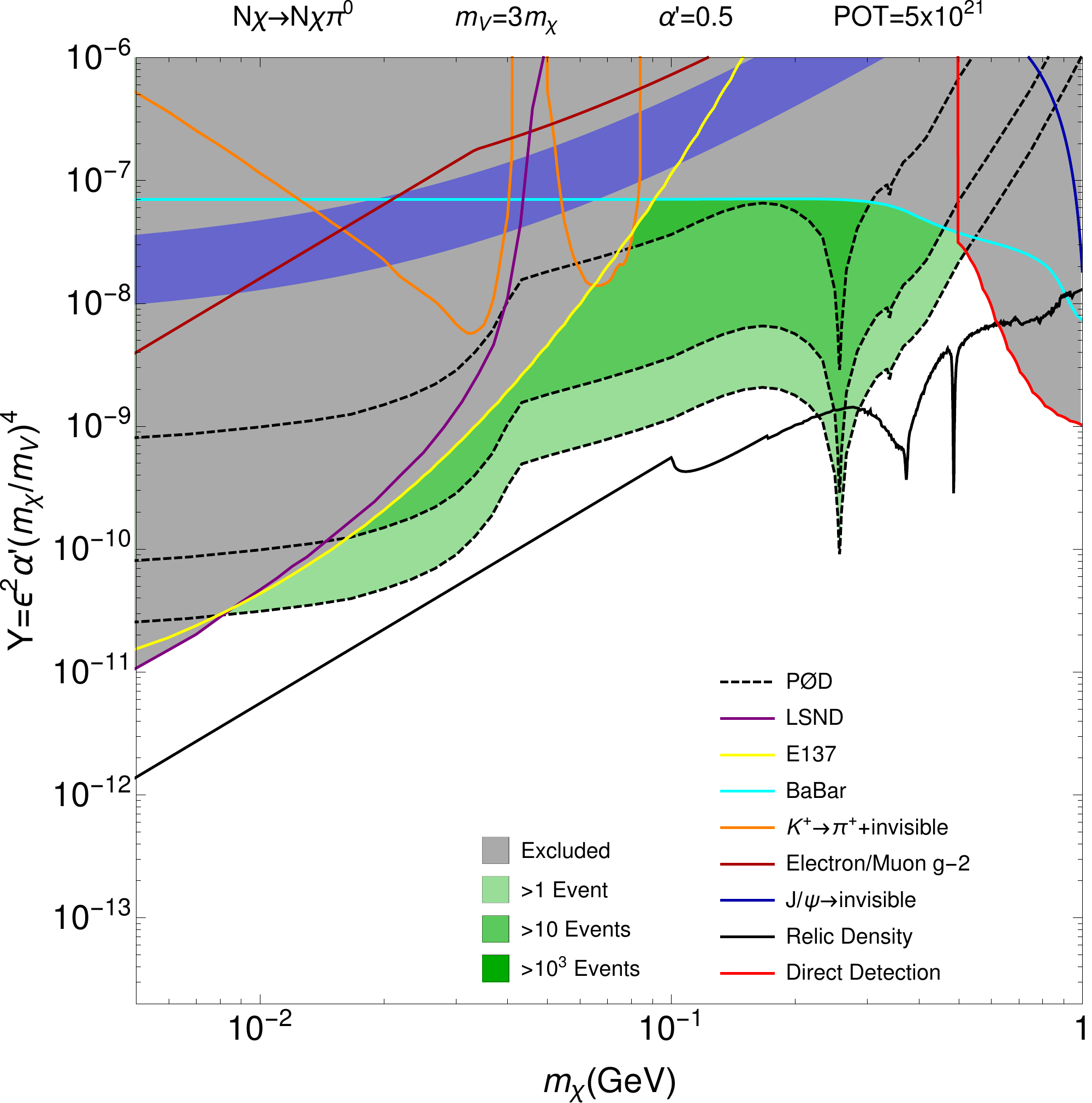}\hspace*{0.4cm}\includegraphics[width=0.43\textwidth]{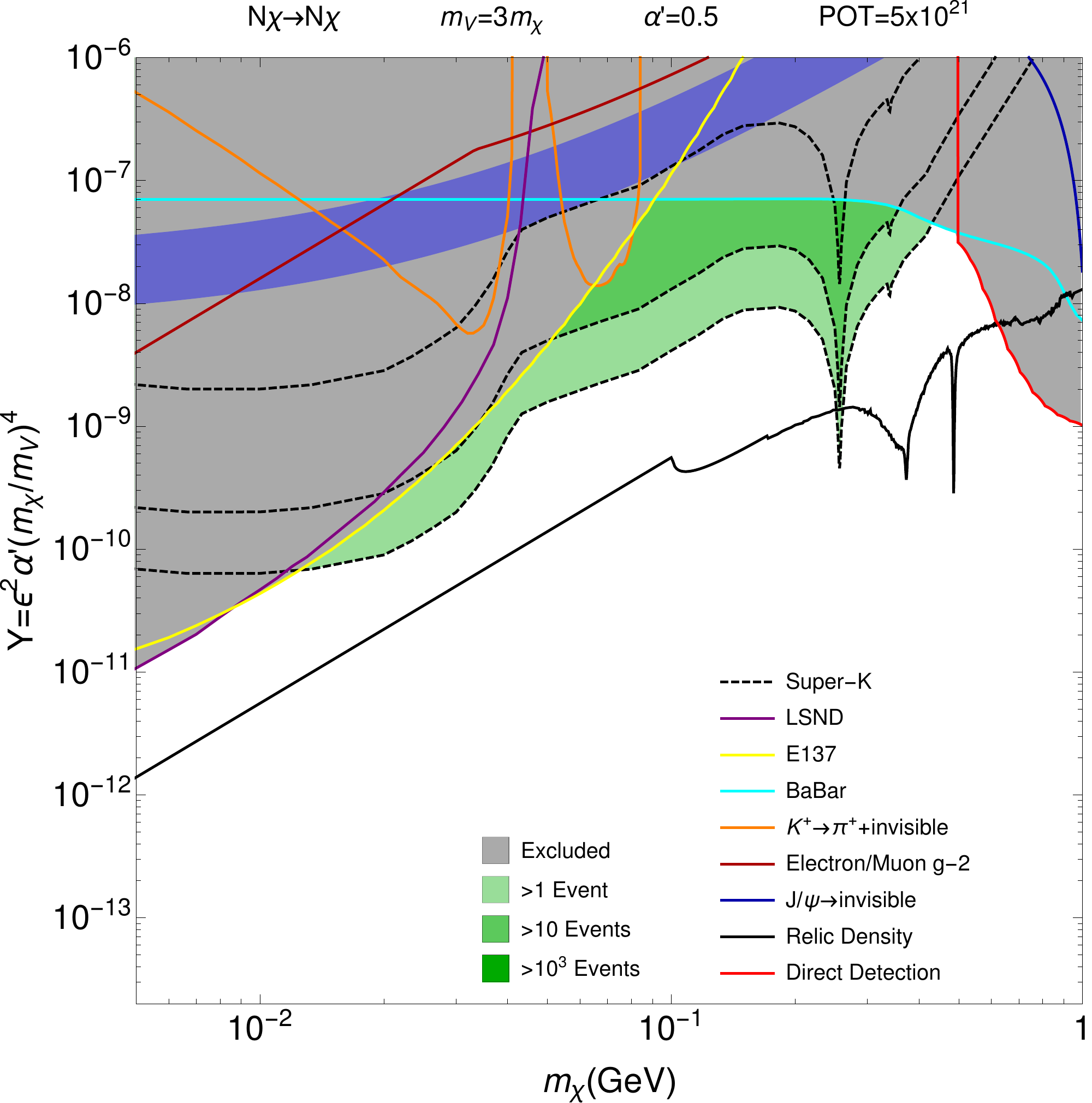}}
 \vspace*{0.4cm}
 \centerline{\includegraphics[width=0.43\textwidth]{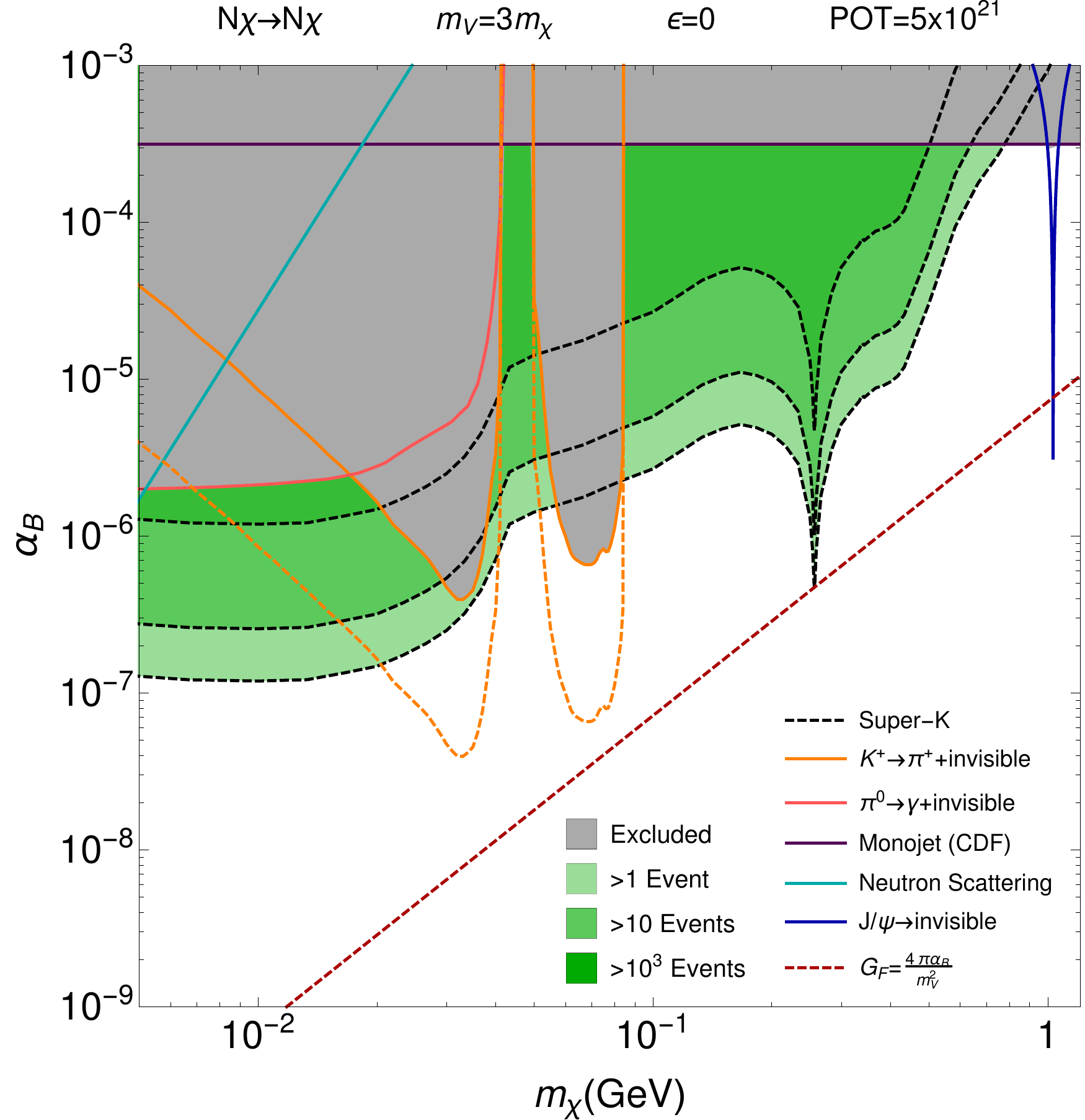}}
 \caption{Plots showing the T2K yield  (ND280 on the left, Super-K on the right and below) of light dark matter scattering events in various channels.}
 \label{fig:T2K}
\end{figure}

\begin{figure}[t]
\centerline{\includegraphics[width=0.43\textwidth]{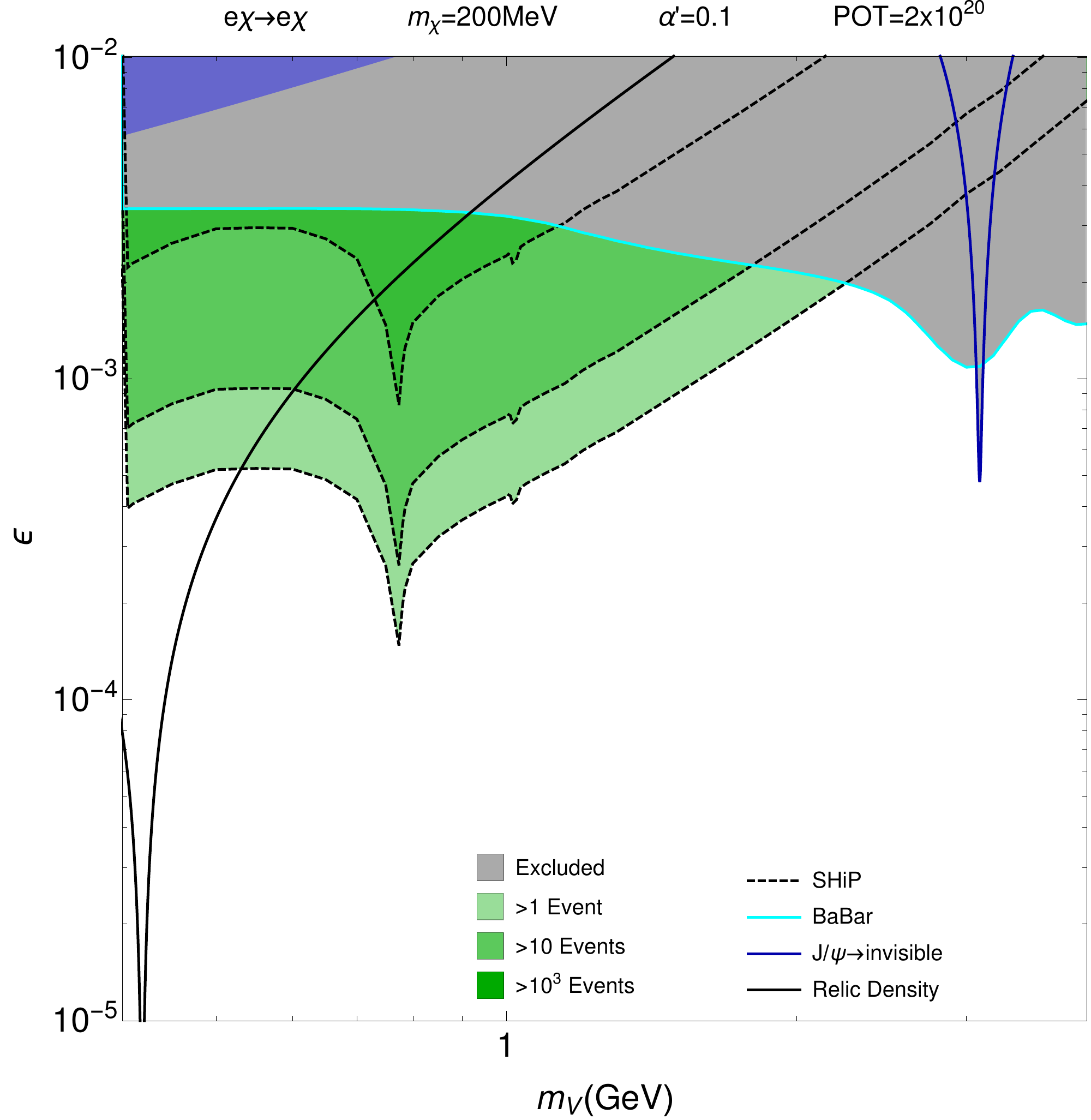}\hspace*{1cm}\includegraphics[width=0.43\textwidth]{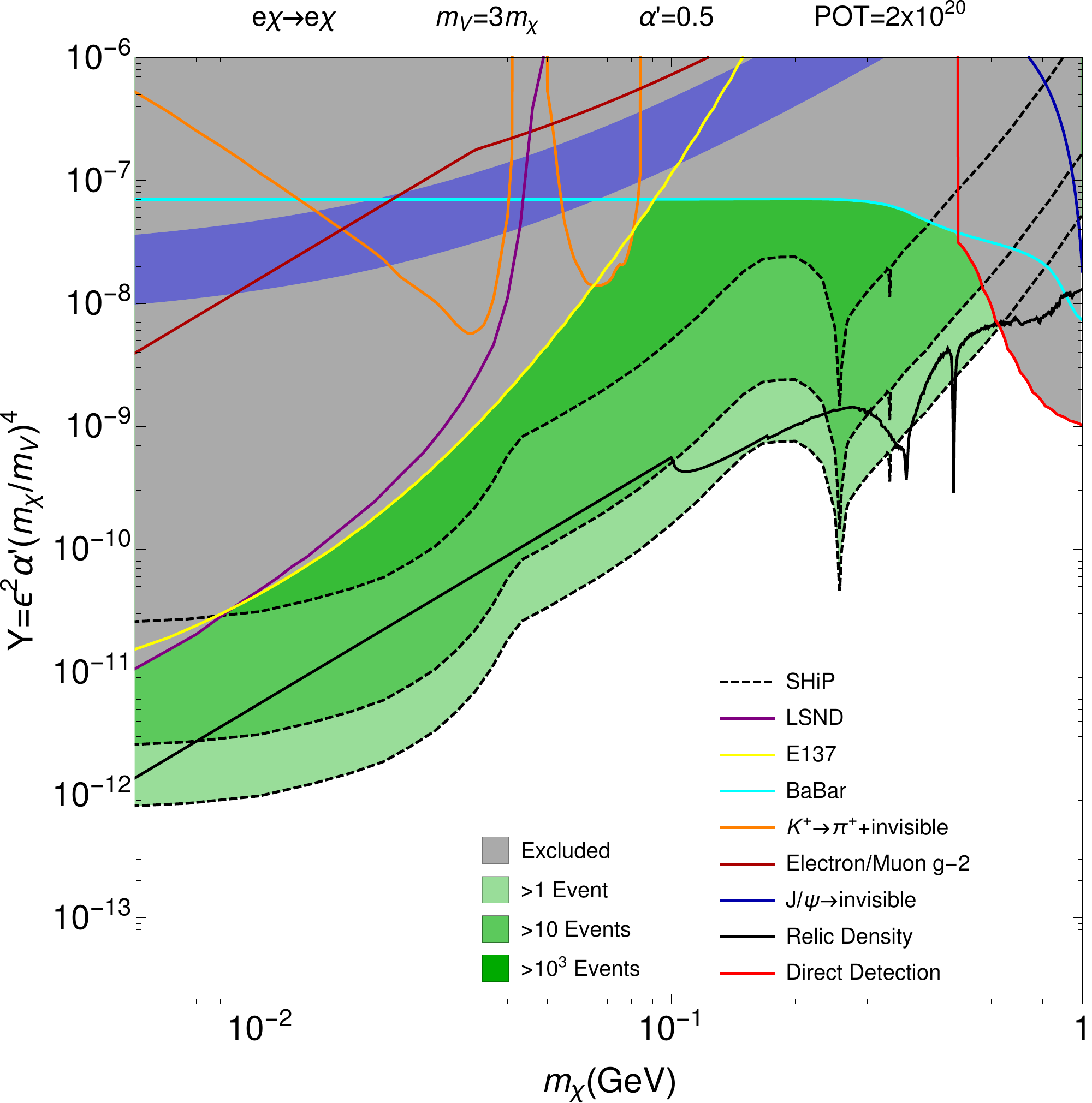}}
\vspace*{0.4cm}
 \centerline{\includegraphics[width=0.43\textwidth]{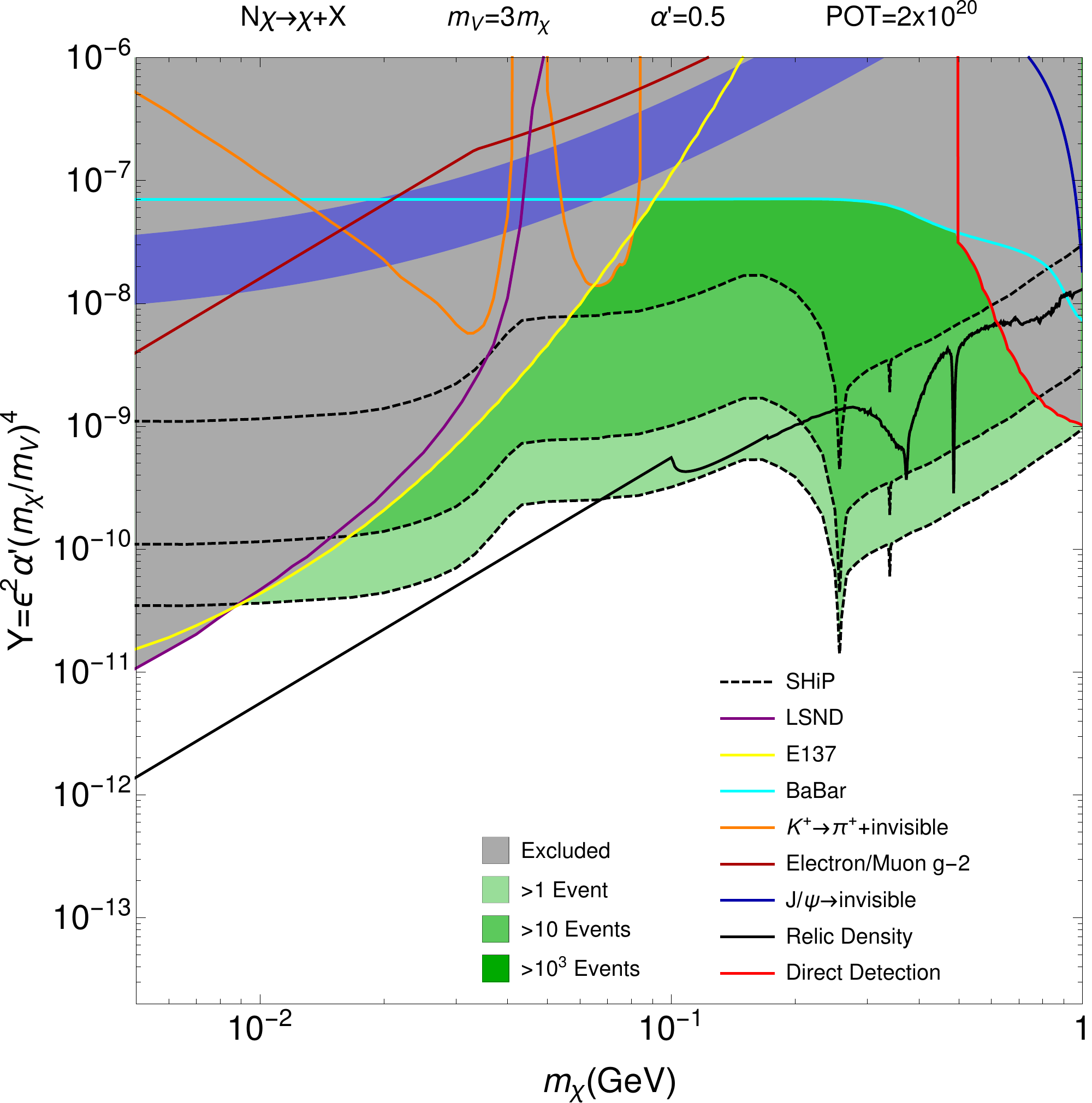}\hspace*{1cm}\includegraphics[width=0.43\textwidth]{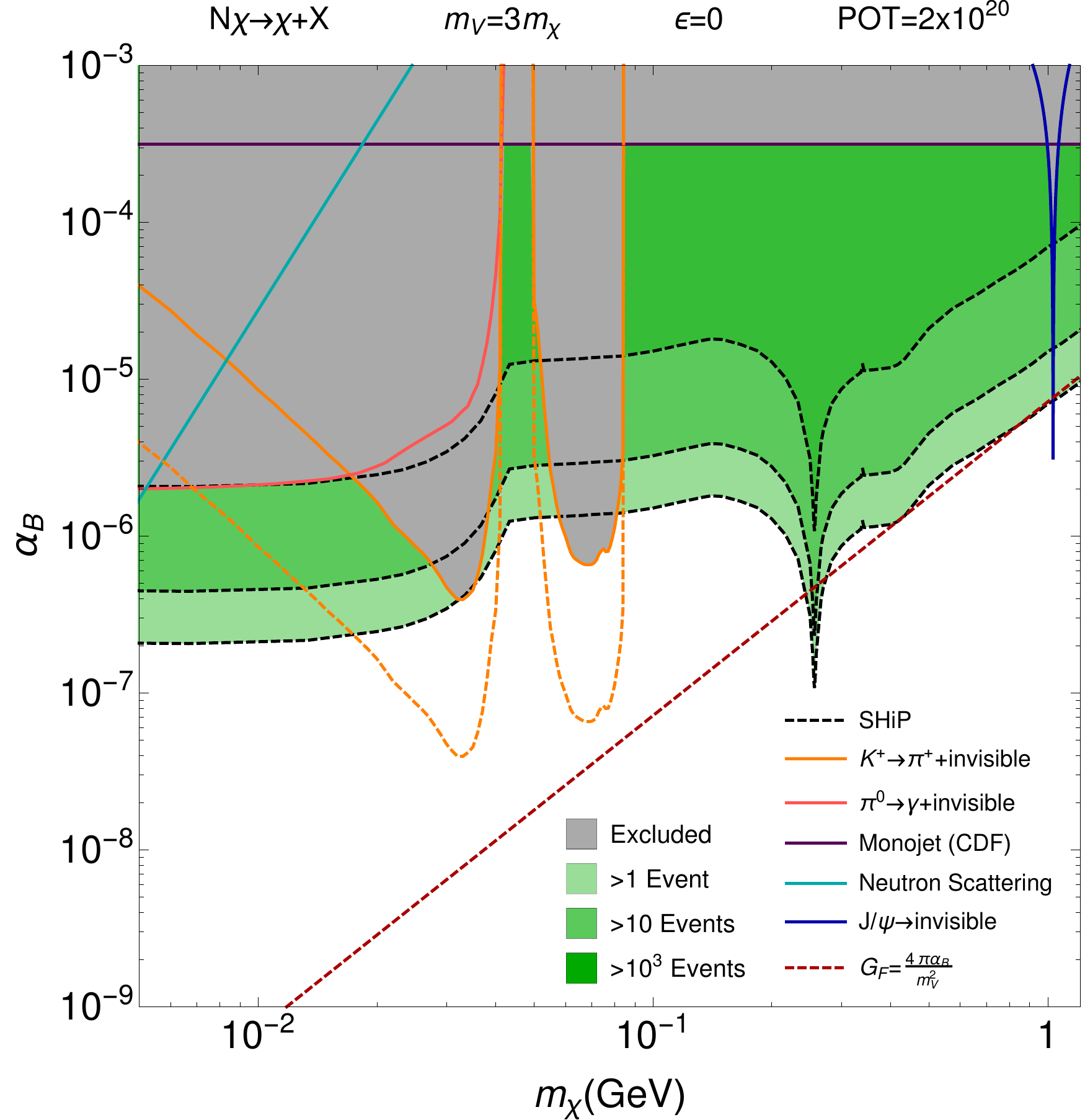}}
 \caption{Plots showing the SHiP yield of light dark matter scattering events in various channels.}
 \label{fig:SHiP}
\end{figure}

Our results are exhibited in a number of parameter space plots, which include the existing constraints summarized above. The majority of these plots for the vector portal model make use of the following slice \cite{Izaguirre:2015yja} through the parameter space: $Y$ vs $m_\chi$ at fixed $m_V/m_\chi=3$, where
\be
 Y\equiv \ep^2 \al' \left(\frac{m_\ch}{m_V}\right)^4, \label{Y}
\ee
which captures the essential parameter scaling of the annihilation and scattering cross sections, and assists the comparison with direct detection sensitivity (see \cite{Izaguirre:2015yja}). By convention the plots use the choice $\al'=0.5$, which is relatively large from a model-building perspective. However, it is used simply to provide a conservative view of the full parameter reach of each experiment, in comparison to the relic density curve, which sets the lowest values of $Y$ for which the model is cosmologically viable. Note that the relic density curve remains fixed in $Y$, as parameters are varied, while the other constraints generally scale to lower values of $Y$ with decreasing $\al'$ \cite{Izaguirre:2015yja}. (For comparison with earlier results, Fig.~\ref{fig:MB1} shows the MiniBooNE sensitivity using the alternate parameter slice $\epsilon$ vs $m_V$ at fixed $m_\chi$ and $\al'$.)

The signal yields at MiniBooNE, T2K and SHiP are summarized below and exhibited in Figs.~\ref{fig:MB1}--\ref{fig:SHiP}.
\begin{itemize}
\item {\bf MiniBooNE} - The MiniBooNE collaboration took data during a dedicated beam dump run in 2013/14 \cite{Dharmapalan:2012xp}, and as detailed below we consider a number of possible scattering signatures for both vector and baryonic portal interactions. The sensitivity contours are shown in Figs.~\ref{fig:MB1} and \ref{fig:MB2}. 
\begin{enumerate}
\item {\it Elastic nucleon scattering}. We use the following cuts on nucleon recoil energy: $E_R \in [0.35,1]\,\mathrm{GeV}$.
\item {\it Elastic electron scattering}. This channel has the advantage that a stringent forward angle cut on the scattered electron can significantly reduce the neutrino background. We therefore use a cut on the electron scattering angle of $\cos \theta_e > 0.99$. While no explicit cut on the recoil energy is imposed, few low energy recoils survive the cut on $\cos \theta$. 
\item {\it Quasi-elastic single pion production}. This inelastic channel has a slightly reduced overall yield, but the signature is detected with high efficiency, and there is a considerable reduction in the cosmogenic and neutrino background. It is therefore possible that it will prove to be the most effective search channel for nucleon scattering at MiniBooNE.
\item {\it Future progress with SBND}. In Fig.~\ref{fig:SBND}, we also show the projected sensitivity for a future beam dump proposal,\footnote{We thank Richard Van de Water for discussions and for providing the SBND parameters.} SNBD, using the same Booster beam-line at Fermilab, but with a 112 ton LArTPC detector at 110\,m. 
\end{enumerate}
\item {\bf T2K} - For the T2K experiment, we consider both quasielastic single pion production in the P\O D, part of the ND280 near detector, and elastic scattering of nucleons in the far detector, Super-K, where timing cuts are highly effective at reducing beam-related backgrounds. The sensitivity contours are shown in Fig.~\ref{fig:T2K}.
\begin{enumerate}
\item {\it Elastic nucleon scattering at Super-K}. The angular acceptance at the far detector, Super-K, is necessarily suppressed but a compensating enhancement in background reduction is possible through timing cuts. The time structure of the bunches in each beam spill can be used to cleanly separate beam-related backgrounds. Recent analyses of neutral current neutrino scattering, via observation of de-excitation $\gamma$'s from quasi elastic scattering on Oxygen \cite{Abe:2014dyd} provides the benchmark technique that could be used for a dedicated light dark matter analysis.
\item {\it Quasi-elastic single pion production at ND280(P\O D)}. We also show results for single pion production in the ND280 near detector (specifically the P\O D) that is specifically designed for high efficiency detection of these signatures. The overall yield is very high, but the neutrino scattering background is also large and further analysis is required to determine if the final sensitivity may be better than a search using Super-K.
\end{enumerate}
\item {\bf SHiP} - For the proposed SHiP facility at the CERN SPS, we consider elastic scattering off electrons, and also DIS showering off nucleons in the detector. The experiment is a high energy beam dump, with a relatively large angular acceptance for the detector of interest here, which is foreseen for detection of $\ta$-neutrino scattering, placed in front of a larger tracking detector than can detect charged particles.
\begin{enumerate}
\item {\it Elastic electron scattering}. The high 400 GeV beam energy at the SPS leads to a relatively high average DM momentum, so that almost all scattering events will induce showering. It has been estimated that neutrino scattering background level would be $\sim$300 events for the expected POT.\footnote{We thank W. Bonivento for discussions concerning the neutrino background at SHiP.}
\item {\it Deep inelastic scattering}. The high beam energy necessarily implies that most hadronic scattering events in the detector will be deeply inelastic. Thus, for SHiP we only consider DIS scattering in the form $\chi + N \rightarrow \chi + X$, and present the corresponding yield contours in Fig.~\ref{fig:SHiP}.
\end{enumerate}
\end{itemize}

\subsection{Discussion}

In all cases the plots exhibited in the previous subsection indicate the event yield, using three contours showing 1, 10 and 1000 events. In this subsection we comment on features of these yield curves, and also the important question of backgrounds, specifically from neutrino scattering, and various potential mitigation strategies.

\begin{itemize}

\item {\it Features of the dark matter yield}. It is notable that dark matter electron scattering becomes particularly efficient at low masses, due to the corresponding kinematics. Nucleon scattering becomes more efficient for higher masses, and particularly in the neighbourhood of the $\rho/\omega$ resonance. 

As noted earlier, there is enhanced production in the resonance region, although the timelike form factor used does not resolve the resonance shape particularly well. It is also worth remarking that the resonance occurs at different masses in the production and relic density contours. In the plots of $Y$ vs $m_\chi$, the production resonance appears at $m_\chi=m_V/3 \sim m_\rho/3$ via bremsstrahlung, but at $m_\chi \sim m_\rho/2$ (shifted down slightly by thermal kinetic energy) in the relic density contour associated with annihilation in the early universe. In practice, the latter resonance presumably contributes to production as well, but is not captured by the current approximation which is therefore still somewhat conservative. It is worth noting that modelling of production remains the major source of uncertainty in estimating the event rate. While data-driven checks provide confidence in the rates associated with pseudoscalar meson decay at the 30-40\% level, the rates at higher vector masses are less certain.

For comparison the plots also show the recently obtained direct-detection limits from CRESST-II \cite{Angloher:2015ewa}, which is the first direct nucleon scattering exclusion contour  to push down to masses below 1 GeV. It is important to emphasize that the contour is shown for illustration, but the comparison is not precise, because for all parameter points above the relic density contour, the model under-produces dark matter. 

\item {\it Mitigation of the neutrino scattering background}. The attempt to use proton fixed target facilities, dedicated to studying neutrino physics, necessarily requires dealing with beam-related backgrounds from neutrino scattering. The signatures we have discussed mimic those of neutrinos, for which the detectors are well-suited. However, with elastic scattering event rates of ${\cal O}(10^5)$, the current precision in neutrino scattering cross sections is not sufficient to pursue simple counting experiments, Fortunately, these backgrounds can be mitigated in various ways, that were discussed in detail in the MiniBooNE proposal \cite{Dharmapalan:2012xp}. These include use of timing cuts, for heavier dark matter candidates, use of distinct scattering kinematics, or removal of the meson decay volume by running the experiment as a beam dump. The latter approach is particularly effective in reducing the number of neutrinos produced, which arise primarily from charged pion and kaon decays in the decay volume. This is the approach used by MiniBooNE in its recent beam dump run. It of course has the potential disadvantage that it cannot be carried out simultaneously with neutrino physics analyses (unless multiple detectors are used, on- or off-axis). Timing cuts, combined with knowledge of the fine structure of the pulsed beam, can be used to reduce the background significantly. A clear example would be the use of the Super-K far detector at T2K, where timing cuts could cleanly isolate dark matter scattering events from neutrinos. Scattering kinematics can also be particularly useful, and can be controlled by taking the detector off axis (see e.g.~\cite{Coloma:2015pih}). Perhaps the most efficient use is for electron scattering, where a stringent forward angle cut significantly reduces the neutrino background. 

While these strategies will clearly be important for providing competitive sensitivity, assessing the precise level will require further work for most of the facilities investigated in this paper. However, the estimates obtained in \cite{Dharmapalan:2012xp} for a run of $2\times 10^{20}$ POT at MiniBooNE, in beam dump mode, provide some benchmarks. For nucleon elastic scattering, this is likely at the ${\cal O}(100)$ event level, and we note that quasielastic $NC\pi^0$ scattering may actually provide better sensitivity, at the ${\cal O}(10)$ event level. A tight forward angle cut on electron scattering appears to be highly efficient, and could provide sensitivity at the few event level \cite{Dharmapalan:2012xp}. For the other experiments, we note that the reduced event rate at Super-K may be more than compensated for by the efficiency of timing cuts. The background to DIS at SHiP is expected to be sizeable, but remains to be studied in detail.

\end{itemize}

\section{Concluding Remarks}
\label{sec:outlook}

Astrophysical and cosmological evidence for dark matter presents a compelling argument for new physics. Thermal relic particles provide conceptually one of the simplest model classes, and there are significant efforts underway to search for its direct and indirect signatures. In this paper, we have provided further evidence that proton fixed target experiments, e.g. those associated with the long-baseline neutrino program, are well suited to testing the sub-GeV mass range for thermal relic (or WIMP) dark matter.  In the remainder of this concluding section, we comment on a number of issues impacting searches of this kind.

\begin{itemize}
\item {\it Model-independent presentation of search results}. The searches discussed in this paper necessarily consider dark matter models with light mediators, which cannot be studied in a fully model-independent manner. We have considered broad mediation classes, that are the least constrained by other data, as providing useful benchmarks. In practice, the experimental scattering signatures are chosen to mimic those of neutrinos, which provides a significant background. It follows that, in the most general terms, the signature is anomalous neutral current-like scattering, or alternatively the observation of an anomalous ratio of neutral to charged current scattering of neutrinos in the detector.\footnote{We thank Richard van de Water and Rex Tayloe for discussions of model-independent signatures.} This is a quantity that can be determined with higher precision that the actual event rate estimates for specific dark matter models, and thus will be an important output from any experimental search. Limits presented in this form can then be applied to constrain specific light DM models.

\item {\it Alternate detection strategies}. We have focussed on proton fixed target facilities as a means of probing thermal relic dark matter candidates in the MeV-GeV mass range. Other potential strategies include the use of electron beam facilities, e.g. at JLab \cite{Battaglieri:2016ggd}, which have a lower neutrino background, and also direct detection searches with lower recoil thresholds, e.g. by searching for electron scattering \cite{Essig:2011nj,Essig:2012yx} (with potential improvements shortly from XENON100). In addition, collider searches for missing energy, mass or momentum can also provide important sensitivity to this regime, and mono-photon searches at Belle II \cite{Essig:2013vha} are likely to be highly effective for the vector portal. We expect that a combination of all approaches will be valuable in exploring this range of dark matter parameter space. (See \cite{Alexander:2016aln} for a recent survey of search channels.)

\item {\it Sensitivity to other models}. Finally, it is important to emphasize that, while we have focussed on elastic and inelastic scattering signatures, these experiments will also be sensitive to a range of other signatures of light hidden sectors. For example, ``leptophilic" models ({\em e.g.} based on gauged lepton flavour $L_i$)
are also accessible at the experiments discussed here. The multitude of $\pi^0$'s produced which fragment into the photons, and further to $e^+e^-$, automatically 
creates a lepton beam being dumped into the target. This allows a probe of light states coupled to the electrons in an alternative manner to electron beams.  

\end{itemize}

\section*{Acknowledgements}

We would like to thank Brian Batell, in particular, for initial collaboration on the simulation code documented in the Appendix and related discussions. We would also like to thank Walter Bonivento, Ranjan Dharmapalan, Andrei Golutvin, Martin Hoferichter, Akira Konaka, Corina Nantais, Hiro Tanaka, Rex Tayloe, Tyler Thornton, Richard Van de Water, and other members of MiniBooNE, T2K and SHiP for many helpful discussions. The work of  C.-Y.C., P.dN., M.P. and A.R. is supported in part by NSERC, Canada, and research at the Perimeter Institute is supported in part by the Government of Canada through NSERC and by the Province of Ontario through MEDT. 

\vfil\pagebreak

\centerline{\bf APPENDIX: {\tt  BdNMC} Simulation Tools}
\vspace*{0.2cm}

\centerline{Patrick deNiverville\footnote{The software was developed by building on some initial components designed in collaboration with Brian Batell. His support, along with members of the MiniBooNE Coolaboration, is gratefully acknowledged.}}

\appendix

\section{Introduction}
\label{sec:sim_intro}

The {\tt BdNMC} software was developed to model light dark matter production and downstream scattering at a range of fixed target experiments, and can be applied flexibly to various proton beam energies, production targets and detector geometries. It can now be configured for a wide array of different fixed target geometries, and over a dozen production channels at a variety of energies. In addition, it supports four possible scattering interactions between hidden sector dark matter and the material in fixed target neutrino detectors. The code may be run in one of two ways:
\begin{enumerate}
 \item The \code{bin/BdNMC} bash script. This will compile the code and execute it using the setup described in the \code{parameter.dat} file located in the main BdNMC directory. Alternatively, the user may supply the filepaths of one or more parameter files as arguments, and the script will execute them sequentially.
 \item Compile the code with \code{make} in the \code{build} directory, and execute it as \code{./build/main} from the main BdNMC directory. Optionally, a filepath to a parameter file may be supplied as an argument. If no filepath argument is supplied, \code{parameter.dat} is used by default.
\end{enumerate}
The BdNMC code is freely available on github at \url{https://github.com/pgdeniverville/BdNMC/releases} - please cite as P.~deNiverville et al. (this paper).

In this standalone Appendix, we will document the most important features of the BdNMC software and its usage. We begin with a simple example set-up and and overview of code.  A guide to formatting the parameter files used to customize a run of the code follows in App.~\ref{sec:parameter}, with documentation of all of the options available. Following this are further sections detailing the structure of the code and the simulation loop. These are not required reading for use of the code, but should be of some assistance if you wish to introduce modifications. App.~\ref{sec:dmprod} covers the dark matter four-vector generation process, App.~\ref{sec:signal} briefly looks at the scattering signal calculation, App.~\ref{sec:simloop} provides a schematic of the simulation loop and finally App.~\ref{sec:Output} documents the formatting of the simulation's output. Some of the numerical techniques are documented in App.~\ref{sec:numerics}, and App.~\ref{sec:detector} covers modelling of the detector geometry.

\subsection{An Example Experiment}
\label{ssec:example}

Fixed target neutrino experiments generate neutrino beams by impacting high intensity proton beams into thick targets. The interactions of the protons with the nuclear material of the target produce a range of secondary particles, such as charged pions and kaons, whose decays (facilitated by an extended decay volume) produce a neutrino beam. The interactions of these neutrinos are detected some distance away through their scattering off nucleons and electrons. This setup also provides an opportunity to search for light sub-GeV dark matter, produced e.g. through neutral meson decay, with similar scattering signatures to neutrinos in the detector. We provide a very simple example setup for one such experiment in Fig.~12 and Table~II, and will refer to this setup for examples while introducing the code. Note that while we call the experiments fixed target neutrino experiments in this section, our discussion could also be applied to a beam dump experiment such as SHiP.

\begin{figure}[!b]
\label{fig:experiment_example}
\includegraphics[width=0.8\textwidth]{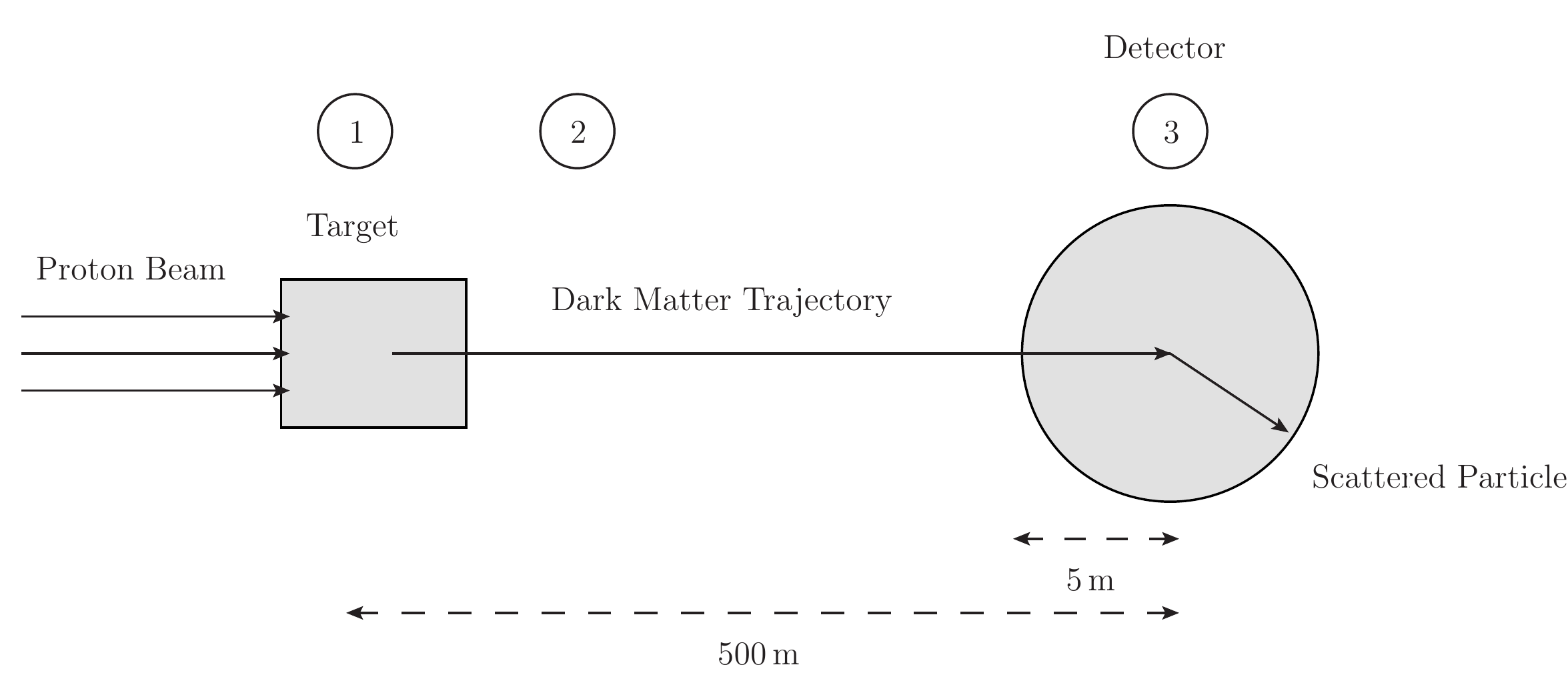}
\caption{\footnotesize A diagram of dark matter production at simplified fixed target neutrino experiment with geometry and beamline similar to that found at MiniBooNE. The detector is a sphere with a radius of 5\,m filled with CH$_2$, located 500\,m from the centre of the target. Most of the 500\,m between the target and the detector is expected to be filled with dirt and other dense materials. Dark matter and neutrinos are produced through the interactions of an 8.9 GeV proton beam impacting on the target.}
\end{figure}

\begin{table}[!thb]
\centering
\begin{tabular}{|l|l|l|l|}
\hline
\multicolumn{4}{|c|}{Parameters of the Example Experiment} \\
\hline
Protons on Target & $2\times10^{20}$ & Beam Energy & 8.9\,GeV \\
\hline
Production Channel & $\pi^0\to\chi\chi\gamma$ & $\pi^0$'s per POT & 0.9 \\
\hline
Production Distribution & Sanford-Wang & Signal Channel & $N\chi \to N\chi$\\
\hline
\end{tabular}
\label{tab:example_experiment}
\caption{Alongside Fig.~12, this table provides a minimal set of parameters describing a simple fixed target experiment and the hidden sector dark matter scenario to be tested.}
\end{table}

\begin{table}[!htb]
\centering
\begin{tabular}{|l|l|l|l|}
\hline
\multicolumn{4}{|c|}{Example Model Parameters} \\
\hline
$\epsilon$ & $10^{-3}$ & $\alpha^\prime$ & 0.1\\
\hline
$m_\chi$ & $10$\,MeV & $m_V$ & 100\,MeV \\
\hline
\end{tabular}
\label{tab:example_model}
\caption{A sample set of model parameters for use with our example experimental setup.}
\end{table}

In Sections~3 and 4, we discussed the channels via which these experiments could both produce and subsequently detect a beam of dark matter particles. In our example, interactions between the target and the proton beam could lead to production of $\pi^0$'s, whose radiative decays lead to the production of $V$'s that rapidly decay to a pair of dark matter particles. This occurs on a sufficiently short time scale that we can assume that the initial $\pi^0$ (and often the intermediate $V$) do not propagate a significant distance before decaying, and the chain of decays occurs within the target. The dark matter particles then propagate along straight-line trajectories, a small subset of which will intersect with, and scatter within, the neutrino detector producing a detectable signal.

The primary goal of the code is to calculate the number of dark matter scattering events that would be observed over the lifetime of an experiment. 
The code simulates the production and scattering of dark matter using Monte Carlo methods, sampling 
the underlying distributions (e.g. the momentum and angle distribution of the produced $\pi^0$'s, or the differential cross section of neutral current elastic dark matter nucleon scattering). 
An advantage of generating individual dark matter production and scattering events, is that the code should prove straightforward to integrate with the simulations maintained by experimental collaborations. For example, using a beamline simulation for the production of pions, instead of generating pions from a phenomenological parameterization, would allow the use of a list of pions supplied by the experimental collaboration, improving the accuracy of the signal calculation.

We now move to a high level overview of the components of the code.

\subsection{Overview}
\label{ssec:overview}

A schematic flow-diagram of the code appears in Fig.~13, and the various components are summarized below. Note that the physical DM production, propagation and detection steps 1, 2 and 3 occur at the physical positions indicated by the circled numbers in Fig.~12.

\tikzstyle{setup} = [rectangle, draw, fill=yellow!20, 
    text width=10em, text centered,minimum height=4em]
\tikzstyle{block} = [rectangle, draw, fill=blue!20, 
    text width=10em, text centered, minimum height=4em]
\tikzstyle{line} = [draw, -latex']
\tikzstyle{class} = [rectangle, draw, fill=red!40, node distance=10em,
    minimum height=3em, rounded corners]
\tikzstyle{file} = [ellipse, draw, fill=green!20, node distance=10em, minimum height=2.5em]

\begin{figure}[!h]
\centering
\begin{tikzpicture}[node distance = 2cm, auto]
	\node [setup] (main) {build/main};
	\node [setup,below of=main] (main-init) {A. Initialize run variables (App.~\ref{sec:parameter})};
	\node [class, left of=main-init] (parameter) {\code{Parameter}};
	\node [file, node distance=15em, left of=main] (parameter_card) {\code{parameter.dat}};
	
	\node [block, below of=main-init] (prodcalc) {1. Calculate production rate (Sec.~\ref{sec:prod}, App.~\ref{sec:dmprod})};

	\node [block, below of=prodcalc] (dm_gen) {2. Generate Dark Matter Particles (Sec.~\ref{sec:prod}, App.~\ref{sec:dmprod})};
	\node [block, below of=dm_gen] (scatter_gen) {3. Generate DM-Detector Interactions (Sec.~\ref{sec:scatter}, App.~\ref{sec:signal})};
	\node [block, below of=scatter_gen] (loop_condition) {Test if nevents$\geq$\\ \textbf{samplesize}};
	\node [setup, below of=loop_condition] (sum_out) {B. Write total signal (App.~\ref{sec:signal_rate})};
	\node [setup, node distance=12em, right of=loop_condition] (record) {C. Record Scattering Events (App.~\ref{sec:Output})};
	
	\node [node distance= 10em, right of=dm_gen] (dm_prod) {};
	\node [class, above of=dm_prod, node distance=2em] (dist) {\code{Distribution}};
	\node [class, below of=dm_prod, node distance=2em] (dmgen) {\code{DMGenerator}};

	\node [node distance= 10em, left of=scatter_gen] (dm_scatter) {};
	\node [class, above of=dm_scatter, node distance=2em] (det) {\code{Detector}};
	\node [class, below of=dm_scatter, node distance=2em] (scat) {\code{Scatter}};

	\node [file, node distance=5em , below of=sum_out] (sum) {\code{summary.dat}};
	\node [file, node distance=12em, right of=sum] (events) {\code{Events/events.dat}};

	\path [line] (main) -- (main-init);
	\path [line] (main-init) -- (prodcalc);
	\path [line] (prodcalc) -- (dm_gen);
	\path [line, dashed] (parameter) -- (main-init);
	\path [line, dashed] (parameter_card) -- (parameter);
	\path [line, dashed] (sum_out) -- (sum);
	\path [line, dashed] (dm_gen) -- (dist);
	\path [line, dashed] (dist) -- (dmgen);
	\path [line, dashed] (dmgen) -- (dm_gen);
	\path [line] (dm_gen) -- (scatter_gen);
	\path [line] (scatter_gen) -- (loop_condition);
	\path [line, dashed] (scatter_gen) -- (det);
	\path [line, dashed] (det) -- (scat);
	\path [line, dashed] (scat) -- (scatter_gen);
	\path [line] (loop_condition) -- node  {\code{true}} (sum_out);
	\path [line] (scatter_gen) -- (record);
	\path [line, dashed] (sum_out) -- (sum);
	\path [line, dashed] (record) -- (events);
	
	\path [line] (loop_condition) -| node [near start] {\code{false}} ([xshift=-1cm] dm_scatter.west) |- (dm_gen);
\end{tikzpicture}
\caption{Schematic outline of the simulation code.}
\end{figure}
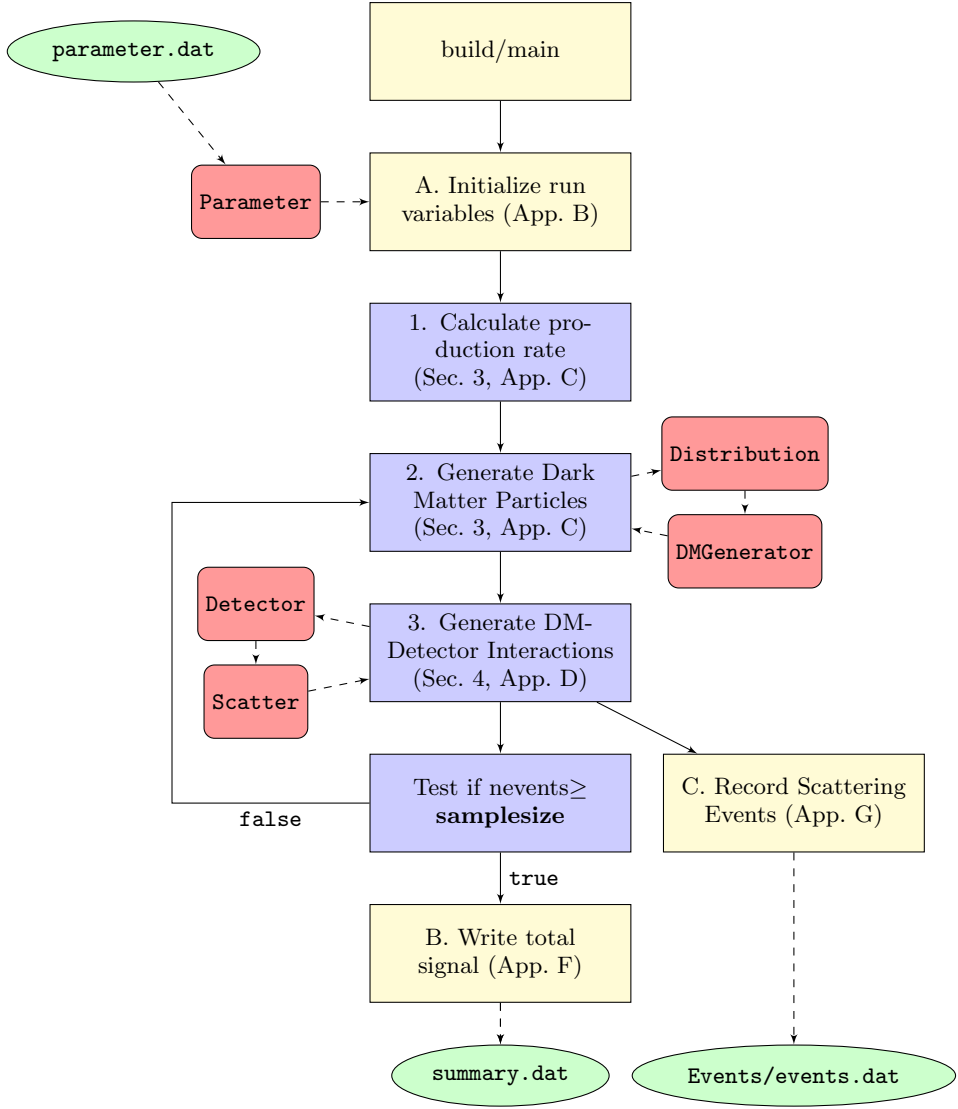

\begin{description}
 \item[A] The simulation is run by calling the \code{build/main} program. It accepts a single argument, a file path denoting the location of a parameter file. If no argument is provided, a default parameter file, \code{parameter.dat}, is loaded instead. The parameter file contains a list of parameters describing the desired run configuration. A comprehensive list of the arguments that may be provided in the parameter file will be given in App.~\ref{sec:parameter}, and an example parameter card for the example setup is provided at the end of that section. All of the valid parameter arguments are stored as variables in a \code{Parameter} object. After reading the parameter file, the code initializes variables describing the run (e.g. the number of events to generate, the output file to which events are to be written), the physics model (e.g. the mass of the dark matter candidate and vector mediator) and a \code{Detector} object. For the example experiment, the \code{Detector} would be a sphere with a radius of 5\,m located 500\,m from the center of target (taken as the coordinate origin).

 \item[1] The next step is that a list of production channels is created, each element comprising an initial momentum distribution of primary beam products, \code{Distribution}, and a \code{DMGenerator} to decay those particles into dark matter pairs. In addition, a \code{Scatter} object is initialized to handle the interactions between the dark matter and the detector material. At this time, we calculate the total number of dark matter particles that is expected to be produced for a number of Protons on Target (POT) through each production channel in the list. The exact equations differ by production channel, and are described in App.~\ref{sec:dmprod}.
 
 In the example experiment, the list of production channels only has one element, for $\pi^0$ decays. This requires creating a Sanford-Wang \code{Distribution} for an 8.9 GeV proton beam and a \code{DMGenerator} to simulate the decay chain $\pi^0\to \gamma V \to \gamma \chi \bar \chi$. The total number of dark matter particles produced would then be equal to
 \begin{equation}
   N_\chi = \mathrm{\textbf{POT}} \times \mathrm{\textbf{pi0\_per\_POT}} \times \mathrm{Br}(\pi^0 \to \chi \chi^\dagger \gamma),
 \end{equation}
where \textbf{pi0\_per\_POT} is the number of $\pi^0$'s produced per POT. For a MiniBooNE-like experiment, this is approximately 1. The \code{Scatter} object is created to handle neutral current-like elastic scattering between a nucleon and a dark matter particle with no cuts imposed on the outgoing energy and angle of the nucleon.
 
 \item[2] The simulation loop begins by selecting a production channel from the list generated in the previous step. Each channel has a probability of being chosen equal to fraction of the total number of dark matter particles that would be produced through the channel. The \code{Distribution} corresponding to that production channel generates the four-momentum of a meson or $V$ produced by the interactions of the proton beam with the target. This is accomplished via rejection sampling using a differential particle production cross-section or by iterating through a pre-generated list of particle four-momenta. This initial particle is decayed by the \code{DMGenerator}, outputting a pair of dark matter particle four-momenta. The length of the intersection between these dark matter trajectories and the detector is calculated (see App.~\ref{sec:detector}), and if either are found to be non-zero, then the initial meson or $V$ and the intersecting dark matter particle(s) four-momenta are stored in a list and the code continues to step 3. More commonly, no intersection is found, and we return to the beginning of step 2 to choose another production channel.

 In the example, there is only one production channel to choose from, $\pi^0$ decay. The \code{Distribution} generates a momentum and an emission angle for a $\pi^0$ from the Sanford-Wang distribution using rejection sampling (see App.~\ref{ssec:reject}). This initial 4-momentum is handed off to the \code{DMGenerator}, which decays this particle into a photon and a $V$, and then decays the $V$ into two dark matter particles. A list of the initial $\pi^0$ and all the products is returned. All of these processes are assumed to occur at the origin, which defaults to the centre of the target. Only the dark matter particles will propagate outward. The two dark matter trajectories are passed to the spherical \code{Detector} object's \code{Ldet} member function, which uses the algorithm described in App.~\ref{ssec:detsphere} to check for the length of the intersection between their trajectories and the detector geometry.
 
 \item[3] Once a dark matter four-momentum is found to intersect with the detector, it is passed to the \code{Scatter} object's \code{probscatter} member function. The code first determines whether a dark matter particle scatters at all by calculating the ratio of the interaction probability of this particular dark matter particle to the maximum possible interaction probability,
 \begin{equation}
  R = \frac{\sigma(E_i) \times L_i \times n}{\mathrm{MAX}(\sigma \times L \times n)},
 \end{equation}
  where $\sigma(E_i)$ is the interaction cross section for an incident dark matter particle with some energy $E_i$, $L_i$ is the length of the intersection between the dark matter trajectory and the detector, and $n$ is the number density of the particles with which the dark matter is interacting. The denominator is the largest interaction probability encountered thus far. A uniform random number $u\in[0,1]$ is generated, and if $u<R$, no interaction occurs and the dark matter particle is discarded. If an interaction does occur, then an acceptance-rejection algorithm is used to generate a final state four-momentum by sampling from a differential distribution of the interaction cross section. This particle is added to the list generated in step 2, and the number of recorded interactions is incremented.
 
  Continuing with our specific example, dark matter particles that intersect with the detector are passed on to the \code{Scatter} object, which in this case determines first whether an NCE nucleon dark matter scattering event occurs, followed by the generation of an outgoing nucleon trajectory by selecting an energy from the differential nucleon dark matter scattering cross section through rejection sampling (The scattering angle is uniquely determined by the outgoing energy). If a scattering event occurred, the outgoing nucleon is added to list of particles generated in step 2.
 
 \item[B] After the requested number of sample signal events have been generated, the total event rate is calculated as detailed in App.~\ref{sec:signal_rate}. This data, along with other details of the run described in App.~\ref{sec:Output}, is appended to the summary file.
 
 \item[C] For each successful interaction, the four-momentum of the recoil particle is recorded in an events file with the format described in App.~\ref{sec:Output}.
 
\end{description}

\section{The Parameter File}
\label{sec:parameter}

The parameters of the experimental setup and the dark matter scenario to be studied are set in a parameter file supplied to the main code at the beginning of the run. Parameters are supplied in \textbf{$<$parameter\_name$>$} \code{$<$value$>$} pairs, where valid values may be keywords, real numbers or integers. There is some basic error handling to catch improperly formatted parameters and notify the user, but as it has not been extensively tested, parameter files should be written with care. The simulation uses meters, seconds and GeV, except where otherwise noted, and all parameters should be supplied in these units. Any line that begins with \# is treated as a comment, and is ignored by the parser. Note that many parameters possess default values that are used if they are not explicitly defined.

\subsection{Metaparameters}
\label{ssec:sim_meta}

These describe details of the run itself, as well as where to record the results.

\begin{description}
	\item[burn\_max] The number of scatterings to perform during burn-in. Default is 1000.
	\item[burn\_timeout] The simulation will perform \textbf{burn\_max}$\times$\textbf{burn\_timeout} burn-in loops before terminating the burn-in phase and moving on to the simulation run. This should probably be combined with a \textbf{max\_trials} to ensure that a run will terminate in a reasonable amount of time. Default is $2\times10^4$.
	\item[max\_trials] An optional upper limit on the number of times the main simulation loop will run as counted by the \code{trials} variable. Default is -1, which causes the code to act as if there is no maximum.
	\item[output\_file] The filepath to which the event summary should be written when the simulation is set to comprehensive mode.
	\item[output\_mode] This determines what should be written to files during and after the run. Can be set to \textbf{comprehensive} or \textbf{summary}. Defaults to \textbf{comprehensive}.
	\item[run] Every run is labelled by a string. If no run name is supplied, an integer run number will be generated from the current time.
	\item[samplesize] The number of scattering events to generate before terminating the run. The code will throw an error if this is not set.
	\item[seed] Choose the seed to supply to the random number generator. A seed based on the current time is generated if no seed is supplied.
	\item[summary\_file] The filepath to which a summary of the run should be appended.
\end{description}

\subsection{Experiment setup}
\label{ssec:exp_set}

Parameters describing the target and proton beam of the experiment.

\begin{description}
	\item[atomic\_mass\_target] Not currently used for any physics.
	\item[beam\_energy] The total energy of the proton beam. Defaults to 8\,GeV.
	\item[dm\_energy\_resolution] The step size to use when creating an interpolation function for the scattering cross section. Should be varied before starting a large batch of runs to check if it sufficiently small. Defaults to 0.1\,GeV, which may not be small enough for low energy experiments like LSND.
	\item[efficiency] This is an overall factor which accounts for the percentage of signal events the experiment successfully detects and for any differences between the detector's fiducial volume and the geometry used in the simulation. Defaults to 1.0.
	\item[e\_num\_target] The number of electrons per atom in the target. Not currently used.
	\item[pi0\_per\_POT] The number of $\pi^0$'s expected per proton on target. As many of the other meson production estimates are scaled relative to that of the $\pi^0$, it is important to have a good estimate of this quantity. Defaults to 0.9, which is appropriate for MiniBooNE, with beam on-target.
	\item[max\_dm\_energy] Used in creating interpolation functions for scattering cross sections. This can be safely set equal to the beam energy as an upper limit, at the cost of slowing the initialization of the simulation slightly. Defaults to \textbf{beam\_energy}.
	\item[max,min\_scatter\_angle] Cuts on the scattering angle of the outgoing visible particle (e.g. nucleon, electron) in an interaction event. This defaults to $2.1\pi$ radians and 0.0 radians respectively, which amounts to no cut on the angle. 
	\item[max,min\_scatter\_energy] Place cuts on the energy of the outgoing visible particle (e.g. nucleon, electron) in an interaction event. Defaults to $10^9$\,GeV and 0.0\,GeV, respectively. 
	\item[n\_num\_target] The number of neutrons per atom in the target. Defaults to 0.
	\item[n\_density\_target] The number density of atoms in the target. Defaults to 0.
	\item[p\_num\_target] The number of protons per atom in the target. Defaults to 0.
	\item[POT] The total number of protons on target over the duration of the experiment. This value must be supplied or an error will be thrown.
	\item[proton\_target\_cross\_section] The total scattering cross section of protons on the target. Used to normalize the $V$ production rate for partonic $V$ production. Defaults to 0.
	\item[signal\_channel] Which scattering channel to use. Choose one from: \textbf{NCE\_nucleon}, \textbf{NCE\_nucleon\_baryonic}, \textbf{NCE\_electron} and \textbf{pion\_inelastic}\footnote{Note that coherent nucleon scattering is not currently supported in the code. This was added in a fork of the code specifically developed for a study of coherent dark matter nucleus scattering in \cite{deNiverville:2015mwa}, and was not folded back into the primary code. This should be added in a later update.}.
	\item[target\_length] The length of the target in meters. Default is 0.
	\item[timing\_cut] The length of the time delay required to register as an event. This currently calculates a timing efficiency
	\begin{align}\label{eq:timing}
 t_{\mathrm{efficiency},i} &=\left\{ \begin{array}{ll}  1 & t_{\mathrm{delay},i} \ge t_\mathrm{cut}, \\
                       \frac{t_\mathrm{cut}-t_{\mathrm{delay},i}}{t_\mathrm{cut}} & t_{\mathrm{delay},i} < t_\mathrm{cut}, \end{array} \right.
\end{align}
	where $t_{\mathrm{delay},i}$ is the delay between the travel time of a neutrino moving between the target and the detector at $c$ and a dark matter particle. The mean of \ref{eq:timing} over all dark matter particles $i$ produced by a given production channel is used to calculate the timing efficiency of that channel, and the total signal is multiplied by this efficiency to determine the event rate after timing cuts. The default value is 0.0, which results in a timing efficiency of 1.0.
\end{description}

\subsection{Model}

The model itself is determined largely by the choice of production and signal channels. A few overall parameters must be set no matter which production or signal channel is chosen.

\begin{description}
	\item[alpha\_D] The dark sector coupling strength $\alpha^\prime=\frac{{e^\prime}^2}{4 \pi}$, or in a baryonic model, $\alpha_B$.
	\item[dark\_matter\_mass] The mass of the dark matter candidate $\chi$
	\item[dark\_photon\_mass] The mass of the $V$ mediator.
	\item[epsilon] The mixing constant $\epsilon$ between the $V$ mediator and the photon.
	\item[kappa] See \textbf{epsilon}.
\end{description}

\subsection{Production Channels}
\label{ssec:prod_param}

Any number of production channels may be specified in the parameter file. A production channel declaration begins with \textbf{production\_channel} $<$\code{channel name}$>$, where the production channels currently supported have the following names (cf. Sec.~\ref{sec:prod}):
\begin{itemize}
\item \textbf{pi0\_decay}, \textbf{pi0\_decay\_baryonic},
\item \textbf{eta\_decay}, \textbf{eta\_decay\_baryonic},
\item \textbf{omega\_decay}, \textbf{omega\_decay\_baryonic},
\item \textbf{rho\_decay},
\item \textbf{phi\_decay}, \textbf{phi\_decay\_baryonic},
\item \textbf{pi-minus\_capture},
\item \textbf{parton\_production}, \textbf{parton\_production\_baryonic},
\item \textbf{V\_decay}, \textbf{V\_decay\_baryonic}.  (These channels apply for bremsstrahlung production - see below.)
\end{itemize}
Note that \textbf{rho\_decay}, \textbf{omega\_decay}, and \textbf{omega\_decay\_baryonic} should not be used with either of the proton bremsstrahlung production channels, as this would result in double counting due to the significant overlap in $V$ production sources. All non-$\pi^0$ decay channels require that \textbf{meson\_per\_pi0} be defined. This quantity is dependent both on the originator meson and the beam energy.

Following the production channel declaration, an optional production distribution can also be specified using \textbf{production\_distribution} $<$\code{distribution name}$>$. Several of these distributions accept additional parameters, which must follow the \textbf{production\_distribution} declaration in the parameter file. We now list the currently accepted distributions along with any additional parameters they accept. 
\begin{itemize}
	\item \textbf{pi0\_sanfordwang} (Utilizing the Sanford-Wang differential distribution for pion production at MiniBooNE energies)
 	\begin{description}
 	\item[distribution\_parameter\_file] This is the filepath to a parameter file listing new values for variables \textbf{c1} through \textbf{c9} and \textbf{d1} through \textbf{d9}, as well as the \textbf{Beam\_Energy} (which overrides any \textbf{beam\_energy} listed in the main parameter file) and \textbf{maximum\_pi\_momentum}. This should be formatted in the same manner as the initial parameter file. Note that the \textbf{c} parameters are for the $\pi^+$ fit, and the \textbf{d} parameters are for the $\pi^-$ fit, where the variable names correspond to those found in the MiniBooNE paper on the topic \cite{sanfordwang_miniboone}. This file is parsed in the same manner as the main parameter file and any parameter names that it does not expect will be saved with their values, but otherwise ignored.
 	\item[sanfordwang\_parameter\_file] Alternative parameter name for \textbf{distribution\_parameter\_file}.
 	\end{description}
 	\item \textbf{k0\_sanfordwang} (Utilizing the Sanford-Wang differential distribution for kaon production at MiniBooNE energies)
 	\begin{description}
 		\item[distribution\_parameter\_file] This is the filepath to a parameter file listing new values for variables \textbf{e1} through \textbf{e8} for the $K0$ fit in \cite{sanfordwang_miniboone}, where we have switched \textbf{c} from the paper for \textbf{e}. Otherwise handled identically to \textbf{pi0\_sanfordwang}.
 		\item[sanfordwang\_parameter\_file] Alternative parameter name for \textbf{distribution\_parameter\_file}.
 	\end{description}
	\item \textbf{particle\_list} (A list of particle 4-vectors in place of a production distribution)
	\begin{description}
		\item[particle\_list\_file] The file path to a text list of particle four-momenta in the format \code{px py pz E}. Optionally, each can be appended with an additional four-vector describing the space-time starting position for each particle, resulting in the format \code{px py pz E x y z t}. Note that the z-axis is oriented parallel to the beam direction.
		\item[particle\_list\_position] Supply true to indicate to the parser that particle positions should be expected. Any other argument will default to false.
	\end{description}
	\item \textbf{burmansmith} (Utilizes the Burman-Smith distribution for pion production at LSND energies) This channel requires some target parameters to be set, specifically \textbf{p\_num\_target} and \textbf{n\_num\_target}.
	\item \textbf{bmpt} (Utilizes the general pion distribution determined in \cite{Bonesini:2001iz} for a range of multi-GeV beam energies) This channel also requires \textbf{p\_num\_target} and \textbf{n\_num\_target} to be set. 
	\item \textbf{parton\_V}, \textbf{parton\_V\_baryonic} (Requires externally generated data for $V$-production at the parton level)
	\begin{description}
		\item[parton\_V\_neutron\_file] A file path to a text list of pairs of $V$ momenta and differential $V$ production cross sections $\frac{d\sigma_{pn\to V+X}}{dE_V}$ listed in microbarns per GeV. Note that the $V$ momenta must be supplied with a constant step size.
		\item[parton\_V\_proton\_file] As \textbf{parton\_V\_neutron\_file}, but for $\frac{d\sigma_{pp\to V+X}}{dE_V}$.
	\end{description}
	\item \textbf{proton\_brem}, \textbf{proton\_brem\_baryonic}. (Utilizes the bremsstrahluncg production mode discussed in Sec.~\ref{sec:prod}) These channels should be paired with \textbf{V\_decay} and \textbf{V\_decay\_baryonic} respectively, as they produce on-shell $V$'s which decay isotropically.
	\begin{description}
		\item[ptmax] The maximum transverse momentum which a produced $V$ mediator may possess. The minimum is assumed to be 0.
		\item[zmin] The minimum value of $z = \frac{p_{V,z}}{P}$, where $p_{V,z}$ is the momentum of the $V$ parallel to the $z$ axis, and $P$ is the total momentum of a beam proton incident on the target. See below for further details on choosing these parameters. 
		\item[zmax] The maximum value of $z$, defined as in the \textbf{zmin}.   
	\end{description}
\end{itemize}

\subsection{Detector Parameters}
\label{ssec:det_param}

The detector parameters describe the location, size, and material of the detector. In the current version of the code, only one detector may be declared. The detector declaration begins with \textbf{detector} $<$shape$>$, where $<$shape$>$ can be \textbf{sphere} or \textbf{cylinder}. After this command, the parameter parser will expect the dimensions of the detector to be supplied, as well as a list of materials which reflect its composition. It will accept further commands describing the detector until it encounters a non-detector variable, at which point it will return to regular parsing. For a spherical detector, the center of the detector needs to be supplied using \textbf{x-position}, \textbf{y-position}, and \textbf{z-position}, as well as the \textbf{radius}. For a cylindrical detector, the \textbf{length} must also be defined, as well as its orientation. The orientation defaults to the length of the cylinder aligned parallel to the z-axis, and it is rotated to other orientations by an angle $\theta$ by rotating about the x-axis, and then an angle $\phi$ by rotating about the y-axis, supplied using \textbf{det-theta} and \textbf{det-phi}, respectively. 

A material is defined by providing a \textbf{material} $<$name$>$, and then providing the \textbf{number\_density} in particles per cm$^3$ (note that this is one of the few areas where we do not use meters), \textbf{proton\_number}, \textbf{neutron\_number}, \textbf{electron\_number} and \textbf{mass} in GeV.

\subsection{Parameter File Example}
\label{ssec:param_example}

We provide below a simple parameter card for the MiniBooNE-like example experiment detailed in App.~\ref{ssec:example}. 

\begin{lstlisting}[language=bash,frame=single]
#Parameter Card
#All masses should be provided in GeV, all lengths in meters.
#Lines preceded by a # are ignored by the parser.

#Model Parameters
epsilon 1e-3
dark_matter_mass 0.01
dark_photon_mass 0.1
alpha_D 0.1

#Run parameters
POT 2e20
pi0_per_POT 0.9
samplesize 2000
beam_energy 8.9


#Production Parameters
production_channel pi0_decay
production_distribution pi0_sanfordwang

#Scattering Parameters
signal_channel NCE_Nucleon

output_mode comprehensive

#Where to write events and summary information. 
output_file Events/events.dat
summary_file Events/summary.dat

#Detector Parameters
detector sphere
x-position 0.0
y-position 0.0
z-position 500.0
radius 5.0

#Material parameters
material Carbon
number_density 3.63471e22
proton_number 6
neutron_number 6
electron_number 6
mass 11.2593

material Hydrogen
number_density 7.26942e22
proton_number 1
neutron_number 0
electron_number 1
mass 0.945778
\end{lstlisting}

\section{Dark Matter Production}
\label{sec:dmprod}

Dark matter production involves the generation of an initial particle from a production distribution (handled by the \code{Distribution} class), and its decay to dark matter and subsequent propagation to a detector (handled by the \code{DMGenerator} class). As multiple channels often contribute to dark matter production simultaneously, a list of production channels is prepared, each with its own \code{Distribution} and \code{DMGenerator}, and the number of dark matter particles produced by each channel, \code{vnum[i]} (see App.~\ref{ssec:dmgen}), is calculated in advance. This process can be broken into a few steps:
\begin{itemize}
 \item A production channel $i$ is chosen randomly from the previously mentioned list with a probability
 \begin{equation}
   Prob(i) = \frac{\mathrm{\code{vnum[i]}}}{\mathrm{\code{vnumtot}}},
 \end{equation}
  where \code{vnumtot}$=\sum_i$\code{vnum[i]}.
 \item An initial particle must be generated from a sample distribution that describes particle production in proton-target collisions. This is handled by the \code{Sample\_Particle} member function supplied by \code{Distribution}, which assigns a \code{Particle} object a four-momentum drawn from an internal production distribution using rejection sampling, or from a list of four-momenta supplied from a text file. This initial particle becomes the first element of a list of particles.
 \item The list of particles containing the initial particle created by the \code{Distribution} is passed to the \code{DMGen} member function of the \code{DMGenerator} object. This function will decay the initial and all subsequently produced particles until the four-momenta for dark matter particles are generated. These dark matter four momenta are checked for intersection with the detector using the \code{LDet} member function of the \code{Detector} defined during initialization. If non-zero intersection is found, all of the produced particles are added to the supplied list of particles and the function returns \code{true}, indicating that the simulation should now attempt to generate a scattering probability.
\end{itemize}

\subsubsection{The \code{Particle} Class}

The \code{Particle} class encodes all of the salient details about a particle and provides a number of methods to calculate and set useful quantities. The most important variables are:
\begin{itemize}
	\item \code{px, py, pz, E}, the four-momentum.
	\item \code{m}, the particle mass
	\item \code{name}, a string identifying the particle. During the simulation loop, the code will look for particles named ``DM'' in order to propagate them to the detector and attempt to scatter them.
	\item \code{origin\_coords, end\_coords}, two \code{double[4]} arrays which record the origin point and time of the particle and where and when it was destroyed, either by decaying or scattering. The position is stored in the elements 0 to 2, and the time in element 3. This is only truly important for the dark matter at the current time, as \code{end\_coords} records when and where in the detector it scattered.
\end{itemize}
The most important computations include:
\begin{itemize}
	\item \code{void Lorentz(Particle\& part)}, perform a Lorentz boost on the \code{Particle} from the rest frame of \code{part} to the lab frame.
	
	\item \code{void Set\_Time(double t)}, calculate the \code{end\_coords} by propagating from \code{origin\_coords} in the direction of the four-momentum of the \code{Particle} for \code{t}-\code{origin\_coords[3]} seconds.
	
	\item \code{void ThreeMomentum(double PX, double PY, double PZ)}, set the three-momentum (\code{px, py, pz}) of a \code{Particle} with mass \code{m} to the supplied momenta, and then calculates the energy \code{E}.
\end{itemize}

\subsection{Production Distributions}
\label{ssec:proddist}

\subsubsection{The \code{Distribution} Class}

All of the distributions in the code inherit the \code{Distribution} class, which requires the implementation of a single function:

\begin{itemize}
	\item \code{sample\_particle(Particle\& part)} generate a four-vector for an initial particle produced through the interactions of the proton beam and the target, ideally according to a distribution function or a supplied list of four-vectors. The implementation of this is entirely dependent on the distribution selected. An estimate of the maximum value of the distribution is required for rejection-sampling, and as this is normally not known before run-time, each distribution will perform a short burn-in run by generating $\sim 10^4$ sample particles upon initialization. The \code{sample\_particle} method automatically updates its estimate of the maximum value of the production cross section each time it generates an acceptable four-momentum\footnote{This does not apply to Particle Lists, which simply iterate through the list of particle 4-momenta that have been supplied.}.
\end{itemize}

In addition, as it is not guaranteed that a $\code{Distribution}$ object will know the mass of the particle it is generating, this can be indicated at a later time by using the \code{set\_mass(double)} member function.

\subsubsection{Pion Distributions}

The Sanford Wang and BMPT distributions both generate a pair of candidate angle $\theta_i$ and momentum $p_i$ variables using a uniform random number drawn from the kinematically allowed regime. The differential production cross section $\sin(\theta)\frac{d^2\sigma}{d\theta dp_\pi}(\theta, p_\pi)$ is compared with ${\rm MAX}\left(\sin(\theta)\frac{d^2\sigma}{d\theta dp_\pi}\right)$ for use in rejection sampling. Once a particle has been accepted, an angle $\phi$ is generated by drawing a uniform random variable from the range $[0, 2\pi)$. The Burman-Smith distribution is used in a similar manner, but samples $\pi$ kinetic energy instead of momentum.

\subsubsection{Direct (partonic and bremsstrahlung) V production}

The partonic $V$ distribution produces $V$ mediators directly from quark level interactions in the colliding nuclei rather than through intermediate particles. This calculation requires knowledge of parton distribution functions that are not included in the simulation, as outlined in Sec.~\ref{ssec:parton}, following \cite{deNiverville:2012ij}. The required input comprises two csv files containing data points of $\frac{d\sigma_{pN\to V}}{dp_V}$, for proton-proton and proton-neutron collisions with a given beam energy and $V$ mass. The \code{parton\_V\_gen} class constructs interpolation functions based on these files, and uses acceptance rejection sampling to generate $V$ momenta. The total $V$ production cross section $\sigma_{pN\to V+X}$ can be calculated by summing over the numerical data. As described earlier in Sec.~\ref{ssec:parton}, at tree-level each $V$ is created with a momentum parallel to the proton momentum (i.e. along the z-axis).

The \textbf{proton\_brem} distribution samples $V$ four-momenta from the proton bremsstrahlung distribution (\ref{eq:bremrate}), where the maximum value is known to be 
\begin{equation}
	\mathrm{MAX}\left(\frac{d^2 N_V}{dz dp^2_\perp}(z,p_\perp)\right)=\frac{d^2 N_V}{dz dp^2_\perp}(\mathrm{\textbf{zmin}},0).
\end{equation} 

\subsubsection{Particle Lists}

One final option is to use an explicit particle list. Rather than sampling from a distribution, this reads a list of four-vectors from a text file, and stores it in a \code{list}. Each time \code{sample\_particle} is called, it iterates one step through the list, constructs a particle based on the four-momentum, and returns it. A position and timing distribution can also be supplied by appending a four-vector to each four-momentum in the supplied text file. This is especially useful when particles are supplied by a beamline simulation, as different portions of the target and surrounding materials are likely to exhibit different momentum distributions.

\subsection{Dark Matter Generation}
\label{ssec:dmgen}

\subsubsection{The \code{DMGenerator} Class}

Once a \code{Distribution} has returned an initial particle, a \code{DMGenerator} object is tasked with simulating the chain of decays which will result in the production of dark matter. Each dark matter production channel is implemented through a class derived from \code{DMGenerator}. The most important member variables of \code{DMGenerator} are:
\begin{itemize}
	\item \code{mv, mx, kappa, alphaD}, the parameters describing the hidden sector dark matter scenario.
	\item \code{chan\_name}, a string storing the name of the production channel. As an example, the $\pi^0\to\gamma\chi\bar\chi$ channel is called ``pi0\_decay''.
	\item \code{branchingratio}, the probability that the decay process will occur. For direct $V$ production channels this will be nearly 1 so long as invisible decays dominate.
\end{itemize}

Classes that are derived from \code{DMGenerator} must implement two member functions:

\begin{itemize}
	\item \code{bool DMGen(list<Particle>\& plist, function<double(Particle)> det\_int, Particle\& inital\_part)} uses the \code{inital\_part} supplied by \code{Distribution} to generate a pair of dark matter particles through a chain of decays. The produced dark matter particles (always a pair in the current code) are checked for intersection with the detector by \code{det\_int}, the function \code{Ldet} supplied by the \code{detector} class (see App.~\ref{sec:detector}). If a non-zero intersection is recorded, all of the particles involved in the production of the dark matter particle are saved, and the function returns \code{true}.
	\item \code{void Evaluate\_Branching\_Ratio()} calculates the branching ratio for the production channel and stores it in \code{branchingratio}. This is called automatically on initialization of \code{DMGenerator}.
\end{itemize}

The \code{DMGenerator} must be able to calculate the branching ratio for dark matter production, though the exact chain of processes is highly dependent on the production channel. Each production distribution is responsible for returning a branching ratio or some equivalent value which can be used to calculate \code{vnum}, the total number of $V$ mediators produced by \textbf{POT} protons that will decay into dark matter particles. If the $V$ is produced on-shell, we have
\begin{equation}
	\mathrm{\code{vnum}} = N_V \times \mathrm{Br}(V \to \chi \chi^\dagger),
\end{equation}
where $N_V$ is the number of $V$ bosons generated. This is generalized appropriately for decays involving of-shell mediators.

\subsubsection{Meson Decays}

The most thoroughly studied production channel is that of pseudoscalar meson decay, in which a pseudoscalar meson $X$ decays radiatively as $X\to \chi \bar \chi + \gamma$. These production channels should be paired with one of the meson production distributions: Sanford-Wang, BMPT or Burman-Smith. For the on-shell case, in which $m_V>2 m_\chi$ and $m_V<m_X$, the decay is performed in two steps, first by decaying the meson, $X\to V + \gamma$, and then decaying the mediator, $V \to \chi \bar \chi$. The total branching ratio is a product of that for the two sub-processes
\begin{equation}
{\rm Br}(X\to \chi \bar \chi + \gamma) = {\rm Br}(X \to V + \gamma){\rm Br}(V \to \chi \bar \chi),
\end{equation}
where 
${\rm Br}(V \to \chi \bar \chi) \approx 1$ for large $\alpha^\prime$. The total number of on-shell $V$ particles, \code{vnum}, is given by
\begin{equation}
\label{eq:nv_meson}
 	\mathrm{\code{vnum}} = {\rm Br}(X \to \chi \bar \chi \gamma) \times \mathrm{\textbf{meson\_per\_pi0}} \times \mathrm{\textbf{pi0\_per\_POT}} \times  \mathrm{\textbf{POT}},
\end{equation}
where $\mathrm{\textbf{meson\_per\_pi0}}=1$ for $\pi^0$. Each decay is isotropic in the parent particle's rest frame, and the momentum of the daughter particles is determined by
\begin{equation}
\label{eq:trianglefunc}
	\lambda(m_1,m_2,m_3) = \frac{1}{2 m_1} \sqrt{m_1^4+m_2^4+m_3^4-2 m_1^2 m_2^2 - 2 m_2^2 m_3^2 - 2 m_3^2 m_1^2},
\end{equation}
where $m_1$ is the parent particle, and $m_{2,3}$ are the daughter particles. After performing the $X\to V +\gamma$ decay, the $V$ is Lorentz boosted from the $X$ rest frame to the lab frame, and this process is repeated for the $V$ and its daughter $\chi \bar \chi$ particles.

For off-shell decays, the branching ratio ${\rm Br}(X\to \chi \bar \chi + \gamma)$ does not possess an analytical form, and must be integrated numerically. The branching ratio is calculated as
\begin{equation}
\int_{4 m_\chi^2}^{m_X^2} ds \frac{d {\rm Br}}{ds}(X \to V^* + \gamma \to \chi \bar \chi + \gamma),
\end{equation}
where $s$ is the center of mass energy of the $V$, and the integral is performed using Sinh-Tanh quadrature. At this point it should be noted that the previous paragraph on the handling of the on-shell case was not quite correct, as near threshold the off-shell contributions become competitive with the on-shell expression. As a result, the complete off-shell branching ratio is always calculated, and if off-shell corrections are found to be sufficiently large (as defined by ${\rm Br}(X\to \chi \bar \chi + \gamma) \ge 1.3 \times {\rm Br}(X \to V + \gamma){\rm Br}(V \to \chi \bar \chi)$), the off-shell machinery will be used. We also note that, in this approximation, the mesons are treated as elementary but a form factor could be incorporated to account for the virtuality dependence.

The simulation of the off-shell decay uses rejection sampling on $\frac{d^2 {\rm Br}}{ds d\theta}(X \to V^* + \gamma \to \chi \bar \chi + \gamma)$ to generate an ($s$, $\theta$) pair, where $\theta$ is the angle between the $\chi$ and the z-axis in the rest frame of the $V$. The z-axis of this rest frame is aligned such that it is parallel to the $V$'s momentum in the $X$'s rest frame. In order to boost to the $X$ rest frame, we first boost along the z-axis to match the $V^*$ momentum, and then rotate the z-axis to be parallel to the $V$ momentum in the $X$ rest frame.

The simulation of vector meson mixing is quite similar to that of the pseudoscalar meson decays, but the calculation of \code{vnum} is slightly different,
\begin{equation}
\label{eq:nv_meson2}
 	\mathrm{\code{vnum}} = {\rm Br}(X \to \chi \bar \chi) \times \mathrm{\textbf{meson\_per\_pi0}} \times \mathrm{\textbf{pi0\_per\_POT}}   \times \mathrm{\textbf{POT}}.
\end{equation}
The kinematics of this process are simulated as an off-shell vector meson $X$ oscillating into an on-shell $V$, which then decays normally into a $\chi \bar \chi$ pair. This channel is normally paired with a pion production distribution to generate a somewhat reasonable set of three-momenta from which to sample. Note that this channel overlaps with, and has now been replaced by, bremsstrahlung production which also receives a significant contribution from resonant vector meson mixing.

\subsubsection{Direct (partonic and bremsstrahlung) V Production}

The handling of partonic $V$ production is quite a bit simpler, as we have only to decay the vector particle supplied by the distribution. The only complication is that the decay is not isotropic, but is instead drawn from\footnote{In the case of fermionic dark matter, this would be $g\left(\hat \theta \right) = \frac{3}{8}\left(1+\cos^2\hat \theta\right)$.}
\begin{equation}
	g(\hat \theta) = \frac{3}{4}(1-\cos^2(\hat \theta)),
\end{equation}
where $\hat \theta$ is the angle between the dark matter momentum and the beam direction. The $V$ production rate is calculated using the assumption that experiments are designed to have the majority of their beam protons interact with target material, and so $N_V$ should be proportional to the ratio of the total proton-material cross section to the $V$ production cross section,
\begin{equation}
\label{eq:part_vnum}
N_V = \frac{\sigma_{pA\to V+X}}{\sigma_{pA}} \times \mathrm{\textbf{POT}},
\end{equation}
where $\sigma_{pA}$ is the total proton scattering cross section with material of mass number $A$.

Proton bremsstrahlung is handled in a similar manner, but the $V$ mediators are decayed to dark matter isotropically. The total number of $V$'s produced is calculated as
\begin{equation}
\label{eq:brem_vnum}
N_V = \mathrm{\textbf{POT}} \int_0^{\mathrm{\textbf{ptmax}}^2} dp_\perp^2 \int_{\mathrm{\textbf{zmin}}}^{\mathrm{\textbf{zmax}}} dz  \frac{d^2 N_V}{dz dp^2_\perp},
\end{equation}
where $\frac{d^2 N_V}{dz dp^2_\perp}$ is defined in (\ref{eq:bremrate}).

\section{Signal Channels}
\label{sec:signal}

Once a dark matter particle's four-momentum has been found to intersect with the experiment's detector, it is passed on to the \code{Scatter} object through the \code{probscatter} member function. The \code{Scatter} object determines whether an interaction occurs, and if so, generates the final state particles produced by the scattering.

\subsubsection{The \code{Scatter} class}

Dark matter signal calculations are handled by the \code{Scatter} class. Each signal channel is implemented as a class derived from \code{Scatter}, and they all share the following variables 

\begin{itemize}
	\item \code{MDP, mdm, alD, kap}\footnote{This naming convention is not consistent with \code{DMGenerator}.}, the model parameters.
	\item \code{pMax}, the maximum scattering probability encountered.
	\item \code{Escatmax, Escatmin, min\_angle, max\_angle}, kinematic cuts on outgoing visible particles (e.g. nucleon, electron). The range of acceptable angles defaults to $[0,2\pi)$ unless otherwise indicated.
\end{itemize}

Scattering is handled by:

\begin{itemize}
	\item \code{probscatter(shared\_ptr<detector>\& det, list<Particle>\& partlist, list<Particle>::iterator\& DMit)}, where \code{det} is a pointer to a \code{detector} object, \code{partlist} is the list of particles produced by \code{DMGenerator} (see App.~\ref{ssec:dmgen}) and \code{DMit} is an iterator pointing to the dark matter particle that is being checked for scattering. Should a scattering occur, a final state particle is generated and inserted into \code{partlist} as the element following the dark matter. More details are provided below.
\end{itemize}

For NCE nucleon scattering and inelastic $\pi^0$ production, the total interaction cross section includes one or more integrals which must be evaluated numerically. Upon initialization these \code{Scatter} objects will generate an interpolation function of its interaction cross section on both neutrons and protons over $E_\chi\in[0,$\textbf{max\_dm\_energy}$]$, where $E_\chi$ is the incident dark matter energy. Any cuts on the recoil energy for the NCE nucleon channels are implemented as limits on the range of integration during this step. This is unnecessary for NCE electron scattering, as a closed form exists for its scattering cross section. In addition, an interpolation function for the maximum value of the differential cross section is found for both neutron and proton scattering over the same incident dark matter energy range using the minimization algorithm described in App.~\ref{ssec:minimization}. 

The probability that an elastic scattering will occur for dark matter particle $\chi_i$ is given by 
\begin{equation}
\label{eq:probscat}
p_i = \sigma_{\chi N,e \to \chi X}(E_{\chi,i})\times L_{{\rm Det},i} \times n_{N,e},
\end{equation}
where $\sigma_{\chi N,e \to \chi X}$ is the integrated interaction cross section, $L_{{\rm Det},i}$ is as defined in (\ref{eq:detlength}) and $n_{N,e}$ is the number density of nucleons or electrons. The formula for inelastic pion production is similar,
\begin{equation}
\label{eq:probscat_pion}
p_i =\frac{2}{3} \sigma_{\chi N \to \chi \Delta}(E_{\chi,i})\times L_{{\rm Det},i} \times n_{N},
\end{equation}
where the factor of $2/3$ reflects the branching ratio of $\Delta\to\pi^0$ as opposed to $\Delta\to\pi^+$. An interaction is deemed to have occurred if the inequality $p_i > u \times$\code{pmax}, where as before $u$ is a uniform random number drawn from $[0,1]$.

Once a scattering has occurred, a final state particle observable by the detector must be generated. In every signal channel, an outgoing energy is generated through rejection sampling on the differential scattering cross section, with the envelope set by the maximum value of the differential cross section. This maximum is calculated for all possible dark matter energies during initialization. Should it pass, a scattering angle $\theta$ relative to the incoming dark matter's momentum is generated for the scattered particle, followed by $\phi \in [0,2\pi)$. While this particle is ready for output in the case of elastic scattering, there is one additional step in the inelastic $\pi^0$ channel: the decay $\Delta\to\pi^0\gamma$. This is handled in the same manner as the $\pi^0\to\gamma V$ decay channel, but with different masses for the parent and massive daughter particles. Note that all scatterings are handled as if the dark matter was travelling parallel to the $z$-axis, all output particles are rotated to the lab frame after the scattering occurs.

For the elastic scattering channels, the particles generated through this process are guaranteed to pass any energy cuts as those cuts are included in the calculation of the total scattering cross section. This is not the case for inelastic $\pi^0$ production, as the end state is not as deterministic, and so any cuts on the outgoing $\pi^0$ momentum must be checked separately. In all cases, angular scattering cuts are checked afterwards by the \code{main} code before being accepted as scattering events.

\section{The Simulation Loop}
\label{sec:simloop}

Before the simulation loop proper begins, a burn-in run is conducted for each production channel to estimate \code{pmax}. This is very similar to the simulation loop, but the end state interaction results are not generated, and the variable \code{nburn} is incremented each time. The burn-in run is conducted until \code{burntrials} is equal to \code{BURNMAX}. We provide a schematic view of the simulation loop here.

\begin{enumerate}
	\item Set \code{trials}=0, \code{nevents}=0, \code{ninteractions = zeros(chan\_count)}.
	\item While \code{nevent} $\leq$ \code{samplesize}:
	\begin{enumerate}
		\item \code{trials}++.    
		\item Generate a uniform random number \code{vrnd} $\in [0,$\code{vnumtot}$]$.
		\item Set $i=0$, \code{scatterswitch=False}.
		\item While $i<$ \code{chan\_count}:
		\begin{enumerate}
			\item If \code{vrnd} $<$ \code{vnum$[i]$} then break
			\item Else set \code{vrnd} = \code{vrnd} - \code{vnum$[i]$}.
		\end{enumerate}
		\item Initialize \code{Particle} \code{part} with four-momentum \code{p}=0.
		\item Set the four-momentum of \code{part} using \code{Distribution}'s \code{Sample\_Particle} method.
		\item Initialize an empty \code{list} of \code{Particles} \code{partlist}. 
		\item Append \code{part} to \code{partlist}.
		\item Generate a list of decay product \code{Particles} of \code{part} using \code{DMGenerator}$_i$'s \code{DMGen} member function and store them in \code{partlist}.
		\item For each dark matter \code{Particle} $j$ in \code{partlist}:
		\begin{enumerate}
			\item If \code{Ldet}$_j == 0$ then continue
			\item Else
			\begin{enumerate} 
				\item Simulate an interaction using \code{Scatter}'s \code{probscatter} method.
				\item If probscatter returns true, then insert the end state \code{Particle} generated by \code{probscatter} after $j$ in \code{partlist}, and set \code{scatterswitch=True}.
			\end{enumerate}
		\end{enumerate}
		\item If \code{scatterswitch}, then write all the particles in \code{partlist} to \textbf{output\_file}, \code{nevent}++ and \code{ninteraction}$_i$++.
	\end{enumerate}
	\item end while loop
\end{enumerate}

\section{Calculating the Number of Signal Events}
\label{sec:signal_rate}

Signal events are particle interactions which pass all experimental cuts and are successfully identified (the latter condition is represented by efficiency factors). The number of signal events reported by the code is an estimate of the total number of signal events that the experiment would observe given some number of protons on target, \textbf{POT}. The number of signal events is different from the number of sample events requested, as these are mere representative events generated to estimate the total signal. The signal rate from a given production channel $i$ is calculated as
\begin{equation}
	\mathrm{\code{signal\_events}}[i] = \frac{\mathrm{\code{ninteractions}}[i]}{\mathrm{\code{trials}}} \times \mathrm{\code{vnumtot}} \times \mathrm{\code{pmax}} \times \mathrm{\textbf{efficiency}},
\end{equation}
where \code{ninteractions} and \code{trials} are discussed in App.~\ref{sec:simloop}, \code{vnumtot} is found in App.~\ref{ssec:dmgen}, \code{pmax} is discussed in App.~\ref{sec:signal} and \textbf{efficiency} is a parameter mentioned in \ref{ssec:exp_set}. The total signal rate is given by the sum
\begin{equation}
	\mathrm{\code{total\_signal\_events}} = \sum_i^{\mathrm{\code{chan\_count}}-1} \mathrm{\code{signal\_events}}[i].
\end{equation}
This total signal number is required to estimate an experiment's discovery potential for a given new physics scenario.

\section{Output}
\label{sec:Output}

In all modes, the simulation will append a summary of the results to the file path supplied for \textbf{summary\_file} after completing a run. Each entry begins with \code{Run} \textbf{run}, and is followed by a number of lines equal to \code{chan\_count} with the results for each channel in the format: 

\vspace{0.1cm}
\textbf{channel\_name} \textbf{dark\_photon\_mass} \textbf{dark\_matter\_mass} \code{\code{signal\_events}}$[i]$ \textbf{epsilon} \textbf{alpha\_D} \textbf{signal\_channel} \textbf{POT} \textbf{Efficiency} \textbf{samplesize} \code{vnum}$[i]$

\vspace{0.1cm}
The summary output terminates with one final line with the format:
\vspace{0.1cm}

\code{Total} \textbf{dark\_photon\_mass} \textbf{dark\_matter\_mass} \code{total\_signal\_events} \textbf{epsilon} \textbf{alpha\_D} \textbf{signal\_channel} \textbf{POT} \textbf{Efficiency} \textbf{samplesize} \code{vnumtot}
\vspace{0.1cm}

The output expected for our example experiment from App.~\ref{ssec:example} is provided here:

\begin{lstlisting}[language=bash,frame=single,basicstyle=\footnotesize]
Run 1470677009
pi0_decay 0.1 0.01 4137.09 0.001 0.1 NCE_nucleon 2e+20 1 2000
Total 0.1 0.01 4137.09 0.001 0.1 NCE_nucleon 2e+20 1 2000 
\end{lstlisting}

As a further example, we show a few runs for NCE nucleon scattering with \textbf{alpha\_D}=0.1, \textbf{epsilon}=$10^{-3}$, \textbf{dark\_matter\_mass}=5\,MeV and \textbf{dark\_photon\_mass}=0.4\,GeV:

\begin{lstlisting}[language=bash,frame=single,basicstyle=\footnotesize]
Run 1446004457
pi0_decay 0.4 0.005 0.00365517 0.001 0.1 NCE_nucleon 2e+20 1 5000
eta_decay 0.4 0.005 0.996236 0.001 0.1 NCE_nucleon 2e+20 1 5000
proton_berm 0.4 0.005 0.0154329 0.001 0.1 NCE_nucleon 2e+20 1 5000
Total 0.4 0.005 1.01532 0.001 0.1 NCE_nucleon 2e+20 1 5000
Run 1446086551
pi0_decay 0.4 0.005 0.00195789 0.001 0.1 NCE_nucleon 2e+20 1 5000
eta_decay 0.4 0.005 0.93881 0.001 0.1 NCE_nucleon 2e+20 1 5000
proton_brem 0.4 0.005 3.95397 0.001 0.1 NCE_nucleon 2e+20 1 5000
Total 0.4 0.005 4.89474 0.001 0.1 NCE_nucleon 2e+20 1 5000
Run 1446086747
pi0_decay 0.4 0.005 0.00120486 0.001 0.1 NCE_nucleon 2e+20 1 5000
eta_decay 0.4 0.005 0.490378 0.001 0.1 NCE_nucleon 2e+20 1 5000
proton_brem 0.4 0.005 1.51652 0.001 0.1 NCE_nucleon 2e+20 1 5000
Total 0.4 0.005 2.0081 0.001 0.1 NCE_nucleon 2e+20 1 5000
\end{lstlisting}

In \textbf{summary} mode, the summary file is the only output produced, discounting various status reports written to the terminal. While in \textbf{comprehensive} mode, the code also writes a report of all the particles which took part in each event to \textbf{output\_file}. This file begins with the run name in the format \code{Run} \textbf{run}. Each event output is preceded by \code{event nevent}, where \code{nevent} is the number of events at the time which it was recorded. Next, each particle in the \code{partlist} for the given event writes a report of its four-momentum and, if it was the final endstate particle, the position where the signal interaction occurred in the format [\code{particle\_name px py pz E x y z t}] where \code{x y z t} are the optional position and time coordinates only shown for the signal particle. The particle reports are followed by the line \code{endevent nevent}.

Once again, we provide a sample output:

\begin{lstlisting}[language=bash,frame=single,basicstyle=\footnotesize]
Run 1462311643
event 1
pion           0.285398       -0.0544237     5.54369        5.55294        
V              0.204215       -0.0118137     4.14206        4.14832        
DM             -0.00377294    -0.00148947    0.658276       0.658307       
proton         -0.185011      0.235048       0.134491       0.993941       
-2.82029       -1.11338       492.064        1.64143e-06    

endevent 1

event 2
pion           0.0252685      0.113951       1.771          1.77997        
V              0.039564       0.105479       1.71365        1.72026        
DM             -0.00213871    -0.000317347   0.733531       0.733551       
proton         0.205311       -0.208192      0.118041       0.989841       
-1.42464       -0.21139       488.618        1.6299e-06     

endevent 2

event 3
pion           -0.0597918     -0.0244405     5.17368        5.17585        
V              -0.0199172     -0.0266187     4.24083        4.24214        
DM             -0.0127331     -0.0416011     4.10137        4.1016         
neutron        -0.305361      -0.15944       0.0761656      1.00362        
-1.53362       -5.01059       493.984        1.64785e-06    

endevent 3
\end{lstlisting}

In this example, two of the scattering events occurred between dark matter and a proton, while the last one was a rare scattering off a neutron. All three originated with relatively low energy $\pi^0$'s.



\section{Numerical Techniques}
\label{sec:numerics}

\subsection{Rejection Sampling}
\label{ssec:reject}

The primary method used to sample cross-sections or momentum distributions is rejection sampling, also known as acceptance-rejection (see e.g. \cite{casella}). To draw samples from a distribution $f(\mathbf{x})$, we find an easily sampled enveloping distribution $g(\mathbf{x})$ for which
\begin{equation}
\label{eq:reject1}
g(\mathbf{x}) \ge \epsilon f(\mathbf{x})\, \forall \mathbf{x},
\end{equation}
where $\epsilon$ is some real number chosen such that this condition is satisfied. We then execute the following loop:
\begin{enumerate}
\item Generate a uniform random number $u \in [0,1]$ and an array of random numbers $\mathbf{x}$ from $g$.
\item If $u> \epsilon f(\mathbf{x})/g(\mathbf{x})$, return $\mathbf{x}$.
\item Else goto to step 1.
\end{enumerate}
The simplest version of this algorithm, which is employed by the simulation code, is to set $g(\mathbf{x}) = {\rm MAX}\left(f(\mathbf{x})\right)$, which allows us to generate $\mathbf{x}$'s using uniform random numbers over the domain of $f$. While crude, this sampling regime is quite effective for distributions of low dimensionality that are not sharply peaked. The code does sometimes encounter issues with peaked distributions, and a more adaptive sampling algorithm could be implemented to tackle them more efficiently if required, but it has thus far not been a significant issue.

In our example experiment, we use rejection sampling to generate a momentum $p_{\pi^0}$ and an emission angle $\theta_{\pi^0}$ for a $\pi^0$ created in the target. For the Sanford-Wang distribution, this would be done by generating a random $p_{\pi^0}\in[0,7]\,\mathrm{GeV}$ and $\theta_{\pi^0} \in [0,\pi/2]\,\mathrm{rad}$, as well as a uniform random number $u\in[0,1]$. We then check if
\begin{equation}
\label{eq:reject2}
SW(p_{\pi^0},\theta_{\pi^0})>u\times SW_\mathrm{max},
\end{equation}
where $SW_\mathrm{max}$ is the maximum possible value of the Sanford-Wang distribution. If this condition holds true, the $p_{\pi^0}$ and $\theta_{\pi^0}$ that were found are returned\footnote{We may also want to generate an azimuthal angle $\phi\in[0,2\pi]$ at this point}. If it fails, one starts again with a new set of numbers $p_{\pi^0}$, $\theta_{\pi^0}$, and $u$, and repeats until \ref{eq:reject2} is satisfied.

\subsection{Function Extrema}
\label{ssec:minimization}

Rejection-sampling requires knowledge of the extrema of the sampled functions. This is often difficult to predict ahead of time. For example, the maximum value of the dark matter interaction with the detector is dependent on detector geometry, position and composition, beam energy, dark matter production mode and distribution, and experimental cuts on the recoil momentum and angle of an outgoing particle. In these cases the relevant maxima are initially set to zero, and then updated over time. A short burn-in period, where numbers, particles, or entire scattering events are generated but not used, precedes regular running to obtain a realistic estimate of the maximum of the distribution.

There are other cases, particularly in sampling the differential scattering cross sections, where this is far less feasible, as the differential scattering cross section's shape and maximum vary wildly with the incident dark matter energy. For one dimensional distributions such as differential cross sections $\frac{d\sigma}{dE_R}$, the simulation code employs Golden Section Search \cite{press_etal:1992}. This is not the fastest routine available, but it is straightforward, makes few assumptions about the shape of the function, and is sufficiently efficient to avoid noticeably slowing the simulation. Note that this is a minimization algorithm, so we will be minimizing the value of $f(x=E_R) = - \frac{d\sigma}{dE_R}$, and setting $f(x)=0$ for all $x$ outside of the kinematically allowed domain of the differential scattering cross section. 

In our example case, we could use this algorithm to find the maximum value of the differential NCE nucleon dark matter scattering cross section\footnote{For this cross section, the maximum always occurs for minimum recoil, so while it is possible to determine numerically, it is not necessary. This algorithm is much more useful for NCE electron dark matter scattering or inelastic $\pi^0$ production, which both possess minima not located at the extreme of the kinematically allowed regime.}. This maximum value can then be used with rejection sampling by \code{Scatter} to generate a recoil energy for a scattered nucleon.

It should be noted that the minimum must be bracketed before a Golden Section minimization can be attempted, that is, we find two points $a$ and $c$ for which there exists some $b$ such that $a<b<c$ and $f(b)<f(a),f(c)$. The bracketing algorithm accepts two valid $x$ values, and then begins taking ever growing steps downhill until it successfully brackets the minimum, using a parabolic fit to target likely points for a minimum, or failing that, to choose the size of the next step. The golden section search itself takes the three points returned by the bracketing algorithm and repeatedly finds new brackets for the minimum by bisecting the search region into a pair of golden sections. The algorithm terminates once it has narrowed the range to a set tolerance. These algorithms only find a local minimum, but as the functions of interest are all concave and possess only a single minimum, this is not an issue.

\subsection{Integration Techniques}
\label{ssec:int_tech}

There is frequently a need to perform numerical integrals for functions which are difficult or impossible to integrate analytically. For example, we do not possess an analytic expression for the total NCE nucleon-dark matter cross section,
\begin{equation}
\sigma_{N\chi\to N\chi}(E_\chi) = \int dE_r \frac{d\sigma_{N\chi\to N\chi}}{dE_r}(E_\chi),
\end{equation}
where $E_r$ is the recoil energy of the nucleon and $E_\chi$ is the energy of the incident dark matter particle. The total cross section is required to calculate the probability of dark matter interacting with the neutrino detector.

One-dimensional integrals in the code are handled using Sinh-Tanh Quadrature \cite{Bailey:2010aj}, also called double exponential integration. This integration scheme is particularly good for dealing with singularities at the edges of the integration regime, which were encountered when performing integrals to calculate ${\rm Br}(\pi^0 \to V^*_B \gamma \to \chi \bar \chi \gamma)$. The technique remaps an integral of a function $f(x)$ on $x \in [-1,1]$ to an integral on $g \in (-\infty, \infty)$, where $g(x) = \tanh(\pi/2 \sinh)$, and then breaking it into an Euler-Maclaurin sum. This is quite effective, largely because the transformation tends to reshape the integrand into a bell curve. The change of variable goes as follows
\begin{equation}
	F = \int_{-1}^1 f(x)dx = \int_{-\infty}^\infty f(g(t)) g^\prime(t) dt \approx h \sum_{j=-N}^N w_j f(x_j),
\end{equation}
where $x_j = g(hj)$ and 
\begin{equation}
w_j = g^\prime(hj) = \frac{\frac{h\pi}{2}  \cosh(hj)}{\cosh^2\left(\frac{\pi}{2}\sinh(jh)\right)}.
\end{equation}
In the simulation code itself, this is adapted somewhat to increase $N$ and decrease $h$ until a specified fractional accuracy goal is achieved. In the code we continue to iterate until $\left|\frac{F_{n+1}-F_n}{F_{n+1}}\right| \le 10^{-3}$, where $n$ and $n+1$ indicate two sequential estimates of the integral.

Two dimensional integrals are handled using Simpson Cubature (see \cite{Engeln-Mullges:1996:NAC:232512}). The algorithm itself is a long series of sums and not terribly enlightening, and so will not be reproduced here. This algorithm is currently only used to calculate the total production rate of $V$'s in proton bremsstrahlung.

\section{The Detector Class}
\label{sec:detector}

The \code{detector} class exists to hold details about the detector's physical characteristics (the parameters discussed in App.~\ref{ssec:det_param}), and to return the length of the intersection between a ray representing a particle's trajectory and the detector geometry, called \code{Ldet}. We discuss the algorithms in detail for each case in the following sections.

\subsection{Spherical Geometry}
\label{ssec:detsphere}

For this algorithm, $\mathbf{p}_\chi$ is the dark matter three-momentum, $R_{\rm det}$ is the radius of the fiducial volume\footnote{This defines the volume in which to accept signal events. Events originating too close to the outer shell of the detector are often difficult to resolve efficiently.} of the detector and 
\begin{equation}
\mathbf{o} = \mathbf{r}_{\rm det} - \mathbf{r}_\chi, 
\end{equation}
where $\mathbf{r}_{\rm det}$ is a vector pointing from the origin to the center of the detector, and $\mathbf{r}_\chi$ is a vector from the origin to the creation point of the dark matter particle. If a position distribution is not supplied to modify the momentum distribution, then $\mathbf{r}_\chi = \mathbf{0}$. We also define a few scalar quantities
\begin{align}
A =& \mathbf{p}_\chi \cdot \mathbf{p}_\chi,\\
B =& -2 \mathbf{o} \cdot \mathbf{p}_\chi,\\
C =& \mathbf{o} \cdot \mathbf{o} - R_{\rm det}^2.
\end{align}
These scalars are the coefficients of the quadratic equation $A x^2 + B x + C = 0$ that is solved to determine the points of intersection between the dark matter trajectory and the spherical shell of the detector. These intersection points exist if the following conditions are satisfied:
\begin{align}
&B^2 - 4AC >0,\\
&A\ne0
\end{align}

We now define
\begin{equation}
L_{\pm} = \frac{-B\pm\sqrt{B^2-4AC}}{2A},
\end{equation} 
where $L_{-(+)}$ is the entrance (exit) point along the dark matter trajectory. These points are stored for later use in generating an event position inside the detector. One final ambiguity exists in that the particle's trajectory extends both forwards and backwards in time. If an intersecting particle's three-momentum were reversed, this algorithm would still find an intersection. We can check for this by verifying that $L_\pm \ge 0$. If either is less than zero, we set it to 0. The length of the intersection is given by
\begin{equation}
\label{eq:detlength}
L_{\rm det} = (L_+ - L_-)\sqrt{A}.
\end{equation}

\subsection{Cylindrical Geometry}
\label{ssec:detcylinder}

The geometry of a cylindrical detector is more complicated to handle, but uses many of the tricks employed in App.~\ref{ssec:detsphere}. When creating a cylindrical detector object, a new vector $\mathbf{l}$ is stored which points from the center of the detector to the center of one of the circular faces,
\begin{align}
l_x &= {\rm length}/2 \cos(\mathrm{\textbf{det-phi}}) \sin(\mathrm{\textbf{det-theta}}),\\
l_y &= {\rm length}/2 \sin(\mathrm{\textbf{det-phi}}) \sin(\mathrm{\textbf{det-theta}}),\\
l_z &= {\rm length}/2 \cos(\mathrm{\textbf{det-theta}}),
\end{align} 
where we are using the variables defined in App.~\ref{ssec:det_param}.

There are four possible intersection points for a line through a cylinder: one through each face and two possible crossings on the circular surface. We will check each in turn, and append each intersection found to a list of crossing points $L$. The algorithm terminates and returns a length of intersection as soon as two points of intersection are found.

We begin by checking for intersection with the faces of the cylinder. This amounts to calculating the intersection point with a plane parallel to the face, and then checking that it is within $R_{\rm det}$ of the center of the face. If $\mathbf{p}_\chi \cdot l = 0$, then the trajectory is parallel to the cylinder faces, and this step can be skipped. The points of intersection with planes parallel to the cylinder are given by
\begin{equation}
B_\pm = \frac{(\mathbf{o}+\mathbf{l}) \cdot \mathbf{l}}{\mathbf{p}_\chi \cdot \mathbf{l}}.
\end{equation}
We next find the vector pointing from the center of each circular face to the point of intersection on the plane
\begin{equation}
r_{\rm intersect,\pm} = B_\pm \mathbf{p}_\chi \mp \mathbf{l} - \mathbf{o},
\end{equation}
and check that $r_{\rm intersect,\pm}^2 \le R_{\rm det}^2$. If so, then $B_\pm$ is added to the list of intersection points.

Calculating the points of intersection with the circular surface is very similar to the spherical case, in that we define three coefficients of a quadratic equation
\begin{align}
	X =& -(\mathbf{p}_\chi \cdot \mathbf{l})^2 + \mathbf{p}_\chi^2 \mathbf{l}^2,\\
	Y=& 2(\mathbf{p}_\chi \cdot \mathbf{l}) (\mathbf{o} \cdot \mathbf{l}) -2 (\mathbf{o} \cdot \mathbf{p}_\chi) \mathbf{l}^2,\\
	Z =& -(\mathbf{o} \cdot \mathbf{l})^2 - R_{\rm det}^2 \mathbf{l}^2+\mathbf{o}^2 \mathbf{l}^2,
\end{align}
and solving the quadratic equation $XC^2+YC+Z=0$ one finds $C_\pm = \frac{-Y\pm \sqrt{Y^2-4XZ}}{2X}$. If the argument of the square root is imaginary, no intersection occurs and we skip to the next step. To check that the intersection occurred in the region of the surface bounded by the two circular faces, we define
\begin{equation}
	D_\pm = \frac{C_\pm \mathbf{p}_\chi \cdot \mathbf{l} - \mathbf{o} \cdot \mathbf{l}}{\mathbf{l}^2}.
\end{equation}
If $|D_\pm|<1$, then an intersection occurs, and the point $C_\pm$ is added to the list of intersection points.

The final step of this process is almost identical to that of the spherical case. Each member of $L$ is set to 0 if negative and the entrance and exit points are defined as
\begin{align}
L_{-} =& {\rm MIN}(L[0],L[1]),\\
L_{+} =& {\rm MAX}(L[0],L[1]).
\end{align}
Substituting $L_\pm$ into (\ref{eq:detlength}) returns the length of intersection.

\subsection{Generating Interaction Positions}
\label{ssec:pos_gen}

The distribution of positions of interaction events within the detector can also be useful information experimentally. As these dark matter events occur very rarely, the event rate is roughly equal at any given point along the trajectory, and an interaction point can be generated by selecting a random number $u \in [L_-,L_+]$. The four-vector of the position and time of the interaction event can then be calculated using
\begin{align}
t_{\rm int} &= \frac{u\|\mathbf{p}_\chi\|}{\|\mathbf{v}_\chi\|} + t_0,\\
\mathbf{r}_{\rm int} &= (t-t_0)\mathbf{v}_\chi + \mathbf{r}_\chi,
\end{align}
where $t_0$ is the creation time of the dark matter particle and $\mathbf{v}_\chi$ the velocity.

\subsection*{Concluding Remarks}

This concludes the documentation for the {\tt BdNMC} software package. The code is available for download from github at \url{https://github.com/pgdeniverville/BdNMC/releases}, where further updates will be available. The code is maintained by Patrick deNiverville (pgdeniverville@gmail.com).

\bibliography{NCpi}
\end{document}